\documentclass[aps,prd,showpacs,amssymb,floatfix,nofootinbib]{revtex4}
\usepackage{graphicx,amsmath,subfigure}
\usepackage[usenames]{color}

\newcommand {\PN}{{\cal PN}}
\newcommand {\MBH}{{\cal{M}_{\bullet}}}
\newcommand {\Msun}{\ensuremath{M_{\odot}}}

\newcommand {\Nstar}{\ensuremath{{\cal{N}_{\star}}}}

\newcommand {\RS}{\ensuremath{R_{\mathrm{S}}}}
\newcommand {\kms}{\ensuremath{\mathrm{km\,s}^{-1}}}

\newcommand {\nb}{{\sc Nbody }}

\def\be{\begin{equation}}
\def\ee{\end{equation}}
\def\bea{\begin{eqnarray}}
\def\eea{\end{eqnarray}}
\def\ltsima{$\; \buildrel < \over \sim \;$}
\def\simlt{\lower.5ex\hbox{\ltsima}}
\def\gtsima{$\; \buildrel > \over \sim \;$}
\def\simgt{\lower.5ex\hbox{\gtsima}}
\def\etal{{\it et al.}}

\def\lesssim{\mathrel{\hbox{\rlap{\hbox{\lower4pt\hbox{$\sim$}}}\hbox{$<$}}}}
\def\gtrsim{\mathrel{\hbox{\rlap{\hbox{\lower4pt\hbox{$\sim$}}}\hbox{$>$}}}}
\def\alt{\mathrel{\hbox{\rlap{\hbox{\lower4pt\hbox{$\sim$}}}\hbox{$<$}}}}
\def\agt{\mathrel{\hbox{\rlap{\hbox{\lower4pt\hbox{$\sim$}}}\hbox{$>$}}}}
\def\PRD{{\it Phys. Rev.} D~}
\def\Msun{M_\odot}
\def\PR{{\it Phys. Rev.}}
\def\CQG{{\it Class. Quantum Grav.}}

\newcommand {\trlx}{\ensuremath{t_\mathrm{rlx}}}
\newcommand {\tcoll}{\ensuremath{t_\mathrm{coll}}}
\newcommand {\tGW}{\ensuremath{\tau_\mathrm{GW}}}
\newcommand {\SgrA}{{Sgr~A$^\ast$}}
\newcommand{\beq}{\begin{equation}}
\newcommand{\eeq}{\end{equation}}
\newcommand{\beqs}{\begin{equation}\begin{split}}
\newcommand{\eeqs}{\end{split}\end{equation}}
\newcommand{\partd}[2]{\frac{\partial{#1}}{\partial{#2}}}

\newcommand{\erf}[1]{(\ref{#1})}

\def\gta{\ifmmode {\mathbin{\lower 3pt\hbox   
    {$\,\rlap{\raise 5pt\hbox{$\char'076$}}\mathchar"7218\,$}}}
    \else {${\mathbin{\lower 3pt\hbox
    {$\rlap{\raise 5pt\hbox{$\char'076$}}\mathchar"7218\,$}}}
    $}\fi}
\def\lta{\ifmmode {\,\mathbin{\lower 3pt\hbox   
    {$\,\rlap{\raise 5pt\hbox{$\char'074$}}\mathchar"7218\,$}}}
    \else {${\mathbin{\lower 3pt\hbox
    {$\rlap{\raise 5pt\hbox{$\char'074$}}\mathchar"7218\,$}}}
    $}\fi}

\begin{document}

\author{
Pau Amaro-Seoane$^{1}$, 
Jonathan R. Gair$^{2}$, 
Marc Freitag$^{2}$,\\
M. Coleman Miller$^{3}$,
Ilya Mandel$^{4}$, 
Curt J. Cutler$^{5}$,   
Stanislav Babak$^{1}$
}

\affiliation{
$^{1}$Max-Planck Institut f{\"u}r Gravitationsphysik (Albert Einstein-Institut),
Am M{\"u}hlenberg 1,  D-14476 Potsdam, Germany\\ 
$^{2}$Institute of Astronomy, Madingley Road, CB3 0HA Cambridge, UK\\
$^{3}$Department of Astronomy, University of Maryland, College Park, MD 20742-2421, USA
$^{4}$Theoretical Astrophysics, California Institute of Technology, Pasadena, CA 91125, USA\\
$^{5}$Jet Propulsion Laboratory, California Institute of Technology, Pasadena, CA 91109, USA\\
}

\email{pau@aei.mpg.de, jrg23@cam.ac.uk, freitag@cam.ac.uk, miller@astro.umd.edu,
ilya@caltech.edu, curt.j.cutler@jpl.nasa.gov, stba@aei.mpg.de}

\date{\today}

\title{Intermediate and Extreme Mass-Ratio Inspirals --\\
Astrophysics, Science Applications and Detection using LISA}

\begin{abstract}

Black hole binaries with extreme ($\gtrsim 10^4:1$) or intermediate ($\sim
10^2-10^4:1$) mass ratios are among the most interesting gravitational
wave sources that are expected to be detected by the proposed Laser
Interferometer Space Antenna. These sources have the potential to tell us
much about astrophysics, but are also of unique importance for testing
aspects of the general theory of relativity in the strong field regime.
Here we discuss these sources from the perspectives of astrophysics, data
analysis, and applications to testing general relativity, providing both a
description of the current state of knowledge and an outline of some of
the outstanding questions that still need to be addressed. This review
grew out of discussions at a workshop in September 2006 hosted by the
Albert Einstein Institute in Golm, Germany.
\end{abstract}


\maketitle

\tableofcontents
\newpage

\section{Background}
\label{sec.background}


Our understanding of the central regions of galaxies has advanced rapidly during the
past few years, not least due to major advances in high angular resolution
instrumentation at a variety of wavelengths. Observations carried out
with space-borne telescopes, such as the Hubble Space Telescope (HST), and from
the ground, using adaptive optics, have allowed the study of the kinematics of
stars or gas in regions reaching down to milli-pc for the Milky Way and to sub-pc scales for more distant galaxies. One remarkable conclusion is that dark
compact objects, most probably massive black holes (MBHs), with a mass $\MBH
\simeq 10^6-10^9\,\Msun$, are present at the centre of most of the galaxies for which
such observations can be made.  A deep link exists between the central MBH and
the host galaxy. This is exemplified by the discovery of
correlations between $\MBH$ and global properties of the spheroid, the tightest correlation being with
its velocity dispersion, the so-called $\MBH-\sigma$ relation
\citep{FM00,TremaineEtAl02}. The central part of a galaxy, its {\em nucleus},
consists of a cluster of a few $10^7$ to a few $10^8$ stars surrounding the
MBH, with a size of a few pc. The nucleus is understandably expected to play a major
role in the interaction between the MBH and the host galaxy. In the nucleus,
one finds stellar densities in excess of $10^6\,{\rm pc}^{-3}$ and relative
velocities of order a few $100\,\kms$ to a few $1000\,\kms$. In these exceptional
conditions and unlike anywhere else in the bulk of the galaxy, collisional
effects come into play. These include 2-body relaxation, i.e., mutual
gravitational deflections, and genuine contact collisions.

The stars and the MBH interact in two primary ways. Firstly, stars can produce gas
to be accreted on to the MBH, through normal stellar evolution, collisions or
disruptions of stars by the strong central tidal field. These processes may
contribute significantly to the mass of the MBH \citep{MCD91,FB02b}. Tidal
disruptions trigger phases of bright accretion that may reveal the presence of
a MBH in an otherwise quiescent, possibly very distant, galaxy
\citep{Hills75,GezariEtal03}.  Secondly, stars can be swallowed whole if they
are kicked directly through the horizon (referred to as {\em direct plunges}) or
inspiral gradually due to the emission of gravitational waves (GWs). The
latter process, known as an ``{\em Extreme Mass Ratio Inspiral}'' (EMRI) is one
of the main sources expected for the future space-borne GW detector LISA (Laser
Interferometer Space Antenna) \citep{LISA,Thorne98}.  For the last stages of an EMRI
to produce GWs in the frequency domain to which LISA will be most sensitive,
i.e., $0.1$mHz--$100\,$mHz \citep{LHH00}, the mass of the MBH must be
between $\sim 10^4\,\Msun$ and $\sim 10^7\,\Msun$.  Only compact
stars, i.e., white dwarfs, neutron stars, stellar mass black holes or, possibly, the
Helium cores of giant stars can produce EMRI signals detectable at extra-galactic
distances. Main-sequence stars are either not compact enough to withstand the
tidal forces in the vicinity of the MBH or not massive enough to produce waves
of large enough amplitude. Predictions for the expected number of EMRI
detections that LISA will make are rather uncertain but lie in the
range of a few to a few thousand.


On the other hand, numerical simulations of young dense clusters show that
runaway collisions due to mass segregation can produce central stars with
masses $\sim 10^{2-4}\,M_\odot$
\citep{PortegiesZwartMcMillan00,GFR04,PortegiesZwartEtAl04, FreitagEtAl06}.
Such a star might undergo collapse and form a so-called intermediate-mass black
hole (IMBH) with $M\sim 10^{2-4} \Msun$. It has also been proposed that
globular clusters can capture the compact remnant of a zero-metallicity
population III star \citep{AmaroSeoaneEtAl06}.  A cluster harbouring an
IMBH which starts relatively close to the central MBH of the host galaxy will
sink to the centre in a few million years, and will eventually release its central
IMBH due to tidal stripping of the cluster \citep{EbisuzakiEtAl01}.  A
first-order estimate of the event rate of this process leads to a few
detectable coalescences of IMBHs with MBHs per year in the universe
\citep{Miller05} but even if only one of these events occurs during the LISA mission, the
signal-to-noise ratio by the end of the inspiral would be so high
\citep{Hughes01,CutlerThorne02,GHK02} that it would be visible in a time-frequency spectrogram of the LISA data, without having to resort to matched filtering
\citep{Miller05}. The mass ratio of such a merger would typically be $10^{3-4}:1$ --we shall
refer to it as an ``{\em Intermediate Mass Ratio Inspiral}'' (IMRI) for obvious
reasons.

The LISA mission is scheduled to fly in about 10 yrs and critical design
choices which will affect the ability to detect E/IMRIs will be made soon. It
is important to produce robust estimates for the rates and typical orbital
parameters of these events as input for the development of search algorithms. Such search algorithms must have the capability to extract science information out of the complex LISA data stream, which will contain many thousands of overlapping resolvable signals, plus astrophysical backgrounds from millions of more distant sources.. The readiness of data analysis for the LISA mission
will be assessed on an equal footing with the hardware so it is essential that data analysis strategies are finalised in the near future.

Detection of EMRIs with LISA is difficult (in discussions of data analysis, we will
generally use the term EMRIs to refer to both EMRIs and IMRIs since most of the
discussion applies to both types of inspirals, except for those obvious cases
when we discuss differences between the two). A typical signal will be very weak,
lying buried in instrumental noise and in the gravitational wave foreground
created by nearby Galactic white dwarf binaries. The signals are long-lived,
typically being observable for several years prior to plunge, which in
principle allows the EMRIs to be detected by matched filtering. Matched
filtering employs a bank of templates that describes signals with all possible
parameters within the expected range. Unfortunately, the large parameter space
of possible EMRIs makes the number of templates required for such a search
computationally prohibitive. Over the past few years, several alternative
algorithms have been developed --- a semi-coherent matched filtering algorithm,
time-frequency search algorithms and Markov Chain Monte Carlo techniques. The
results are promising, but more work needs to be done before we will have an
optimal algorithm for EMRI detection. The correspondingly higher signal-to-noise ratios of IMRIs make detection of those events somewhat easier, but it is still a challenging task.

The structure of this paper is as follows. In Section~\ref{EMRIAstro}, we
present the various astrophysical circumstances that might lead to the
formation of EMRI sources. The simplest and most studied situation is that of a
dense, spherical galactic nucleus where 2-body relaxation brings compact stars
on to very eccentric orbits around a MBH. Gravitational wave emission may
shrink the orbit and create an EMRI but relaxation can hinder this evolution by
increasing the pericentre distance. We present this standard picture and
several processes that complicate it and make the EMRI rate rather uncertain.
We then mention several other EMRI and IMRI-formation channels which have been
recently proposed but remain to be investigated in more detail.

Predictions for the rates and characteristics of EMRIs rely heavily on numerical
modelling of dynamics of dense stellar systems. In Section~\ref{EMRINum}, we
explain the methods that are in use or have been used in stellar dynamical
studies applied to EMRIs. We outline the limitations of these approaches and
suggest avenues for future numerical work in that field.


In Section~\ref{EMRIDA} we describe the existing algorithms and
outstanding challenges for EMRI/IMRI detection.  The search algorithms require
models of EMRI waveforms. In principle, the extreme mass ratio means that the
waveforms can be computed using black hole perturbation theory. However, this
formalism is not fully developed and will be computationally expensive once it
is. Various alternative waveform models are currently being developed and used
for scoping out EMRI detection. We will describe these various models and
necessary future developments of them in Section~\ref{EMRIModel}.

If we do detect many EMRI/IMRI events, we will be able to do some very
interesting science. EMRI observations will provide measurements of the masses
and spins of black holes to an accuracy which is not accessible by
other astronomical observations. This will tell us about the properties and growth of
black holes in the nearby universe. EMRIs also provide a means to probe general
relativity in the strong-field regime close to astrophysical black holes. The
extreme mass ratio ensures that over many orbits the inspiralling object acts
essentially like a test-body moving in the space-time of the central body. The
emitted gravitational waves encode a map of the space-time. If we can decode
that map, then we will be able to test the belief that massive compact objects
in the centres of galaxies are indeed Kerr black holes. Carrying out this
mapping is difficult, but is the focus of much current research. We summarise
current results in Section~\ref{testGR} and discuss some outstanding questions
in this area.

In Section~\ref{EMRIScience} we discuss the scientific benefits of LISA
observations of EMRI/IMRI events. This section takes the form of answers
to five broad questions that were the focus of discussions at the LISA
EMRI workshop hosted by the Albert Einstein Institute in Golm, Germany in
September, 2006. Finally, in Section~\ref{summ} we provide a summary of
the main topics in the paper.

\section{EMRI Astrophysics}\label{EMRIAstro}

Astrophysical scenarios producing EMRI events and their relative rates are currently rather uncertain, but multiple physical channels are being explored. We will begin by discussing the
``standard" picture of these events, in which a single $10\,M_\odot$ black hole
spirals into a supermassive black hole via gravitational wave emission, and the
only mechanism by which the angular momentum is changed is through two-body relaxation.
Even within this restricted picture there are substantial unknowns about the
rates.  We follow this by discussing possible modifications of this basic
picture, including the effects of resonant relaxation, triaxiality, and mass
segregation.  We then consider different ``non-standard'' processes that can lead to
EMRIs, including tidal separation of black hole binaries and formation or
capture of massive stars in accretion discs.

Our focus will then switch to IMRIs. If even a single such event is observed with
LISA, the strength of the signal could be enough to yield unique information
about strong gravity, in a relatively model-independent way \cite{Miller05}.
However, given that the existence of IMBHs has not yet been established with
certainty, the unknowns for this scenario are even more daunting than they are
for EMRIs.  In the second part of this section we discuss the current state of
knowledge about IMBHs and IMRIs, and outline problems that need to be addressed in the future.

The third portion of this section deals with numerical methods that are being
used to simulate the interactions of black holes in dense nuclear regions. So
far only approximate methods using many simplifying assumptions have been
used to estimate the rates and characteristics of inspirals of stellar-mass
objects into a MBH. We review these approaches, their accomplishments and
limitations. Thanks to the rapid increase in computer power and the
development of new algorithms, it is likely that the direct $N$-body approach
will soon be able to confirm or disprove these approximate results and extend
them. However, exceptionally long and accurate integrations are needed to
account correctly for secular effects such as resonant relaxation or Kozai
oscillations.  These requirements pose new challenges to developers of $N$-body
codes.

\subsection{Inspirals of stellar objects into MBHs}

\subsubsection{The standard picture}
\label{sec.standard_picture}

The first detailed calculations of EMRI rates were performed by \citeauthor{HB95} \cite{HB95} and \citeauthor{SR97} \cite{SR97}.

Their simple picture involved the following assumptions:

\begin{enumerate}

\item The stellar-mass compact objects are treated as single objects.

\item The distribution of stars is spherically symmetric.

\item The only mechanism that can change the angular momentum of the compact objects is
two-body relaxation.

\item There is no mass segregation, hence the compact objects are distributed
spatially the same way as the stars.

\item There is no gas in the nucleus.

\end{enumerate}

With these assumptions,  inside the ``radius of influence"

\begin{equation}
r_{\rm infl}=\frac{G\MBH}{\sigma_0^2}\approx 1~{\rm pc}\left(\frac{\MBH}{10^6\,M_\odot}\right)
\left(\frac{60~{\rm km/s}}{\sigma_0}\right)^2,
\end{equation}

\noindent
within which the central MBH dominates the gravitational field, the relaxation time is

\begin{equation}
t_{\rm rlx}(r)=\frac{0.339}{\ln\Lambda}\frac{\sigma^3(r)}{G^2\langle m\rangle
m_{\rm CO}n(r)}
\simeq 1.8\times 10^8\,{\rm yr}\left(\frac{\sigma}{100\,\kms}\right)^3
\left(\frac{10\,\Msun}{m_{\rm CO}}\right)
\left(\frac{10^6\,\Msun{\rm pc}^{-3}}{\langle m\rangle n}\right).
\end{equation}

Here $\sigma(r)$ is the local velocity dispersion. It is approximately equal to the Keplerian
orbital speed $\sqrt{G\MBH r^{-1}}$ for $r<r_{\rm infl}$ and has
a value $\approx\sigma_0$ outside of it. $n(r)$ is the local number density of
stars, $\langle m\rangle$ is the average stellar mass, $m_{\rm CO}$ is the mass
of the compact object (we take a standard $m_{\rm CO}=10\,M_\odot$ for
stellar-mass black holes), and 

\begin{equation}
\ln \Lambda \equiv \ln\,(p_{\rm max}/p_0)
\label{eq.lambda}
\end{equation}

\noindent
is the Coulomb logarithm. Here $p_{\max}$ is an upper limit for the impact
parameter $p$; $p_0= G\langle m\rangle \sigma^{-2}$ is the value that
corresponds to a deflection angle of $\pi/2$ for stars of masses $\langle
m\rangle$ and relative velocity $\sigma$ \citep{BT87,Spitzer87}. For a
self-gravitating stellar system, $\Lambda\equiv p_{\rm max}/p_0 \simeq \gamma
\Nstar$ where $\Nstar$ is the number of stars and
$\gamma$ is in the range $0.01-0.1$ depending on the mass spectrum
\citep{GH96,Henon75}.  In the vicinity of the BH ($r<r_{\rm infl}$),
$\Lambda \approx \MBH / m_{\star}$ \citep{BW76,LS77}. In any case,
$\ln\Lambda\sim 10$.  
{For typical density profiles, $t_{\rm rlx}$ decreases
slowly with decreasing $r$ inside $r_{\rm infl}$. It should be noted
that the exchange of energy between stars of different masses
---sometimes refered to as dynamical friction in the case of one or a
few massive bodies in a field of much lighter objects--- occurs on a
timescale shorter than $t_{\rm rlx}$ by a factor of roughly $M/\langle
m\rangle$, where $M$ is the mass of a heavy body (e.g.\
\cite{GFR04,FASK06} and references therein).}

{ 
Relaxation redistributes orbital energy amongst stellar-mass objects
until the most massive of them (presumably stellar-mass black holes)
form a power-law density cusp, $n(r)\propto r^{-\alpha}$ with $\alpha
\simeq 1.75$ around the MBH, while less massive species arrange 
themselves into a shallower 
profile, with $\alpha \simeq 1.4-1.5$
\cite{BW76,LS77,DS83,FB02b,ASFS04,BME04a,PMS04,FASK06,HA06b,MS06} (see 
also Section~\ref{sec.EMRIStatMethods}). Nuclei likely to host MBHs in the
LISA mass range ($\MBH\lesssim \mbox{few}\times 10^6\,\Msun$) probably
have relaxation times comparable to or less than a Hubble time, so that
it is expected that their heavier stars form a steep cusp, although
nothing is known about the presence of Bahcall-Wolf cusps in galaxies
aside from the Milky Way (see discussion in
Section~\ref{sec:astro-uncertain}).}

In a spherical potential, at any given time the stars and compact objects in
the nucleus simply orbit the MBH with their semi-major axes and eccentricities
changing slowly, owing to 2-body relaxation. For an EMRI to occur, in this
standard picture, 2-body relaxation has to bring a compact remnant on to an
orbit with such a small pericentre distance that dissipation of energy by
emission of GWs becomes significant.

If the object is on a very eccentric orbit but one for which the timescale for passage through periapse, $t_{\rm peri}\simeq (1-e)^{3/2}P$, is less than $\sim 10^4$s, the source will generate bursts of gravitational radiation in the LISA band each time the object passes through periapse. However, such GW signals consist of bursts which can probably not benefit from coherent signal processing even if they repeat with a periodicity shorter than LISA mission duration. Only if they reside at the Milky Way centre is there a
non vanishing probability for LISA to detect such sources \cite{HFL07}. An extra-galactic source is only likely to be detectable if it radiates continuously in the LISA band. As a rough guide, therefore, a detectable EMRI source must have an orbital frequency higher than about $f_{\rm LISA}=10^{-4}\,$Hz, corresponding to the condition on the semi-major axis $a\lesssim 0.5\,{\rm AU}\, (f_{\rm LISA}/10^{-4}\,{\rm Hz})^{-2/3} (\MBH/10^6\,\Msun)^{1/3}$. As there is no sharp cut-off in the predicted LISA sensitivity curve at $10^{-4}\,$Hz, a strong source might be detectable at a lower frequency.

Not all objects with an inspiral time by GW emission shorter than a Hubble
time will end up as EMRIs. This is because, although relaxation can increase
the eccentricity of an object to very high values, it can also perturb the
orbit back to a more circular one for which GW emission is completely
negligible.  Typically, neglecting GW emission, it takes a time of order $t_{\rm
rlx}\ln (1-e)$ for an orbit to reach a (large) eccentricity $e$ through the
effects of 2-body relaxation. However, the pericentre distance $R_{\rm
p}=a(1-e)$ can be significantly altered by relaxation on a timescale $t_{\rm
rel,p} \simeq (1-e)t_{\rm rlx}$, so the condition for a star to become an EMRI
is that it moves onto an orbit for which the timescale for orbital decay by GW emission,
$\tGW$ (see Eq.~\ref{eq.tGW}) is sufficiently shorter than $(1-e)t_{\rm rlx}$.
If the semi-major axis of the orbit is too large, this condition cannot be
obeyed unless the star actually finds itself on an unstable, plunging orbit,
with $e\ge e_{\rm pl}(a) \equiv 1-4\RS/a$ where $\RS$ is the Schwarzschild
radius of the MBH. The very short burst of gravitational radiation emitted
during a plunge through the horizon can only be detected if originating from
the Galactic centre \cite{HFL07}. Coherent integration of the GW signal for
$>10^4$ cycles with a frequency in LISA band is required for detection of
extragalactic EMRIs. 


The situation for EMRI production in the standard picture is more complicated
than that of tidal disruptions by the MBH (e.g., \cite{Rees88,MT99,SU99,WM04})
or GW bursts from stars on very eccentric orbits \cite{RHBF06,HFL07} because
these processes require a single passage within a well-defined distance $R_{\rm
enc}$ from the MBH to be ``successful''. In such cases, at any distance from
the centre and for any given modulus of the velocity, there exists a ``loss
cone'' inside which the velocity vector of a star has to point for it to pass
within $R_{\rm enc}$ of the MBH \cite{FR76,BW77,LS77,AS01}. In contrast, an
EMRI is a progressive process which will only be successful (as a potential
source for LISA) if the stellar object experiences a very large number of
successive dissipative close encounters with the MBHs \cite{AH03}. There is no
well-defined loss cone for such a situation.

As described above, a source becomes an EMRI when the orbital
period becomes shorter than about $10^4\,$s.  It seems unlikely that the
evolution of such a tight orbit can be significantly affected by other stars or
the ambient gas. It is not so at earlier stages of the inspiral as 2-body
relaxation, experienced mostly at apocentre can easily induce a change in the
pericentre distance large enough to either render GW emission completely
insignificant or, on the contrary, cause a sudden plunge into the MBH
\cite{HB95,HA05}. The condition for successful inspiral is not that the
pericentre distance must be sufficiently small, like for tidal disruptions or
GW bursts, but that the timescale for orbit evolution by emission of GWs (see
Eq.~\ref{eq.tGW}) is sufficiently shorter than the timescale over which 2-body
relaxation can affect the pericentre distance significantly,

\begin{equation}
\tGW < C_{\rm EMRI}\, (1-e)\trlx.
\label{eq.EMRIcond}
\end{equation}

\noindent 
What ``sufficiently shorter'' means is the crux of the problem and is encoded in $C_{\rm EMRI}$, a ``safety'' numerical constant that makes this condition
sufficient ($C_{\rm EMRI}<1$). For a given semi-major axis $a$, one can define
a critical eccentricity $\tilde{e}(a)$ above which GW emission dominates over
orbital evolution due to relaxation and a corresponding time scale
$\tilde{\tau}(a)\equiv \tGW(\tilde{e},a)\equiv C_{\rm EMRI}(1-\tilde{e})\trlx$.
Plunging orbits have $e\ge e_{\rm pl}(a) \equiv 1-4\RS/a$ so EMRIs (as opposed
to direct plunges) can only happen if $e_{\rm pl}(a)>\tilde{e}(a)$. This
defines a critical semi-major axis which is a typical value for an EMRI at the
moment orbital evolution starts being dominated by GW emission,

\begin{equation}
a_{\rm EMRI} 
= 5.3\times 10^{-2}\,{\rm pc}\,C_{\rm EMRI}^{2/3} \times 
\left(\frac{\trlx}{10^9\,{\rm yr}}\right)^{2/3} 
\left(\frac{m}{10\,\Msun}\right)^{2/3}
\left(\frac{\MBH}{10^6\,\Msun}\right)^{-1/3}.
\label{eq.aEMRI}
\end{equation}

\noindent
The corresponding eccentricity is given by 

\begin{equation}
1-e_{\rm EMRI} 
= 7.2\times 10^{-6}\,C_{\rm EMRI}^{-2/3} \times 
\left(\frac{\trlx}{10^9\,{\rm yr}}\right)^{-2/3} 
\left(\frac{m}{10\,\Msun}\right)^{-2/3}
\left(\frac{\MBH}{10^6\,\Msun}\right)^{4/3}.
\end{equation}

\begin{figure}
{\includegraphics[width=7in,angle=0]
{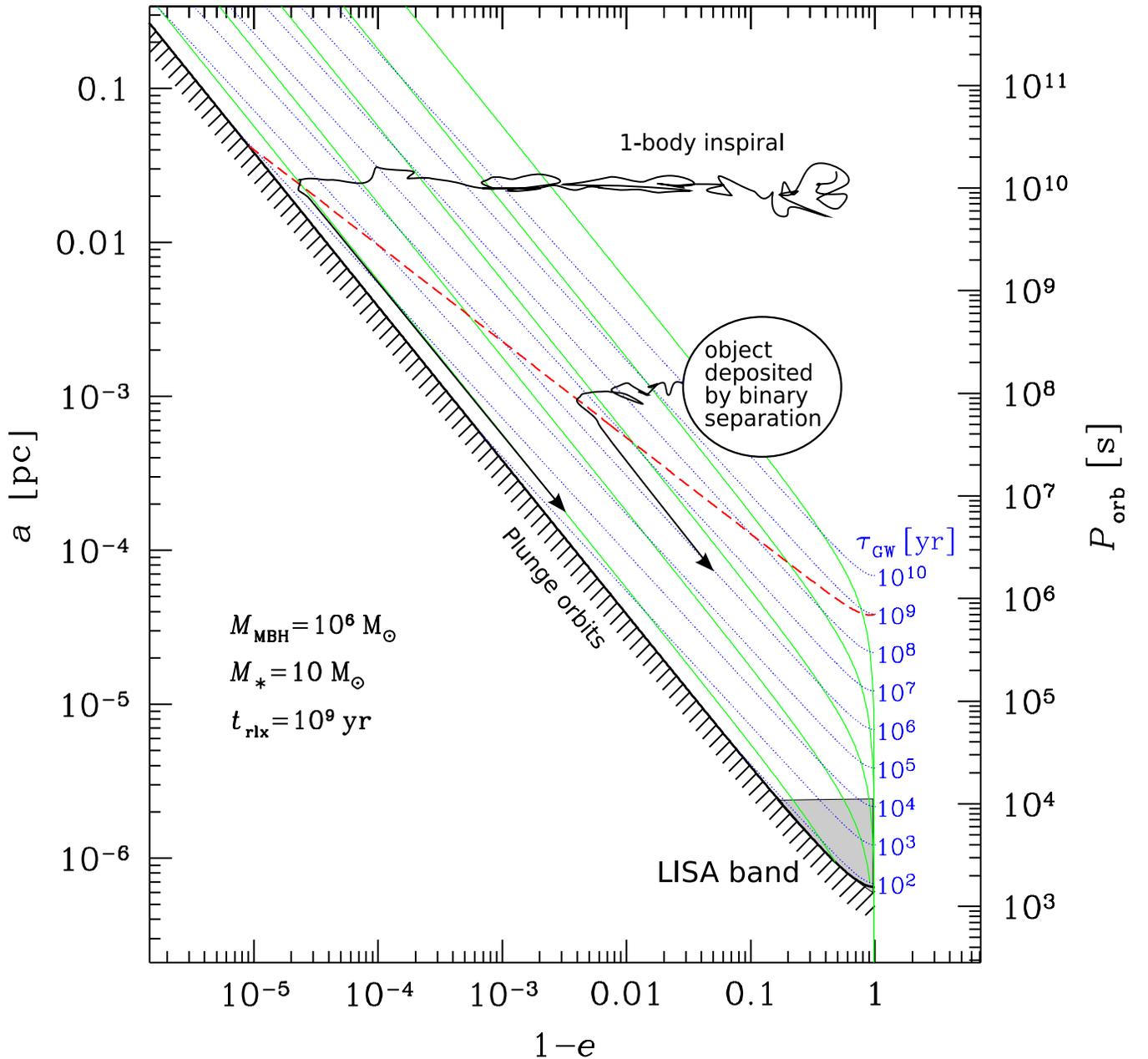}}
\caption{Inspiral trajectories in the semi-major axis, eccentricity plane. The thick diagonal line represents the last stable orbit using effective Keplerian values ($R_{\rm p}\simeq 4\RS$ for $e\ll 0.1$, see \cite{CKP94} for the general relation). The thin diagonal lines (in green in the on-line colour version) show inspiral trajectories due to emission of gravitational waves (GWs) and thin dotted (blue) lines are contours of constant time left until plunge, $\tGW$, as labelled in years on the right \cite{Peters64}. We assume a $10\,\Msun$ stellar black hole orbiting a $10^6\,M_\odot$ MBH on a slowly evolving Keplerian ellipse. The thick (red) dash-dotted line shows $\tilde{e}(a)$, defined by $t_e=\tGW$ (Eq.~\ref{eq.EMRIcond} with $C_{\rm EMRI}=1$) assuming a constant value $\trlx=1$\,Gyr. Below this line, the effects of relaxation on the orbit are negligible in comparison to emission of GWs. We schematically show typical orbital trajectories for EMRIs. 
Stars captured by tidal binary splitting initially have $a$ of order 100-1000\,AU [$5\times(10^{-4}-10^{-3})\,$pc] and $e=0.9-0.99$ \cite{MFHL05}. On a time scale of order $\trlx\ln(1-\tilde{e})^{-1}$, the eccentricity random-walks into the GW-dominated region, leading to a nearly-circular EMRI. If the star has not been deposited by binary splitting but has diffused from large radii or has been captured by GW emission, it will initially have a much larger value of $a$, therefore producing a higher eccentricity EMRI. One sees that stars with $a\gtrsim 5\times 10^{-2}$\,pc can not enter the inspiral domain unless $a$ is first reduced significantly, which takes of order $t_{\rm rlx}$. The grey region is the domain for sources whose orbital frequency is in the LISA band $P_{\rm orb}<10^4\,$s. \label{fig.EMRI}}
\end{figure}

The situation is represented in Fig.~\ref{fig.EMRI} in the semi-major axis,
eccentricity plane. We plot schematically the trajectory for a typical EMRI
evolving according to the standard scenario (labelled ``1-body inspiral'' to distinguish
it from the binary tidal separation scenario discussed later). Initially the
values of $a$ and $e$ random walk due to 2-body relaxation. As it takes of
order $t_{\rm rlx}$ to change $a$ by a factor of 2 but only $(1-e)t_{\rm rlx}$
to change the value of $1-e$ (and hence the periapse), the random walk seems
more and more elongated in the horizontal direction, the smaller the value of
$1-e$. It is much more likely for a star to cross over to the plunging or
GW-dominated region by acquiring a very high eccentricity than by shrinking $a$
significantly. Typically an EMRI ``progenitor'' starts with a semi-major axis
slightly lower than $a_{\rm EMRI}$. It takes on average a time of order $\ln
(1-\tilde{e})^{-1} t_{\rm rlx} \simeq 10 t_{\rm rlx}$ for relaxation to produce
an eccentricity such that GW emission becomes dominant. From that point, the
object will follow a path closer and closer to a pure inspiral (approximated by
Peters equations \cite{Peters64}, see \S~\ref{sec.OrbEvol}).

At larger $a$ values, inspirals are practically impossible because GW emission
is not significant in comparison to relaxation even on plunge orbits. Unless
they first shrink their orbit through 2-body relaxation, these objects will be
swallowed by the MBH on a direct plunge. Inspirals staring with $a\ll a_{\rm
EMRI}$ are rare because, for a density cusp $n\propto r^{-\alpha}$ with
$\alpha\simeq 1.4-1.8$ \cite{BME04a,BME04b,FASK06,HA06b}, the number of stars
per unit $\log(a)$ is roughly ${\rm d}\Nstar/{\rm d}(\log a)\propto a^{(3-\alpha)}$.
Also, as one goes inwards, the value of $\alpha$ is lowered by the
progressively larger plunge loss cone \citep{LS77,ASFS04}. In other words, the
stellar density is reduced there (in comparison to a pure power law) because to
come and populate this region a star has to spend several relaxation times
drifting down in energy while avoiding entering the GW-dominated region and
inspiraling quickly.

Implementing this basic scenario in various ways (see
\S~\ref{sec.EMRIStatMethods}), several authors have estimated the rate at
which stellar remnants are captured by the central MBH, with results between
$\sim 10^{-6}-10^{-8}~{\rm yr}^{-1}$ for a $10^6\,M_\odot$ central black hole
\cite{HB95,SR97,Ivanov02,HA05}.  When combined with the uncertainty in the
number density of massive black holes with $\MBH<{\rm few}\times
10^6\,M_\odot$, the net predicted number of detections that LISA will make spans three orders of
magnitude, from a few to a few thousand events per year.

We note, incidentally, that even in the LISA band (in the final year of
inspiral), the eccentricity of the typical EMRI in the standard picture is high
enough that a large number of harmonics are likely to contribute to the gravitational waves
\cite{Freitag03,BC04,HA05}.  In addition, the orbital plane of the EMRIs is
unlikely to be significantly correlated with the spin plane of the MBH.  These
characteristics are distinct from those in non-standard scenarios (discussed
below), leading to optimism that some aspects of the nuclear dynamics could be
inferred from just a few events.

The word ``capture'' is sometimes used to refer to EMRIs, but this is misleading as, in the
standard picture, stellar objects are not captured by emission of GWs. They are
already bound to the MBH when they are brought into the GW-dominated regime by
2-body relaxation.  A star originally unbound to the MBH, with energy
$\frac{1}{2}v^2$, will be left bound to it by GW emission if it passes with a
pericentre distance smaller than

\begin{equation}
{r_{\rm capt}}\approx 5\RS\left(\frac{m}{10\,M_\odot}\right)^{2/7}
\left(\frac{\MBH}{10^6\,M_\odot}\right)^{-2/7}\left(\frac{v}{100~{\rm km/s}}\right)^{-4/7}.
\end{equation}

\noindent
However, in most cases the star will be left on too wide an orbit. In order to
become an EMRI (rather than experience a direct plunge), the semi-major axis
has to be smaller than a few $10^{-2}\,$pc (see Fig.~\ref{fig.EMRI} and
Eq.\ref{eq.aEMRI}), requiring a passage within a distance

\begin{equation}
{r_{\rm capt, EMRI}}\approx 3\RS\left(\frac{m}{10\,M_\odot}\right)^{2/7}
\left(\frac{\MBH}{10^6\,M_\odot}\right)^{-4/7}\left(\frac{a_{\rm capt}}{0.05~{\rm pc}}\right)^{2/7}.
\end{equation}

\noindent 
Therefore for $\MBH\ll 10^6\,\Msun$ there is a possibility of
capturing unbound (or loosely bound) stars directly on to EMRI
orbits. To our knowledge, the contribution of this channel to EMRI
rates has not been estimated but is probably small because it is
present only for the lowest-mass MBHs in the LISA range.

\subsubsection{Modifications of the standard picture}

There are a number of additional effects now being included in EMRI rate estimates
that could alter the rate significantly.  In particular, processes that
systematically alter the energy distribution of black holes, or change the rate
of angular momentum diffusion, could have a substantial impact.

{\it Mass segregation.}---In galactic nuclei with black holes of mass $\MBH\lta
10^6\,M_\odot$, relaxation times for stellar-mass black holes are much less
than the age of the universe \cite{Lauer98,FB05,FASK06,Merritt06,HA06b}.  This
suggests that they would have had time to sink in the central potential.
Estimates of the effect of this mass segregation indicate that the black holes
have time to form a subcluster of radius $\sim 10^{-2}$~pc, which is small
enough that their orbits are not perturbed into a rapid plunge
\cite{MEG00,FASK06}. The LISA detection rate is therefore not reduced by
plunges.

{\it Triaxiality.}---Galactic nuclei need not be precisely spherical,
particularly if the galaxy has undergone a major merger in the past.  Indeed,
triaxiality can persist much longer than was once thought, even in the presence
of a supermassive black hole \cite{HolleyBockelManEtAl02,PM02}.  If the
potential is non-axisymmetric, it means that an individual black hole need not
conserve its angular momentum from orbit to orbit (although of course the
overall angular momentum is conserved).  This non-conservation means that the
pericentre distance can change radically in a dynamical time, and some triaxial
potentials favour orbits that are centrophilic.  At first sight, it might appear
that this effect could increase the LISA detection rate dramatically.  However,
there are two effects that mitigate this.  First, the dominant
contributors to the LISA inspiral rate are black holes with semimajor axes of
$\sim 10^{-2}$~pc, much smaller than the radius of influence
\cite{Freitag03,HA05}.  It is difficult to see how substantial deviations from
spherical symmetry could persist in this region.  Second, triaxiality implies a
greatly enhanced rate of angular momentum diffusion.  This implies that the
pericentre distance can be reduced significantly in a single orbit, but also
means that the distance can be increased again in the next orbit.  The loss
cone for single-passage processes such as tidal disruptions and direct plunges
is replenished more efficiently, increasing the rate of such events. However,
in the standard picture, EMRIs {\em must} occur in the empty loss-cone regime
for them to be gradual, i.e., to avoid premature plunge \cite{HA05}. Although
triaxiality might increase the rate of binary tidal separations (see below),
what is required to produce more EMRIs with single stellar objects is a process
that allows rapid diffusion to small pericentres, but slows down once those
pericentres are reached.  One such process is resonant relaxation, which we now
discuss.

{\it Resonant relaxation.}---Over times long compared to an orbital time,
orbits of different stellar-mass black holes affect each other as if they were
one-dimensional wires whose mass density at a given radius is proportional to
the time spent there.  The torques exerted by the wires on each other lead to
secular resonances.  The integration over many orbits guarantees that the
orbital semi-major axes remain constant, but the angular momenta can change
substantially \cite{RT96,RI98}.  For EMRIs, an important consequence is
so-called resonant relaxation, which increases the rate of angular momentum
diffusion by a large factor compared to simple two-body scattering.  This might
seem similar to the effects of triaxiality, but there is a crucial difference.
Resonant relaxation is effective because orbital phase relations can be
maintained over many orbits.  If, however, there is a source of precession, the
phase relations are scrambled and angular momentum diffusion is not as
effective.  In practice this means that if the pericentre distance is large,
resonant relaxation operates, but if it is small enough that general
relativistic pericentre precession is significant, resonant relaxation is
reduced.  This is therefore a positive conspiracy: orbits can migrate quickly
to low pericentre, but once there they linger. The current best estimates
suggest that this could increase the overall LISA detection rate by a factor of
a few \cite{HA06}.

\subsubsection{Astrophysical uncertainties}
\label{sec:astro-uncertain}

Before we proceed to describe alternative scenarios, it is worth stopping to
consider  the many uncertainties of astrophysical nature which affect EMRI
rate estimates, even in the most studied and best understood standard picture.

{\it Number density of massive black holes.}---The number of galaxies hosting
MBHs with a mass in the range $10^4-10^7\,\Msun$ is highly uncertain
\cite{BarackEtAl03} as only 3 such objects have so far been detected through
robust kinematic measurements \cite{FF05}, although there does exist more
indirect evidence for their existence based on the activity of low-luminosity
active galactic nuclei \cite{FH03,BHS04,GH04,BGH05,DongEtAl07}. Assuming that
all galaxies harbour a MBH and using the observational distribution of
velocity dispersion or spheroid luminosity combined with the $\MBH -\sigma$
relation or the correlation between $\MBH$ and luminosity
\cite{Mag98,MH03,HR04}, one estimates the density of MBHs in the LISA range to
be a few $10^{-2}\,{\rm Mpc}^{-3}$ in the local universe
\cite{AR02,ShankarEtAl04}. 

{
It has been suggested that most small galaxies have a dense
stellar cluster at their centre {\em instead} of a MBH
\cite{FerrareseEtAl06,WH06}. However, the existence of a MBH with a mass 
compatible with the $\MBH \propto \sigma^\beta$ relation of exponent
$\beta\simeq 4$ is only ruled out observationally in the cases of M~33
and NGC~205
\cite{GebhardtEtAl01,MFJ01,ValluriEtAl05}. The value of $\beta$ is still not 
established very accurately and could be as high as $4.5-5$
\cite{FF05}, making the upper limit on $\MBH$ for NGC~205 compatible with 
the relation (but shifting the MBH mass function to lower values). It
should also be reminded that well-observed galaxies such as the Milky
Way, M~31 and M~32 possess both a MBH of normal mass and a compact
stellar nucleus.
On the other hand, there are theoretical reasons to suspect a deficiency of
low-mass MBHs. Using semi-analytical models for the cosmological
assembly of galaxies, one can study statistically the growth and
possible ejection of MBHs due to the velocity kick, of up to $\sim
4000\,\kms$, imparted by asymmetric emission of gravitational waves when
two MBHs merge
\cite{BakerEtAl07,CLZM07}. These models indicate that, in the
present-day universe, MBHs might be absent or under-grown in a
significant fraction of galaxies with a velocity dispersion below
$100\,\kms$ \cite{Volonteri07}.

For these reasons,} it might be wise not to assume the existence of
a large population of MBHs with $\MBH\lesssim 2\times 10^6\,\Msun$ for conservative
estimates of EMRI detection rates \cite{DeFreitasEtAl06}. This uncertainty will
probably be decreased by the time LISA flies as present and future observations
in the electro-magnetic domain will help constrain the low-mass end of the MBH
population in nearby galaxies, as well as the characteristics of the nuclei in
which they reside. For instance, it is predicted that these smaller MBHs should
dominate the rate of accretion flares resulting from the tidal disruption of
main-sequence or giant stars \cite{WM04}. Tentative detections of such flares
in the UV and X-ray have been reported
\cite{GezariEtal03,HGK04,KomossaEtAl04,Komossa05,EsquejEtAl07} and more robust
detections can be expected in the next few years.

{\it Density profiles of galactic nuclei.}---Direct observational constraints
on the stellar density within the sphere of influence of MBHs with masses in the LISA
range exist only for the Milky Way
\cite{Alexander99,GenzelEtAl03,GhezEtAl03b,EisenhauerEtAl05,GhezEtAl05,MartinsEtAl06,LuEtAl06,PaumardEtAl06,SchoedelEtAl07}
and M~32 \citep{Lauer98}. Even in these cases the interpretation of the
(projected) surface brightness or number counts in terms of stellar density is a
difficult and uncertain matter, mostly because these observations are dominated
by bright stars whose distribution probably does not trace that of the compact
remnants \cite{Alexander2005}. Furthermore, there is no deep reason why all nuclei harbouring a MBH
of a given mass should have the same stellar density or why there should be
scaling laws for the parameters of galactic nuclei with different $\MBH$ (the
correlations between $\MBH$ and the structure of the host galaxy, mentioned
above, apply at scales much larger than the influence radius).

{\it Populations of compact remnants in galactic nuclei.}---In addition to the
role of dynamical mass segregation, important unknowns affecting this aspect
are: (1) the spectrum of masses of new-born stars (known as ``initial mass
function'' or IMF), (2) the history of stellar formation, i.e., the
distribution of ages and metallicities, which strongly affect the populations
of compact remnants, and (3) the theory of stellar evolution.  There are
indications \cite{NS05,StolteEtAl05,BKM07} that the IMF in the central regions
of our galaxy might be richer in massive stars ($M_\ast>5-10\,\Msun$) than in
other environments \cite{Kroupa02,Chabrier03}. If this is a generic situation,
EMRI rates might be larger due to a higher fractional number of stellar BHs.
At the Galactic centre, tens of young massive stars are observed within a few
tenths of a pc of the MBH.  Their unexplained presence reminds us of how little
is understood about stellar formation in galactic nuclei. The prediction of the
number, masses and types of compact remnants in galactic nuclei is also
hindered by uncertainties in stellar evolution. The mass of candidate stellar
BHs is only constrained to be roughly in the range $5-20\,\Msun$
\cite{Orosz03} but the formation of lower- or higher-mass BHs through normal stellar evolution \cite{BSR04} is not excluded. The uncertainties in the
theoretical predictions about the relation between the initial mass of a star
and the mass and type of remnant produced are still very large
\cite{FK01,HFWLH03,ZWH07}. The most recent stellar collapse simulations
indicate that, for solar metallicity or above, very few BHs might be
produced and these could have typical masses below $5\,\Msun$, which would have
dire consequences for EMRI rates. The discovery of a neutron star in a young
cluster \cite{Muno06,MunoEtAl06b} recently showed that not all stars more
massive than $20-30\,\Msun$ become BHs, contrary to what is commonly assumed.

Before LISA starts operating, we might get a better handle on the
population of stellar remnants in our Galactic nucleus through observations and modelling of X-ray point sources \cite{LaycockEtAl05,MunoEtAl05,Bandyopadhyay06,MunoEtAl06,RBH06}.
Most of these are probably compact binaries but even single
stellar BHs can become bright sources when they cross molecular clouds 
\cite{DN06}. However, it is not clear at all how typical the centre 
of our galaxy is and how this information can be used to constrain
models of general nuclei.

{\it Stellar dynamics around a MBH.}---This aspect is covered in
\S~\ref{sec.StellDyn}. In the framework of the (modified) standard scenario,
the difficulties are mostly of a numerical nature: how to implement the
(gravitational) physics in an efficient and realistic way to obtain robust
predictions for EMRIs {\em given an initial galactic nucleus model}.

\subsubsection{Non-standard scenarios}
\label{sec.nonstd}

There are additional ways to place a black hole in a close orbit around an
MBH.  As mentioned above, the distribution of eccentricities and inclinations
in these scenarios differ qualitatively from those in the standard picture.

{\it Binary tidal separation.}---Suppose that stellar-mass black holes in
galactic nuclei are sometimes found in binaries.  If such a binary passes close
enough to a MBH, one member is typically captured by the MBH while the other
is flung out at high speed (indeed, this process is thought to produce
hypervelocity stars in the Galaxy; see
\cite{Hills88,Hills91,YT03,BrownEtAl05,ENHCR05,HHOTB05,GPZS05,Pfahl05}).  The
captured black hole is expected to have a semi-major axis of a few hundred AU
and a pericentre distance of a few to tens of AU, implying that it will
circularise by the time it enters the LISA frequency band \cite{GPZS05,MFHL05}.
The large allowed pericentre distance means that the cross section for this
process is much larger than it is for single-body gravitational wave capture,
and if $\gta 1$\% of black holes are in binaries, the total rate may compete
with the rate from the standard scenario \cite{MFHL05}.  Note that inspirals of
this type are expected to have low eccentricities (see Fig.~\ref{fig.EMRI}),
but arbitrary inclinations relative to the spin axis of the MBH. As for tidal disruption of single stars \citep{FR76}, in a
spherical potential, the largest contribution to the rate of binary separations
comes from binaries with a (circum-MBH) semi-major axis of order the ``critical
radius''.  This corresponds to the situation where in just one orbit,
relaxation can change the direction of the velocity vector by an angle equal to
the aperture of the loss cone. The loss cone is relatively large for binary
separation and the critical radius is well outside the radius of influence of
the MBH, meaning that the rate can be increased significantly by processes able
to replenish loss-cone orbits faster than 2-body relaxation on that scale,
including massive perturbers such as molecular clouds \cite{PHA07} and the
effects of a triaxial potential (see above).

{\it Capture of cores of giants.}---Giant stars venturing very close to the MBH
have their envelope partially or completely stripped
\citep{DSGMG01,DaviesKing05} by the tidal stresses. It is possible that the
dense core of the giant is left on a tight orbit around the MBH and will evolve
into an EMRI. However, detailed hydrodynamical calculations are required to
determine whether the coupling between the diffuse envelope and the core is
strong enough to affect significantly the orbit of the core itself.

{\it Massive star capture or production in disc.}---Another proposal
is related to the presence of massive accretion discs around MBHs.
At distances of $\sim 0.1-1$~pc from the MBH and at typical
accretion rates, these discs can be unstable to star formation
\cite{CZ99,LB03,Goodman03,GT04,ML04,Levin03,Levin06,Nayakshin06}.  If, as in
some calculations, there is a bias towards the production of massive
stars in the disc, they could evolve to become black holes, which are
then dragged in along with the disc matter.  Alternately, massive
stars on orbits that cross the disc could be captured and then evolve
into black holes \cite{SCR91,Rauch95,SubrKaras99,KS01}.  Either way,
in the last year of inspiral the holes would be expected to be in
nearly circular orbits with orbital planes aligned with the MBH.
Rates are highly uncertain as well as the mass of the stellar remnants
formed (which could even be IMBHs). However these events would likely
have a different signature waveform than those of the other two
classes because they should occur on co-rotating, circular orbits
lying in the equatorial plane of the spinning MBH if it has gained a
significant fraction of its mass by accreting from the disc \cite{Bardeen70,KLOP05,VMQR05}. Moreover, there is the exciting possibility that in such a scenario the compact object would open a gap in the disc, which could lead to an optical counterpart to the EMRI event~\cite{Levin06}.

\subsection{Intermediate-mass black holes and inspirals into MBHs}

The existence of stellar-mass black holes ($\sim 5-20\,M_\odot$) and
supermassive black holes ($\sim 10^{6-10}\,M_\odot$) has been confirmed via the
dynamics of stars around them (\cite{Orosz03,FF05} for recent reviews).  In
contrast, intermediate-mass black holes (IMBHs; $\sim 10^{2-4}\,M_\odot$) have
not been dynamically confirmed to exist, because they are rare enough that none
are known in binaries in the Local Group, and their radii of influence are
small enough to make mapping of stellar motions around them very challenging.
The current evidence for them is thus indirect, with the best cases provided by
the ultra-luminous X-ray sources (ULXs), which are bright
$>10^{39-40}$~erg~s$^{-1}$, variable, non-nuclear X-ray sources seen in one out
of every few galaxies (see \cite{MC04,Marel2003} for a review).  There are several strong,
but not conclusive, pieces of evidence that suggest that at least some ULXs are
$M>100\,M_\odot$, including unusually cool accretion discs
\cite{MFMF03,MFM04a,MFM04b}, low-frequency quasi-periodic brightness
oscillations \cite{SM03}, X-ray dominance of the multiwavelength spectrum (note
that if we are instead seeing relativistic jets, we would expect a relatively
flat $\nu F_\nu$ spectrum), and surrounding nebular He~II luminosity suggesting
that the source is quasi-isotropic instead of beamed \cite{PakullMirioni02,PakullMirioni03}.

In the early metal-free universe, it is possible that individual
stars might evolve to several hundred solar mass black holes, where
the high mass is a result of less fragmentation during star formation
and minimal mass loss to winds and instabilities.  In the
current universe, the dynamics of dense stellar regions may allow
runaway collisions of massive stars in young clusters, with subsequent
evolution to an IMBH \cite{EbisuzakiEtAl01,PZMcM02,PortegiesZwartEtAl04,GFR04,OL06,GFR06,
FRB06,FreitagEtAl06,Fregeau06}, 
potentially followed by slower growth
of an IMBH via collision-less interactions and mergers with stellar-mass
compact objects \cite{Tan00,MH02a,MH02b,MT02a,MT02b,GMH04,GMH06,OL06}.

Such clusters, if formed within a few hundred parsecs of the galactic
nucleus, will sink towards the centre due to dynamical friction.  When
the cluster is within a few parsecs, the stars are stripped away by
the galactic tidal field, and the IMBH continues sinking on its own.
There are many uncertainties about the further development, but
analytic \cite{Miller05} and subsequent numerical \cite{PZ06,Mat06}
calculations suggest that the resulting IMBH-MBH mergers should be
detectable with LISA between a few and a few tens of times per year.
As discussed in the data analysis section below, even a single such
event would be strong enough to allow unique tests of the predictions
of general relativity in strong gravity.  It would also shed light on
the elusive IMBHs  themselves, as well as yielding insight into the
dynamics of stellar clusters and galactic nuclei.

\subsection{Expected eccentricity distributions}

\subsubsection{Orbital evolution due to emission of gravitational waves}
\label{sec.OrbEvol}

Consider a binary with component masses $m_1$ and $m_2$, which thus has
total mass $M=m_1+m_2$ and reduced mass $\mu=m_1m_2/M$.  Suppose that
its semi-major axis is $a$ and eccentricity is $e$.  The Peters equations for gravitational wave emission from a Keplerian orbit~\cite{Peters64} give

\begin{equation}
\left\langle \frac{da}{dt}\right\rangle=-\frac{64}{5}\frac{G^3\mu M^2}{
c^5a^3(1-e^2)^{7/2}}\left(1+\frac{73}{24}e^2+\frac{37}{96}e^4\right)
\label{aevol}
\end{equation}

\noindent
and

\begin{equation}
\left\langle \frac{de}{dt}\right\rangle=-\frac{304}{15}e\frac{G^3\mu M^2}
{c^5a^4(1-e^2)^{5/2}}\left(1+\frac{121}{304}e^2\right)\; .
\label{eccevol}
\end{equation}

\noindent
We note that the Peters formalism does not
capture the orbital evolution in the strong-field regime, before
plunge. In particular, for EMRIs around a spinning MBH, a slight {\em
increase} in eccentricity might occur in the late evolution
\cite{GG06}. This does not affect the present discussion. From Eq.\ref{eccevol},
the characteristic time to change the eccentricity is

{\setlength\arraycolsep{2pt}
\begin{eqnarray}
\tau_{\rm GW} & = & \frac{e}{|de/dt|}\approx \frac{15}{304}\frac{c^5a^4(1-e^2)^{5/2}}{G^3\mu M^2} {}
\approx 8\times 10^{17}~{\rm yr}\left(\frac{M_\odot}{\mu}\right)\left(\frac{M_\odot}{M}\right)^2 \left(\frac{a}{1{\rm AU}}\right)^4 {}
(1-e^2)^{5/2}.
\label{eq.tGW}
\end{eqnarray}}

\noindent
Here we neglect the near-unity factor $(1+121e^2/304)$.

We can rewrite this in terms of gravitational wave frequency.  Let us consider
in particular the frequency emitted at pericentre.  If the orbit is
substantially eccentric, then the orbital frequency at that point will be
approximately $\sqrt{2}$ times the circular frequency at that radius (because
the speed is $\sqrt{2}$ times greater than a circular orbit).  If we designate
a maximum gravitational wave frequency $f_{\rm max}$ to be double the frequency
at pericentre, then

\begin{equation}
f_{\rm max}\approx {1\over\pi}\left[2GM\over{(a(1-e))^3}\right]^{1/2}\; .
\end{equation}

\noindent
Therefore 

{\setlength\arraycolsep{2pt}
\begin{eqnarray}
a^4&= &0.75{\rm AU}^4\left(\frac{M}{10^6\,M_\odot}\right)^{4/3}\left(\frac{f_{\rm max}}{10^{-4}~{\rm Hz}}\right)^{-8/3} {}
(1-e)^{-4}, {} 
\end{eqnarray}}

\noindent
and

{\setlength\arraycolsep{2pt}
\begin{eqnarray}
\tau_{\rm GW}&\approx& 6\times 10^2~{\rm yr}\left(\frac{\mu}{10^3\,M_\odot}\right)^{-1}\left(\frac{M}{10^6\,M_\odot}\right)^{-2/3}{}
{}\left(\frac{f_{\rm max}}{10^{-4}~{\rm Hz}}\right)^{-8/3}(1+e)^{5/2}(1-e)^{-3/2}{}\nonumber\\
& & {}\nonumber\\
&\approx& 3\times 10^3~{\rm yr}\left(\frac{\mu}{10^3\,M_\odot}\right)^{-1}\left(\frac{M}{10^6\,M_\odot}\right)^{-2/3}{}
{}\left(\frac{f_{\rm max}}{10^{-4}~{\rm Hz}}\right)^{-8/3}(1-e)^{-3/2}
\end{eqnarray}
}

\noindent
where in the last line we assume a relatively high eccentricity,
so that $1+e\approx 2$.

\subsubsection{Eccentricity of EMRIs}

A classic EMRI, with $M=10^4-10^7\,M_\odot$ and $\mu=1-10\,M_\odot$, could have
a significant eccentricity if (as expected in galactic nuclei) the orbits come
in from large distances, $a>10^{-2}\,$\,pc with $e\gtrsim 0.9999$. Hopman and
Alexander \cite{HA05} made an estimate of the distribution of eccentricities
for one body inspiral and their results showed that it is skewed to high-e
values, with a peak of the distribution at e~0.7, at an orbital period of
$10^4$ s. On the other hand, following a binary separation event (and possibly
the tidal capture of giant's core), the compact star is deposited on an orbit
with semi-major axis of order a few tens to a few hundreds of AU. In this case,
the GW-dominated regime is reached with an eccentricity smaller than 0.99 and
the orbit should be very close to circular when it has shrunk into LISA band.
Such typical orbital evolutions for EMRIS are shown in Fig.~\ref{fig.EMRI}. 

\subsubsection{Eccentricity of IMRIs}





In contrast with stellar binaries, it seems likely that the majority
of mergers between two massive or intermediate-mass black holes occur
because the black holes form separately, then are brought together by
dynamical processes. In particular the favoured scenario for IMRIs
involves the formation of an IMBH in stellar cluster which spirals
into a galactic centre where the cluster is tidally disrupted and
deposits the IMBH on an orbit around the central supermassive black hole.


The result is that mergers between massive black holes very probably require
substantial dynamical interactions as opposed to pure evolution by emission of
gravitational radiation.  These dynamical effects can change the eccentricity.
On their own, they would tend to make the probability distribution ``thermal",
i.e., $P(e)=2e$, which has a mean of $e\approx 0.7$.  A bias towards higher
eccentricities could happen if capture by emission of gravitational radiation
happens preferentially at large $e$. The actual eccentricity is therefore
determined by the competition between dynamical wandering and gravitational
wave circularisation.

To get an estimate of how circular the MBH-IMBH binary will (as a LISA source),
we can compare the time to reduce eccentricity via gravitational radiation with
the time to interact with enough stars to change the eccentricity
significantly.  Obviously, high eccentricity is possible if there is a close
hyperbolic orbit, but we consider the much more likely case that the evolution
is gradual. Here we consider dynamical scenarios in which the primary
interactions are of single objects with the binary of interest; this is
therefore relevant to a MBH-MBH or IMBH-MBH binary, but not to the classic
EMRI scenario in which many low-mass objects interact collectively in the
radius of influence of a MBH. 

To determine the timescale for dynamical eccentricity changes, we note that the
orbital speed at a gravitational wave frequency of $10^{-4}$~Hz is rather high;
roughly 30\,000~km~s$^{-1}[M/(10^6\,M_\odot)]^{1/3}$ for a circular orbit, for
example.  This is much greater than the velocity dispersion $\sigma$ in any
realistic astrophysical environment.  Therefore, the cross section $\Sigma$ for
a binary-single encounter with a closest approach $r_p$ is

\begin{equation}
\Sigma=\pi r_p^2\left[1+2GM/(r_p\sigma^2)\right]\approx
\pi r_p (2GM/\sigma^2)\; .
\end{equation}

If we assume that $r_p<2a$ is required for a significant interaction,
then

{\setlength\arraycolsep{2pt}
\begin{eqnarray}
\Sigma & = &4\pi a\left(\frac{GM}{\sigma^2}\right)\approx 2\times 10^{32}{\rm cm}^2\left(\frac{M}{10^6\,M_\odot}\right)^{4/3}{}
\left(\frac{f_{\rm max}}{10^{-4}{\rm Hz}}\right)^{-2/3}\left(\frac{\sigma}{100{\rm km~s}^{-1}}\right)^{-2}(1-e)^{-1}.{} 
\end{eqnarray}}

\noindent The timescale for a {\it single} interaction is just $\tau_{\rm
int}=1/(n\Sigma\sigma)$, where $n$ is the number density of objects interacting
with the binary.  However, note that for the eccentricity or semi-major axis of
the binary to be affected significantly, the binary must interact with
approximately a mass $\sim M$.  
If the stars
interacting with the binary have average mass $\langle m\rangle$, this means
that the time required for dynamical interactions to change the eccentricity is
$\tau_{\rm dyn}\approx (M/\langle m\rangle)\tau_{\rm int}$.  This yields the
product $n\langle m\rangle$, which is simply the average mass density $\rho$ of
things interacting with the binary, so only this quantity matters and we have

{\setlength\arraycolsep{2pt}
\begin{eqnarray}
\tau_{\rm dyn}&\approx&5\times 10^8~{\rm yr}\left(\frac{\rho}{10^6\,M_\odot~{\rm pc}^{-3}}\right)^{-1}{}
\left(\frac{M}{10^6\,M_\odot}\right)^{-1/3} \left(\frac{f_{\rm max}}{10^{-4}~{\rm Hz}}\right)^{2/3}\left(\frac{\sigma}{100~{\rm km~s}^{-1}}\right)
(1-e).
\end{eqnarray}}

\noindent
The ratio between the circularisation time and the time for dynamical
change is therefore

{\setlength\arraycolsep{2pt}
\begin{eqnarray}
\frac{\tau_{\rm GW}}{\tau_{\rm dyn}}&\approx &6\times 10^{-6}\left(\frac{\mu}{10^6\,M_\odot}\right)^{-1}\left(\frac{M}{10^3\,M_\odot}\right)^{-1/3}{}\nonumber\\
& & {} 
\left(\frac{\rho}{10^6\,M_\odot~{\rm pc}^{-3}}\right) \left(\frac{\sigma}{100~{\rm km~s}^{-1}}\right)^{-1}{}
\left(\frac{f_{\rm max}}{10^{-4}~{\rm Hz}}\right)^{-10/3} (1-e)^{-5/2}. 
\end{eqnarray}}

\noindent
If $\tau_{\rm GW}/\tau_{\rm dyn}\ll 1$ we conclude that
circularisation dominates; if the ratio is the other way, then
circularisation is ineffective and the eccentricity will sample a
thermal distribution.  Therefore, a $10^3\,M_\odot$ black hole
spiraling into a $10^6\,M_\odot$ black hole can not maintain a
significant eccentricity above $\sim 10^{-5}$~Hz as its evolution is
likely to be strongly dominated by gravitational radiation. However, a
slightly eccentric source will exhibit a tell-tale phase evolution 
which could permit the measurement of eccentricities as low as~$\sim 0.01$
\cite{ASF06,CH06}.

\subsection{Numerical stellar dynamics for I/EMRIs}\label{EMRINum}
\label{sec.StellDyn}

\subsubsection{Stellar dynamics for galactic nuclei}

\begin{figure}
\resizebox{\hsize}{!}{\includegraphics[scale=1,clip]
{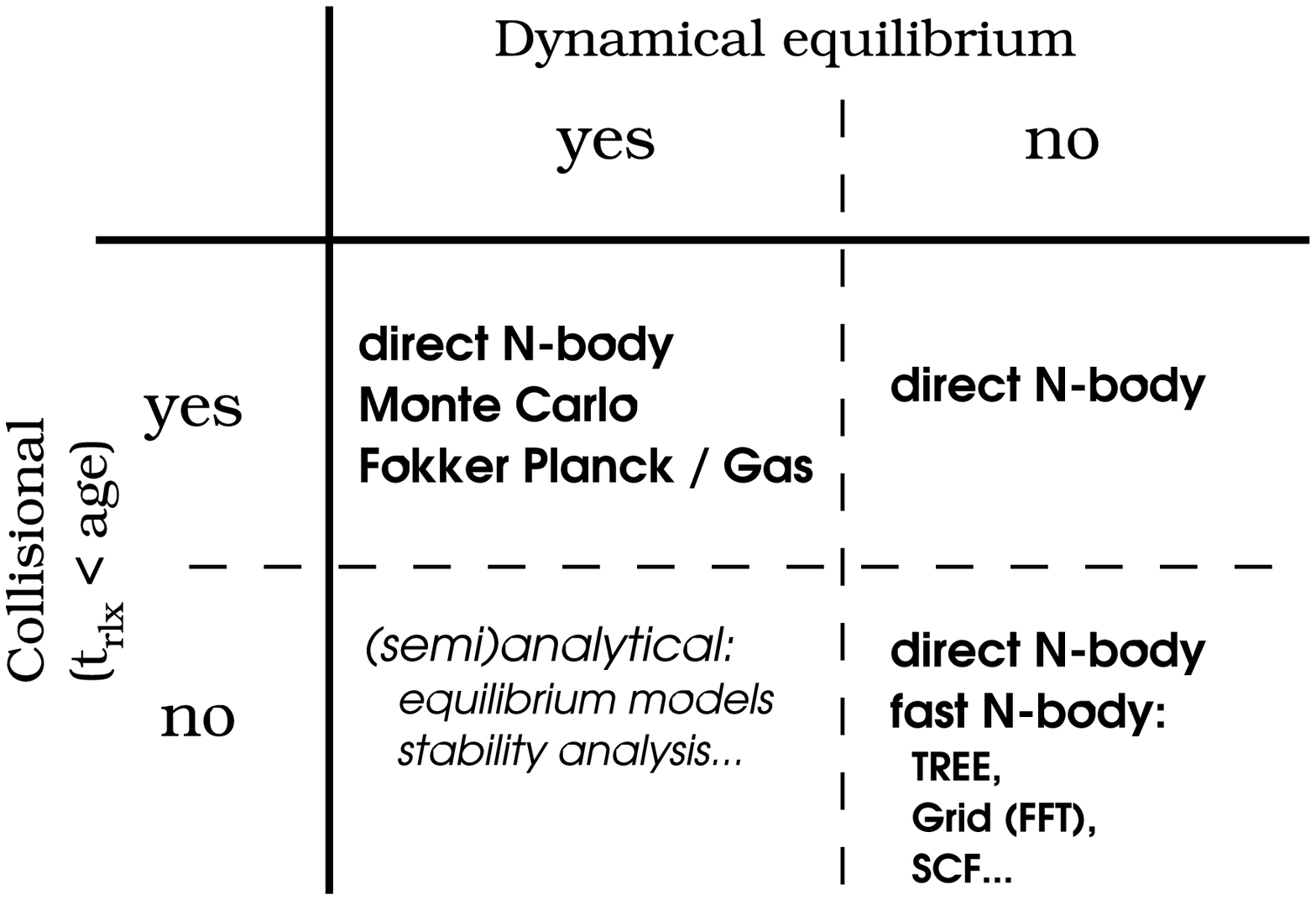}}
\caption{Methods appropriate to the various realms of stellar dynamics.
\label{fig.StellDynRealms}
}
\end{figure}

We can approximately classify the different kinds of techniques employed for studying
stellar dynamics according to the dynamical regime(s) they can
cope with.  In Fig.\,(\ref{fig.StellDynRealms}) we have a classification of
these techniques. (Semi-)analytical methods are generally sufficient only to
study systems which are in dynamical equilibrium  and which are not affected by
collisional (relaxational) processes. In all other cases, including those of
importance for EMRI studies, the complications that arise if we
want to extend the analysis to more complex (realistic)
situations, force us to resort to numerical techniques.

The $N-$body codes are the most straightforward approach from a conceptual
point of view. In those, one seeks to integrate the orbital motion of $N$
particles interacting gravitationally. It is necessary to distinguish between the {\em direct}
$N-$body approaches which are extremely accurate but slow and the fast $N-$body approaches,
which are fast but inaccurate and therefore generally deemed inadequate for studying relaxing systems because relaxation is the cumulative effect of {\em small}
perturbations of the overall, smooth, gravitational potential. Fast $N-$body codes are
usually based on either TREE algorithms \citep{barneshut86} or on an FFT (Fast Fourier
Transform) convolution to calculate the gravitational potential and force
for each particle \citep{superbox} or on an SCF (self-consistent-field)
\citep{Clutton-Brock73,HO92} approach.  We will not describe these numerical
techniques in this section because they have never been used to study E/IMRIs
and the approximations on which they are based make them unsuitable for an accurate study
of such systems, since relaxation plays a role of
paramount importance. Fast $N-$body algorithms can only be employed in
situations in which relaxation is not relevant or over
relatively short dynamical times, such as in studying bulk dynamics of whole galaxies.

On the other hand, if we want to study a system including both collisional effects
and dynamical equilibrium, we can employ direct $N-$body codes or use
faster approaches, like the Monte Carlo, Fokker Planck and Gas methods, which we will describe below.
The only technique that can cope with all physical inputs is the direct $N-$body approach, in which we make no strong assumptions other than that gravity is Newtonian gravity (although nowadays post-Newtonian corrections have also been incorporated, see Section~\ref{sec:directNbody}).

If we neglect capture processes driven by tidal effects, the region from
where we expect most EMRIs to come from is limited to $\sim$ 0.1 pc
around the central MBH. In that zone the potential is totally spherical. 
Non-spherical structures such as triaxial bulges or stellar discs are common on
scales of 100-1000\,pc, and the nucleus itself may be non-spherical. For
example, it could be rotating, as a result of a merger with another nucleus
\citep{MM01} or due to dissipative interactions between the stars and a dense
accretion disc \citep{Rauch95}.  It has been proved that this triaxiality could
boost the disruption rates by order{s} of magnitude
\citep{MerrittPoon04,PoonMerritt04}.  Whilst assuming sphericity will probably
not have any impact on the estimate of capture rates, it is of big relevance for
``tidal processes'', since this is the region in which binary tidal separation and the tidal capture of giant cores will happen (see section
II). For these processes the critical radius is beyond the influence radius of
the central MBH and so triaxiality can probably play an important role.  Due to
insufficient computer power and the limitations of simulation codes
galactic nuclei have so far been modelled only as isolated spherical clusters with purely
Newtonian gravity (i.e. \cite{MCD91,FB02b}). More realistic situations could only be
explored with $N-$body methods or possibly with hybrid codes (Monte Carlo
combined with $N-$body, for instance).

\begin{figure}
\resizebox{\hsize}{!}{\includegraphics[scale=1,clip]
{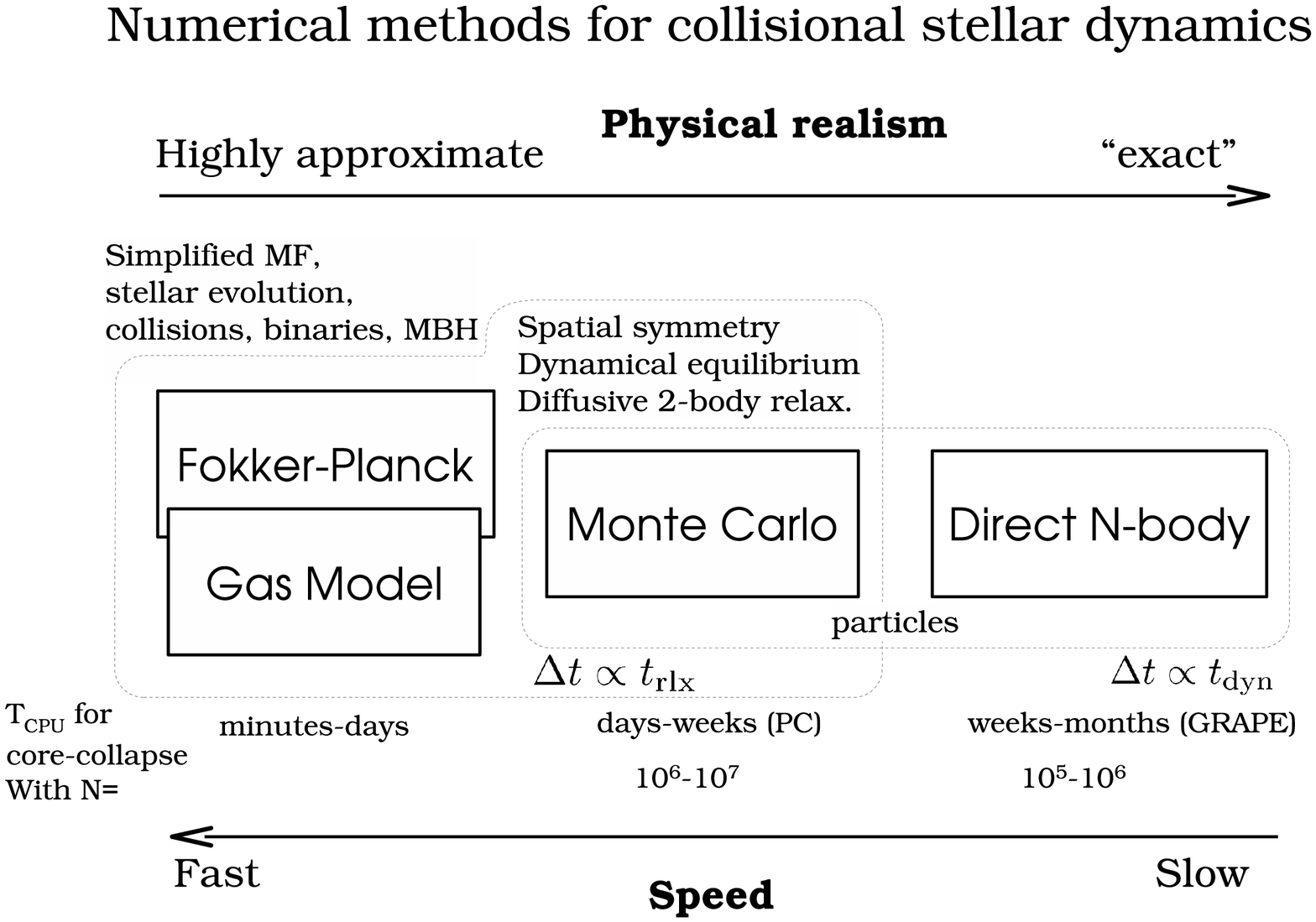}}
\caption{The various methods used to study collisional stellar dynamics.\label{fig.CollStellNumMethods}}
\end{figure}

In Fig.\,(\ref{fig.CollStellNumMethods}) we show a schematic illustration of the current
available codes for stellar dynamics including relaxation. From the
left to the right the {\em physical realism} of the codes increases while the speed decreases.
The two-dimensional numerical direct solutions of the Fokker-Planck
equation~\citep{Takahashi97,Takahashi96,Takahashi95} probably require the least
computational time, but these are followed closely by the gaseous model.
The advantage of these two codes are the
computational time required to perform a simulation (typically of the order of
one minute on a regular PC for a Hubble time) and since they are not
particle-based, the resolution can be envisaged as infinite, so that they are
not limited by the particle number of the system and there is practically no
numerical noise. Nevertheless, although they should be envisaged as powerful
tools to make an initial, fast exploration of the parameter space, the results
give us {\em tendencies} of the system, rather than an accurate answer
\cite{AmaroSeoaneThesis04}. Studying the Astrophysical I/EMRI problem requires
a meticulous characterisation of the orbital parameters, so that approximate
techniques should be regarded as exploratory only~\citep{DeFreitasEtAl06}.

\subsubsection{The Fokker-Planck approach}

\newcommand{\DCf}[1]{\langle{#1}\rangle} 
\newcommand{\DCN}[1]{\{{#1}\}} 
\newcommand{\FCF}[1]{{\cal D}_{#1}} 

Instead of tracking the individual motion of a large number of particles, as
in $N-$body methods, one can attempt to describe a system consisting of a
very large number of stars through the {\em 1-particle phase-space distribution
function} (DF for short) $f(\vec{x},\vec{v},t)$. The best interpretation of $f$
is as a probability density if it is normalised to 1 ---
$f(\vec{x},\vec{v},t)d^3x\,d^3v$ is the probability of finding, at time $t$, any
given particle within a volume of phase space $d^3x\,d^3v$ around the 6-D
phase-space point $(\vec{x},\vec{v})$; the average number of particles in this
volume would be ${\Nstar}f(\vec{x},\vec{v},t)d^3x\,d^3v$. If the particles move
in a common smooth potential $\Phi$, the evolution of $f$ is described by the
collision-less Boltzmann equation \cite{BT87}:

\beq
D_{t}f \equiv 
\partd{f}{t}+\vec{v}\cdot\vec{\nabla}f-\vec{\nabla}\Phi\cdot\partd{f}{\vec{v}}=0.
\eeq

\noindent 
$\Phi$ is obtained from $f$ and a possible external potential $\Phi_{\rm ext}$
(such as produced by a central MBH) from the Poisson equation.

In a real self-gravitating $N$-particle system the potential cannot be
smooth on small scales but has some graininess, i.e., short-term,
small-scale fluctuations, $\Phi_{\rm real}=\Phi+\Delta \Phi_{\rm
grainy}$. Relaxation describes the effects of these fluctuations on
$f$. They arise because a given particle
sees the rest of the system as a collection of point masses rather
than as a smooth mass distribution. Relaxational effects, also known
(somewhat confusingly) as collisional effects, can therefore be seen as
particles influencing each other individually as opposed as to
collectively.  To allow for these effects, a {\em collision
term} has to be introduced on the right hand side of the Boltzmann equation.

The Fokker-Planck (FP) equation is derived by assuming that relaxation is
due to a large number of 2-body gravitational encounters, each of
which leads to a small deflection and occurs ``locally'',
i.e., these affect the velocity of a star without affecting its position. 
This is the basis for Chandrasekhar's theory of relaxation 
\citep{Chandrasekhar60,BT87,Spitzer87}.
Under these asumptions, one can write

\beq
D_{t}f  
= -\sum_{i=1}^{3} \partd{}{v_i}\left[f(\vec{x},\vec{v})\DCf{\Delta v_i}\right] 
+ \frac{1}{2}\sum_{i,j=1}^{3} \frac{\partial^2}{\partial v_i\partial v_j}\left[f(\vec{x},\vec{v})\DCf{\Delta v_i \Delta v_j}\right],
\label{eq:FPEvit}
\eeq

\noindent
where the ``diffusion coefficient'' $\DCf{\Delta v_i}$ is the average
change in $v_i$ per unit of time due to encounters (see
\cite{RMcDJ57,BT87} for a derivation).

{\color{red} From Jeans' theorem \cite{Jeans15,Merritt99}}, for a spherical system
in dynamical equilibrium, the DF $f$ can depend on the phase-space
coordinates $(\vec{x},\vec{v})$ only through the (specific) orbital
binding energy $E$ and angular momentum (in modulus) $J$,
\beq
f(\vec{x},\vec{v}) = F(E(\vec{x},\vec{v}),J(\vec{x},\vec{v})).
\eeq

\noindent
In the vast majority of applications, the Fokker-Planck formalism is applied in the two-dimensional $(E,J)$-space or, assuming
isotropy, the one-dimensional $E$-space rather than the
six-dimensional phase space, through the operation of ``orbit
averaging'' (see
\cite{Cohn79,Cohn80,Cohn85,Spitzer87} amongst others). 

A standard form of the FP equation for an isotropic, spherical system is

\beq
D_t N(E)  \equiv \partd{N}{t} + \partd{N}{E}\left.\frac{dE}{dt}\right|_{\phi} =
-\partd{{\cal F}_E}{E} 
\label{eq.FPiso}
\eeq
where
\beq 
{\cal F}_E=m\FCF{E}F-\FCF{EE}\partd{F}{E}
\eeq
is the flux of particles in the energy
space; $\left.{dE}/{dt}\right|_{\phi}$ is the change of energy due
to the evolution of the potential $\phi$; $N(E)$ is the density of stars in
$E-$space,
\beq
N(E) = 16\pi^2 p(E) F(E)
\label{eq:FtoN}
\eeq 
with $p(E)=\int_0^{r_{\rm max}}r^2v\, dr$. The ``flux coefficients''  are

\beq
\begin{split}
\FCF{E} = &16\pi^3\lambda m_{\rm f} \int_{\phi(0)}^E dE'p(E')F_{\rm f}(E'), \\
\FCF{EE} = &16\pi^3\lambda m_{\rm f}^2 \left[q(E)\int_E^0 dE'F_{\rm f}(E')+\int_{\phi(0)}^E dE'q(E')F_{\rm f}(E')\right],
\label{eq:FluxCoef}
\end{split}
\eeq
\noindent
where $\lambda\equiv 4\pi G^2 \ln \Lambda$. $q(E)=\frac{1}{3}\int_0^{r_{\rm max}} r^2v^3\, dr$.
is the volume of phase space accessible to particles with
energies lower than $E$, and $p(E)={\partial q}/{\partial E}$ \cite{Goodman83}.

We used
an index ``f'' for ``field'' to distinguish the mass and DF of the
population we follow (``test-stars'') from the ``field'' objects. This
distinction does not apply to a single-component system but it is
easy to generalise to a multi-component situation by summing over
components to get the total flux coefficient
\beq
\FCF{E} = \sum_{l=1}^{N_{\rm comp}} {\FCF{E}}_{,l},\ \ \FCF{EE} = \sum_{l=1}^{N_{\rm comp}} {\FCF{EE}}_{,l},
\eeq
where the flux coefficient for component l can written by
replacing the subscript ``f'' by ``$l$'' in Eq.~\ref{eq:FluxCoef}.

We now explain schematically how the FP equation is implemented numerically
to follow the evolution of star clusters.  A more detailed description
can be found in, e.g., \citet{CW90}. In the most common scheme,
pioneered by~\citet{Cohn80}, two types of steps are employed alternately, a
method known as ``operator splitting''.
\begin{enumerate}
\item {\bf Diffusion step}. The change in the
distribution function $F$ for a discrete time-step $\Delta t$ is
computed by use of the FP equation {\em assuming the potential $\phi$ is
fixed}, i.e., setting $D_t N = {\partial N}/{\partial t} =
\left.{\partial N}/{\partial t}\right|_{\rm coll}$. The FP equation
is discretised on an energy grid. The flux coefficients are
computed using the DF(s) of the previous step; this makes the
equations linear in the values of $F$ on the grid points. The
finite-differencing scheme is the implicit \citet{ChangCooper70} algorithm,
which is first order in time and energy.
\item {\bf Poisson step}. Now the change of potential resulting 
from the modification in the DF $F$ is computed and $F$ is modified to
account for the term $\left.dE/dt\right|_\phi$, i.e., assuming $D_t N
= {\partial N}/{\partial t} + {\partial N}/{\partial E}\left.{dE}/{dt}\right|_{\phi} =
0$. This can be done implicitly because, as long as the
change in $\phi$ over $\Delta t$ is very small, the actions of each
orbit are adiabatic invariants. Hence, during the Poisson step, the
distribution function, expressed in terms of the actions, does not change. In practice, an iterative scheme is
used to compute the modified potential, determined implicitly by the
modified DF, through the Poisson equation.
The iteration starts with the values of $\phi$, $\rho$, etc.
computed before the previous diffusion step.
\end{enumerate}

A variant of the FP equation analogous to Eq.~\ref{eq.FPiso} can be written which
allows for anisotropy by taking into account the dependence of $F$ on $J$ and
including a $J$-flux and corresponding flux coefficients
\cite{CK78,Cohn79,Cohn85,Takahashi95,Takahashi96,Takahashi97,DCLY99}.  The
expressions for the flux coefficients are significantly longer than in the
isotropic case and we do not present them here. However, we note that in galactic
nuclei, in contrast to globular clusters, anisotropy plays a key role
because of the existence of a loss cone.

The use of the FP approach to determine the distribution of stars
around a MBH requires a few modifications. First the (Keplerian)
contribution of the MBH to the potential has to be added. 
Several authors have made use of the FP or similar formalisms to study
the dynamics well within the influence radius under the assumption of a
{\color{red} fixed potential 
\cite{BW76,BW77,LS77,CK78,HA06,HA06b,MHB06}, which is an significant 
simplification. The static potential included a contribution for the
stellar nucleus in the last study \cite{MHB06} but was limited to
a Keplerian MBH potential in the other cases.} The presence of the MBH
also constitutes a central sink as stars are destroyed or swallowed if
they come very close to it. This has to be implemented into FP codes
as a boundary condition. \citet{LS77} and~\citet{CK78} have developed
detailed (and rather complex) treatments of the loss cone for the
anisotropic FP formalism. It can be used in a simplified way in an
isotropic FP analysis
\cite{BW77} to obtain a good approximation to the distribution of
stars around a MBH and of the rates of consumption of stars by the
MBH. However, additional analysis is required to determine what fraction of
the swallowed stars are EMRIs and what their orbital properties are
\cite{HA05,HA06b}.

\subsubsection{The gaseous model}

Another way to approximately solve the (collisional) Boltzmann equation is to take velocity moments of it. The moment or order $n=0$
of the DF is the density, the moments of order $n=1$ are bulk velocities and
$n=2$ corresponds to (anisotropic) pressures (or velocity dispersions). This is
analogous to the derivation of the Jeans equation from the collision-less
Boltzmann equation \cite{BT87} but the collision term introduces moments of
order $n+1$ in the equations for moments of order $n$. In the ``gaseous
model''\footnote{\tt{http://www.ari.uni-heidelberg.de/gaseous-model}}, one
assumes spherical symmetry (but not necessarily dynamical equilibrium) and
truncates the infinite set of moment equations at $n=2$.  The system is closed
with the assumption that energy exchanges between stars through 2-body
relaxation can be approximated by an ad hoc (local) heat conduction
prescription \cite{HachisuEtAl78,LBE80}. This reduces the study of the stellar
system to that of a self-gravitating conducting gas sphere. Multi-mass models
have been implemented \cite{LS91,Spurzem92,GS94,ST95} and the
detailed forms for the conductivities have been improved by comparing to direct $N-$body models
(described below). The addition of a central accreting MBH and a treatment for
loss-cone effects was done by~\citet{ASFS04} for the single-mass case (a comprehensive description of
the code is in the appendix of the same work), and by
\citet{AmaroSeoaneThesis04} for a stellar mass spectrum.

\subsubsection{H\'enon-type Monte-Carlo codes}

The Monte-Carlo (MC) numerical scheme is intermediate in realism and numerical
efficiency between Fokker-Planck or gas approaches, which are very fast but
based on a significantly idealised description of the stellar system, and
direct $N-$body codes, which treat (Newtonian) gravity in an essentially
assumption-free way but are extremely demanding in terms of computing time. The
MC scheme was first introduced by H\'enon to follow the relaxational evolution
of globular clusters \cite{Henon71a,Henon71b,Henon73,Henon75}. To our knowledge
there exist three independent codes in active
development and use which are based on H\'enon's ideas. The first is the one written by M.~Giersz (see
\cite{Giersz06} for his most recent published work), which implements many of
the developments first introduced by Stodo{\l}kiewicz \cite{Stodol82,Stodol86}.
Second is the code written by K.~Joshi \cite{JRPZ00,JNR01} and greatly improved
and extended by A.~G\"urkan and J.~Fregeau (see for instance
\cite{FregeauEtAl03,GFR04,FGR06,GFR06}). These codes have been applied to the
study of globular and young clusters. Finally, M.~Freitag developed an MC code
specifically aimed at the study of galactic nuclei containing a central MBH
\cite{FB01a,FB02b,FASK06}. The description of the method given here is based on
this particular implementation.

The MC technique assumes that the cluster is spherically symmetric and
represents it as a set of particles, each of which may be considered as a
homogeneous spherical shell of stars sharing the same orbital and stellar
properties. The number of particles may be lower than the number of stars in
the simulated cluster but the number of stars per particle has to be the same
for each particle.  Another important assumption is that the system is always
in dynamical equilibrium so that orbital time scales need not be resolved and
the natural time-step is a fraction of the relaxation (or collision) time.
Instead of being determined by integration of its orbit, the position of a
particle (i.e., the radius $R$ of the shell) is picked up at random, with a
probability density for $R$ that reflects the time spent at that radius:
$\mathrm{d}P/\mathrm{d}R\propto 1/V_\mathrm{r}(R)$ where $V_\mathrm{r}$ is the
radial velocity. The Freitag scheme adopts time steps that are some small fraction $f$
of the local relaxation (or collision) time: $\delta t(R) \simeq f_{\delta t}
\left(\trlx^{-1} + \tcoll^{-1}\right)^{-1}$.  Consequently the central parts of
the cluster, where evolution is faster, are updated much more frequently than
the outer parts. At each step, a pair of neighbouring particles is selected
randomly with probability $P_\mathrm{selec} \propto 1/\delta t(R)$. This
ensures that a particle stays for an {\em average} time $\delta t(R)$ at $R$
before being updated.

Relaxation is treated as a diffusive process, using the classical
Chandrasekhar theory on which FP codes are also based. The long-term effects on
orbits of the departure of the gravitational field from a smooth stationary
potential are assumed to arise from a large number of uncorrelated, small angle,
hyperbolic 2-body encounters. If a star of mass $M_1$ travels with relative
velocity $v_\mathrm{rel}$ through a homogeneous field of stars of mass $M_2$
with number density $n$ for a time $\delta t$, then in the centre-of-mass reference
frame, its trajectory will be deflected by an angle $\theta_{\delta t}$ with
average values

\begin{eqnarray}
\nonumber \langle \theta_{\delta t} \rangle &=& 0 \mbox{\ \ and}\\
\langle \theta^2_{\delta t}
\rangle &=& 8\pi \ln\Lambda \, G^2 n \left(M_1+M_2\right)^2 \delta t,
\label{eq.thetaMC}
\end{eqnarray}

\noindent 
where $G$ is the gravitational constant and $\ln\Lambda\simeq 10-15$
is the Coulomb logarithm. In the MC code, at each step, the velocities
of the particles of the selected pair are modified as if by a
hyperbolic encounter with deflection angle
$\theta_\mathrm{eff}=\sqrt{\langle \theta^2_{\delta t} \rangle}$. The
particles are then put at random positions on the slightly modified
orbits. As a given particle will be selected many times, at various
positions on its orbit, the MC scheme will integrate the
effect of relaxation over the particle's orbit and over all possible
field particles. Proper averaging is ensured if the time steps are
sufficiently short for the orbit to be modified significantly only
after a large number of effective encounters. The energy is trivially
conserved to machine accuracy in such a scheme because the same
deflection angle $\theta_\mathrm{eff}$ is applied to both particles in
an interacting pair. Only the direction of the relative velocity
vector is changed by $\theta_\mathrm{eff}$.

Using a binary tree structure which allows quick determination and updating
of the potential created by the particles, the self gravity of the stellar
cluster is included accurately. This potential is not completely smooth because
the particles are infinitesimally thin spherical shells whose radii change
discontinuously. Test computations have been used to verify that the
additional, unwanted, relaxation is negligible provided the number of particles is
larger than a few tens of thousands.

Although H\'enon's method is based on the assumption than all departures from
the smooth potential can be treated as 2-body small angle scatterings, it is
flexible enough to incorporate more realism. The dynamical effect of binaries
(i.e., the dominant 3- and 4-body processes), which may be crucial in the
evolution of globular clusters, has been included in various MC codes through
the use of approximate analytical cross-sections \cite{Stodol86,GS00,RFJ01}.
Recently, Fregeau \cite{FGR06,GFR06} has introduced a much more realistic
treatment of binaries by on-the-fly, explicit integrations of the 3- or 4-body
interactions, a brute force approach that is necessary to deal with the full diversity of
unequal-mass binary interactions. This approach was pioneered by
\citeauthor{GS03} \cite{GS03} in a hybrid code where binaries are followed as
MC particles while single stars are treated as a gaseous component. 

The few 2-body encounters that lead to large angle ($> \pi/10$, say)
deflections are usually neglected. In globular clusters, these ``kicks'' have a
negligible imprint on the overall dynamics \cite{Henon75,Goodman83} but it has
been suggested that they lead to a high ejection rate from the density cusp
around a central (I)MBH \cite{LT80}. Kicks can be introduced in the MC code,
where they are treated in a way similar to collisions, with a cross section $\pi
b_\mathrm{l.a.}^2$, where
$b_\mathrm{l.a.}=f_{\mathrm{l.a.}}G(M_1+M_2)v_\mathrm{rel}^{-2}$.
$f_{\mathrm{l.a.}}$ is a numerical factor to distinguish between kicks and
``normal'' small angle scatterings (impact parameter $> b_\mathrm{l.a.}$).
However, recent simulations seem to indicate that such kicks have little
influence on the evolution of a stellar cusp around a MBH \citep{FASK06}.

The MC code is much faster than a direct $N-$body integration: a simulation of
a Milky-Way-type galactic nucleus represented by $10^7$ particles requires between a
few days and a few weeks of computation on a single CPU. Furthermore, with the
proper scaling with the number of stars, the number of stars represented is
independent of the number of particles. A high particle number is obviously
desirable for robust statistics, particularly when it comes to rare events such
as star-MBH interactions. In contrast, because they treat gravitational
(Newtonian) interactions on a elementary level, without relying on any theory
about their collective and/or long-term effects, the results of direct $N-$body
codes can generally be applied only to systems with a number of stars equal to
the number of particles used.

\subsubsection{Applications of Monte-Carlo and Fokker-Planck 
simulations to the EMRI problem}
\label{sec.EMRIStatMethods}

MC and FP codes are only appropriate for studying how collisional
effects (principally relaxation) affect spherical systems in dynamical
equilibrium. These assumptions are probably valid within the radius
of influence of MBHs with masses in the LISA range. Indeed, assuming
naively that the {\SgrA} cluster at the centre of our Galaxy is
typical (as far as the total stellar mass and density is concerned)
and that one can scale to other galactic nuclei using the $M - \sigma$
relation in the form $\sigma = \sigma_{\rm MW} (\MBH/3.6\times
10^6\,\Msun)^{1/\beta}$ with $\beta\approx4-5$
\cite{FM00,TremaineEtAl02}, one can estimate the relaxation time at
the radius of influence to be $\trlx(R_{\rm infl}) \approx 25\times
10^9\,{\rm yr}\,(\MBH/3.6\times 10^6\,\Msun)^{(2-3/\beta)}$.
 
{
Although observations suggest a large spread amongst the values of the
relaxation time at the influence radius of MBHs with similar mass (see,
e.g., Fig.~4 of \cite{MHB06}), most galactic nuclei hosting MBHs less
massive than a few $10^6\,\Msun$ are probably relaxed and amenable to
MC or FP treatment. Even if the age of the system is significantly
smaller than its relaxation time, such approaches are valid as long as
the nucleus is in dynamical equilibrium, with a smooth, spherical
distribution of matter. In such conditions, relaxational processes are
still controlling the EMRI rate, no matter how long the relaxation
time is, but one cannot assume a steady-state rate of diffusion of
stars onto orbits with small pericentres, as is often done in FP codes
(see the discussion in \cite{MM03}, in the different context of
the evolution of binary MBHs).
}

The H\'enon-type MC scheme has been used by Freitag and collaborators to
determine the structure of galactic nuclei \cite{FreitagThesis00,FB02b,FASK06}.
Predictions for the distribution of stars around a MBH have also been obtained
by solving some form of the Fokker-Planck equation
\cite{BW77,MCD91,HA06,HA06b,MHB06} or using the gaseous model
\cite{AmaroSeoaneThesis04,ASFS04}.  These methods have proved useful to
determine how relaxation, collisions, large-angle scatterings, MBH growth,
etc., shape the distribution of stars around the MBH, which is an obvious
prerequisite for the determination of the rate and characteristics of EMRIs. Of
particular importance is the inward segregation of stellar BHs as they lose
energy to lighter objects. This effect, combined with the fact that stellar BHs
produce GWs with higher amplitude than lower-mass stars, explains why they are
expected to dominate the EMRI detection rate \cite{SR97,HA06b}. An advantage of
the MC approach is that it can easily and realistically include a continuous
stellar mass spectrum and extra physical ingredients. However, the first point
might not be critical here as MC results suggest that, for models where all the
stars were born $\sim10\,$Gyr ago, the pattern of mass segregation can be well
approximated by a population of two components only, one representing the
stellar BHs and the other representing all other (lighter) objects \cite{FASK06}.
Furthermore, the uncertainties are certainly dominated by our lack of knowledge
about where and when stellar formation takes place in galactic nuclei, what 
the masses of the stars which form might be, and what type of compact remnants they
become. 

The most recent FP results concerning mass segregation were obtained under the
assumptions of a fixed potential and an isotropic velocity dispersion, with the
effects of (standard or resonant) relaxation being averaged over $J$ at a given
energy. The MC code includes the self-gravity of the cluster so the simulated
region can extend past the radius of influence, allowing a more natural outer
boundary condition. We note that one has to impose a steeper density drop-off
at large radii than what is observed to limit the number of particles to a
reasonable value while keeping a good resolution in the region of influence.
The MC code naturally allows anisotropy and implicitly follows relaxation in
both $E$ and $J$. Anisotropic FP codes for spherical self-gravitating systems
exist \cite{Takahashi96,Takahashi97,DCLY99} but, to our knowledge, none are
currently in use that also include a central MBH. Unique amongst all stellar
dynamical codes based on the Chandrasekhar theory of relaxation is {\sc
Fopax}, a FP code which assumes axial rather than spherical symmetry, thus
permitting the study of clusters and nuclei with significant global rotation
(see~\cite{FSK06} and references therein) and which has recently been adapted to
include a central MBH \cite{FiestasThesis06}.

Determining the EMRI rates and characteristics is a harder challenge for
statistical stellar dynamics codes because these events are intrinsically rare
and critically sensitive to rather fine details of the stellar dynamics around
a MBH. As we explained in \S~\ref{sec.standard_picture}, the main difficulty,
in comparison with, for example, tidal disruptions, is that EMRIs are not
``one-passage'' events but  must be gradual.  The first estimate of EMRI rates
was performed by \citeauthor{HB95} \cite{HB95}. Assuming a static cusp profile,
they followed the evolution of the orbits of test-particles subject to GW emission
(Equations~\ref{aevol} and \ref{eccevol}) and 2-body relaxation introduced by
random perturbations of the energy and angular momentum according to
pre-computed ``diffusion coefficients''. Hopman \& Alexander \cite{HA05} have
used a refined version of this ``single-particle Monte-Carlo method'', as well
as the Fokker-Planck equation, to make a more detailed analysis. It was found
that no more than $\sim10\%$ of the compact objects swallowed by the MBH are
EMRIs, while the rest are direct plunges.

Determination of EMRI rates and characteristics were also attempted
with Freitag's MC code
\cite{Freitag01,Freitag03,Freitag03b}. Despite its present limitations, 
this approach might serve to inspire future, more accurate,
computations and is therefore worth describing in some detail. The MC
code does not include GW emission explicitly (or any other relativistic
effects). At the end of the step in which two particles have
experienced an encounter (to simulate 2-body relaxation), each
particle is tested for entry into the ``radiation-dominated'' regime,
defined by Eq.~\ref{eq.EMRIcond} (with $C_{\rm EMRI}=1$). A
complication arises because the time step $\delta t$ used
in the MC code is a fraction $f_{\delta t}=10^{-3}-10^{-2}$ of the
local relaxation time $\trlx(R)$, which is generally much larger than
the critical timescale defined by the equality $\tGW(e,a) = C_{\rm
EMRI}\, (1-e)\trlx$. In other words, the effective diffusion angle
$\theta_{\rm eff}$ is generally much larger than the opening angle of
the ``radiation cone'', $\tilde\theta\equiv(1-\tilde{e})^{1/2}$. So that the entry of the particle into the radiation cone (corresponding to a possible EMRI) is not missed, it is assumed that, over $\delta t$, the energy of a given
particle does not change. Hence, each time it comes back to a given
distance from the centre, its velocity vector has the same modulus but
relaxation makes its direction execute a random walk with an
individual step per orbital period of $\theta_{\rm orb} = \theta_{\rm
eff} (P_{\rm orb}/\delta t)^{1/2}$. Entry into the unstable or
radiation cone is tested at each of these sub-steps. If the particle
is found on a plunge or radiation-dominated orbit, it is immediately
removed from the simulation and its mass is added to the MBH.

Unfortunately, in addition to this approximate way of treating relaxation on
small time scales, there are a few reasons why the results of these simulations
may be only indicative. One is the way $\trlx$ is estimated, using the
coefficient in front of $\delta t$ in Eq.~\ref{eq.thetaMC}, i.e., an estimate
based on the neighbouring particle. Even if it is correct on average, this
estimate is affected by a very high level of statistical noise and its value
can be far too long in some cases (e.g., when the relative velocity between the
particles in the pair is much larger than the local velocity dispersion). This could lead one to conclude erroneously that a star has reached the
radiation-dominated regime and will become an EMRI. To improve on this one
could base the $\trlx$ estimate on more than one point on the orbit and on more than one
``field-particle'' ({the number of stars within a distance of $10^{-2}$\,pc of
{\SgrA} is probably larger than 1000, so $\trlx$ is a well-defined quantity
even at such small scales}). Another limitation is that GW emission is not
included in the orbital evolution, which forces one to assume an abrupt
transition when $\tau_{\rm GW} = (1-e)\trlx$. Hopman \& Alexander \cite{HA05}
have also shown that a value of $C_{\rm EMRI}$ as small as $10^{-3}$ might be
required to be sure the EMRI will be successful. Furthermore, the MC
simulations carried out so far suffer from relatively poor resolution, with
each particle having the statistical weight of a few tens of stars. To improve
this one would need to limit the simulation to a smaller volume (such as the
influence region) or develop a parallel implementation of the MC code to use
$\sim 10^8$ particles. Finally, these MC simulations did not include resonant
relaxation, an effect which can increase the EMRI rate by of order 10 or
completely suppress it, depending on its strength \citep{HA06}.

\vspace{0.5cm}
\noindent {\it Future of statistical stellar-dynamical simulations for EMRIs}
\vspace{0.3cm}

Despite recent progress, a full understanding of the stellar dynamical
processes leading to EMRIs and a robust prediction for their rate and
characteristics (eccentricities, stellar types and masses etc.) are still
lacking. The key points to address include:

\begin{itemize}
\item Interplay between relaxation and GW emission in the orbital evolution of EMRI ``precursors''. 
\item Importance of resonant relaxation.
\item Dynamics of binaries in a galactic nucleus.
\item Effects of non-sphericity in the potential.
\end{itemize}

\noindent
We now suggest how new or updated statistical stellar-dynamics
methods can contribute to clarify these points.

A self-gravitating, multi-mass, anisotropic Fokker-Planck \cite{MCD91} code,
including a detailed treatment of the loss-cone \cite{CK78} and with additional
terms to include GW-emission and resonant relaxation \cite{Ivanov02,HA05,HA06}
could provide EMRI (and direct-plunge) rates while accounting for mass
segregation at all scales, from the vicinity of the MBH to well beyond the
radius of influence, thus including the whole volume where relaxational effects
might play a role. Contrary to FP codes the gaseous approach does not operate
in $(E,J)$ phase-space where each point correspond to a well-defined orbit, but
in direct space (i.e., the radial coordinate $R$) at each point of which only
statistical quantities (velocity dispersions) are used to represent
approximately the kinematics of all orbits crossing this position. Therefore
the treatment of loss-cone effects can only be done in a much more approximate
way \cite{AmaroSeoaneThesis04,ASFS04}, making it probably unsuitable for determining EMRI rates.
The FP codes evolve distribution functions, therefore relying on the assumption
that any ``interesting'' element of phase-space always contains a significant
number of stars. It is not clear how to apply them to or interpret their
results for situations such as EMRIs, corresponding to rare events where
small-number effects might be crucial. They require that the stellar population
be discretised into a set of ``components'' sharing the same properties ($E$,
$J$, stellar mass and age, etc.) and can therefore not treat accurately a continuous mass
function, a mixed-age population, or binaries.

Monte-Carlo codes are based on a more direct star-by-star (or
particle-by-particle) approach making it much easier to consider
realistically complex stellar populations and to follow in detail the
evolution of individual orbits. Although single-particle MC computations
lack self-consistency, the only application of an H\'enon-type code to
EMRIs suffered from a few shortcomings connected with insufficient
resolution in time or particle number. However there are different
avenues for useful MC work:

\begin{itemize}
\item {\bf Monte-Carlo simulations of binary dynamics.} 
The evolution of binaries in a galactic nucleus is still a virtually
unexplored territory whose study is made urgent both by the
possibility of forming EMRIs (and hyper-velocity stars) through binary
separation and by the observation of X-ray binaries around {\SgrA}
\cite{MunoEtAl05,MunoEtAl05b}. Binary stars can be included in a very realistic way into
H\'enon-type MC codes, using explicit few-body integrations to treat
interactions between a binary and another single or binary star
\cite{GS03,FGR06} or between a binary and the central MBH.
\item {\bf Keplerian Monte-Carlo.} 
Barring tidal processes such as binary separation or the capture of
the core of a giant, the rate of EMRIs is determined by the stellar
dynamics well within the influence radius of the MBH. To follow the
dynamics in this region with both accuracy and computational
efficiency, the development of a specialised MC code is planned. The
potential will be the simple Keplerian contribution of the MBH with
no account for self gravity of the stellar cluster besides
relaxation. It is expected that, in such an external potential, one
does not need to conserve energy exactly as for a self-gravitating
cluster. Therefore, each particle (representing just one star) can
have its own time step, adapted to its orbit. The effects of GW
emission can be taken into account explicitly and the dynamics in or
close to the ``radiative'' cone can be followed on appropriately
short timescales. 
\item {\bf Non-spherical hybrid code.}
Departure from the assumption of sphericity may be important for two
non-standard EMRI channels (see~\S~\ref{sec.nonstd}). The first is the one
involving a dense accretion disc. In this case, the potential might
still be essentially spherical but dissipative interactions with the
disc might cause a flattening of the distribution of stars and a
significant rate of equatorial circular EMRI events. The second
situation is the contribution of binary tidal separations which
(unlike standard EMRIs) can be increased significantly by replenishing
the loss cone outside the sphere of influence. At the present time
{\sc Fopax} is the only statistical stellar-dynamics code which does
not rely on the assumption of spherical symmetry. It is, however,
limited to axially symmetric systems and a particle-based method would
be more suitable for the inclusion of star-disc interaction effects or
binaries. Following a suggestion already made by H\'enon
\citep{Henon71b}, a collision-less $N-$body code could be used as a
``backbone'' orbit integrator and complemented by the pairwise
treatment of relaxation of the Monte-Carlo method. S.~Sigurdsson has
applied a similar idea to the study of globular clusters by combining
the self-consistent-field $N-$body algorithm with FP terms for
2-body relaxation \cite{JSH99}.
  
\end{itemize}

Some possibly important aspects of the dynamics cannot be treated
directly by statistical methods. Important examples are resonant relaxation or the
possible effects of the motion of the MBH (\cite{MBL06} and references
therein). These processes require direct $N-$body simulations for
their study before they can be introduced in an approximate
way into FP or MC codes.


\subsubsection{Direct-summation $N-$body codes}
\label{sec:directNbody}

We finish by considering the direct $N-$body approach~\citep{Aarseth99,Aarseth03,PortegiesZwartEtAl01}.
This is the most expensive method because it involves integrating all
gravitational forces for all particles at every time step, without making any a
priori assumptions about the system. The \nb codes use the improved Hermite
integration scheme as described in~\citep{Aarseth99,Aarseth03}, which requires
computation of not only the accelerations but also their time derivatives.
Since these approaches integrate Newton's equations directly, all Newtonian
gravitational effects are included naturally. More crucial for this subject is
that the family of the direct \nb codes of Aarseth also includes versions in which both {\em
KS regularisation} and {\em chain regularisation} are employed, so that when
particles are tightly bound or their separation becomes too small during a
hyperbolic encounter, the system is regularised (as described first
in~\cite{KS65,Aarseth03}) to prevent dangerous small individual time steps.
This means that we can accurately follow and resolve individual orbits in the
system. Other schemes which make use of a softening in the gravitational forces
(i.e. $1/(r^2+\epsilon^2)$ instead of $1/r^2$, where $\epsilon$ is the
softening parameter) cannot be employed because $\epsilon$ can induce
unacceptable errors in the calculations.  The $N-$body codes scale as $\Nstar^2$, or
$\Delta t \propto t_{\rm dyn}$, which means that even with special-purpose
hardware, a simulation can take of the order of weeks if not months. This
hardware is the GRAPE (short for GRAvity PipE), a family of hardware which acts
as a Newtonian force accelerator.  For instance, a GRAPE-6A PCI card has a peak
performance of 130 Gflop, roughly equivalent to 100 single PCs \citep{GRAPE6A}.
It is possible to parallelise basic versions of the direct \nb codes (without
including regularisation schemes) on clusters of PCs, each equipped with one
GRAPE-6A PCI card. This leads to efficiencies greater than 50\% and speeds in
excess of 2 TFlops and thus the possibility of simulating up to $\Nstar =
2\cdot 10^6$ stars~\citep{HarfstEtAl06}. Nevertheless, when we consider the
situation relevant to an EMRI, in which mass ratios are large and we need to
follow thousands of orbits, the Hermite integrator is not suitable and problems
show up even in the Newtonian regime. Aarseth et al. \cite{Aarseth06,Aarseth03}
summarise different methods developed to cope with large systems with one or
more massive bodies. The problem becomes even more difficult when including
relativistic corrections to the forces when the stellar black hole approaches
the central MBH, because extremely small time-scales are involved in the
integration. Progress is being made in this direction with a recently developed
time-transformed leapfrog method \citep{MA02} (for a description of the
leapfrog integrator see~\citep{MM06}) and the even more promising wheel-spoke
regularisation, which was developed to handle situations in which a very
massive object is surrounded by strongly bound particles, precisely the
situation for EMRIs \citep{Zare74,Aarseth03}. Additionally, one must include
post-Newtonian ($\PN$) corrections in the direct \nb code because secular
effects such as Kozai or resonant relaxation may be smoothed out significantly
by relativistic precession and thus have an impact on the number of captures
\citep{HA06}.

There already exists a version of direct \nb that includes relativistic
corrections at $1\PN$, $2\PN$ (periapsis shifts) and
$2.5\PN$ (energy loss in the form of gravitational wave emission)
\citep{KupiEtAl06}. The authors included perturbations in the {\em KS
regularisation} scheme, so that the forces (actually the accelerations) were
modified by

\begin{equation}
{F}  = \underbrace{{F}_0}_{\rm Newtonian}
+\underbrace{\underbrace{c^{-2}{F}_2}_{1\PN} +
\underbrace{c^{-4}{F}_4}_{2\PN}}_{\rm periapsis~shift} +
\underbrace{\underbrace{c^{-5}{F}_5}_{2.5\PN}}_{\rm GW} +
\mathcal{O}(c^{-6})
\label{eq.F_PN}
\end{equation}

Note that the perturbations do not need to be small compared to the two-body
force \citep{Mikkola97}. The method will be applicable even when the relativistic
terms become comparable to the Newtonian term provided the KS time step is duly
adjusted. Nonetheless, the gauge choice is crucial, since it can introduce an
artificially enlarged cross-section for the pairs if the centre-of-mass of
the system is ``wobbling'' \citep{DD81,Soffel89}. The desired frame is the
centre-of-mass frame, which is equivalent to the centre-of-mass Hamiltonian in the ADM
(Arnowit, Deser and Misner) formalism~\cite{BlanchetIyer03}. The relative accelerations then
have the form

\begin{equation}
\frac{d v^i}{dt}=-\frac{m}{r^2}\Big[(1+{\cal A})\,n^i + {\cal B}\,v^i
\Big]+ {\cal O}\left( \frac{1}{c^7} \right)\;,
\label{eq.Blanchet}
\end{equation}

\noindent
\citep{BlanchetIyer03} where the relative separation of the binary components is
$x^i=y_1^i-y_2^i$, $r=|{\bf x}|$ and $n^i={x^i}/{r}$; ${\cal A}$ and ${\cal
B}$ are given by the expressions (3.10a) and (3.10b) of \cite{BlanchetIyer03}.
Whilst the gauge choice was not a problem for the system studied in~\citep{KupiEtAl06}, since
they were interested in the global dynamical evolution, for
the EMRI problem the centre-of-mass frame (located at the origin of the
coordinates) must be employed. The integration cannot
be extended to velocities higher than $\sim 0.3$c, because at these velocities the post-Newtonian
formalism can no longer be applied accurately. This means that we cannot reach the
final coalescence of the stellar BH with the MBH, but nonetheless the parameter space that can be explored is unprecedented in scope. There has also been encouraging progress including $\PN$ corrections in conjunction with alternative high-accuracy
methods \citep{AmaroSeoaneKupiFregeau06}. We note that it will not be possible to include in N-body codes all the $\PN$ corrections that are required for accurate modelling of the phase evolution of the EMRI during the last few years before plunge. However, the N-body codes are not required in that regime, since the system is then decoupled from the rest of the stellar cluster. Accurate modelling of this regime is required for detection of these systems, and this will be described in Section~\ref{EMRIModel}.


\section{EMRI detection}
\subsection{Data analysis algorithms}
\label{EMRIDA}

A typical EMRI signal will have an instantaneous amplitude an order of
magnitude below the LISA's instrumental noise and (at low frequencies) as many
as several orders of magnitude below the gravitational wave foreground from
Galactic compact binaries. This makes detection a rather difficult problem.
However, the signals are very long lived, and will be observed over more than
$10^5$ cycles, which in principle allows the signal-to-noise-ratio (SNR) to be
built up over time using matched filtering. Estimates of the number of
important parameters in EMRI evolution range from 7 to 15. Even taking a number
at the lower end of this range, the naive expectation is that $N \sim 10^{35}$
templates would be needed to carry out a fully coherent matched filtering
search \cite{emri04}. This is far more than can reasonably be searched with
realistic computing resources. Several alternative approaches to EMRI detection
that will be computationally feasible have been investigated. These will be
able to detect signals with matched filtering SNR $\gtrsim 20$. By comparison,
in a fully coherent search, the SNR required for detection is $~12-14$, to
ensure a reasonable false alarm rate when searching such a huge number of
templates.
 
\begin{figure}
\resizebox{\hsize}{!}{\includegraphics[scale=1,clip] {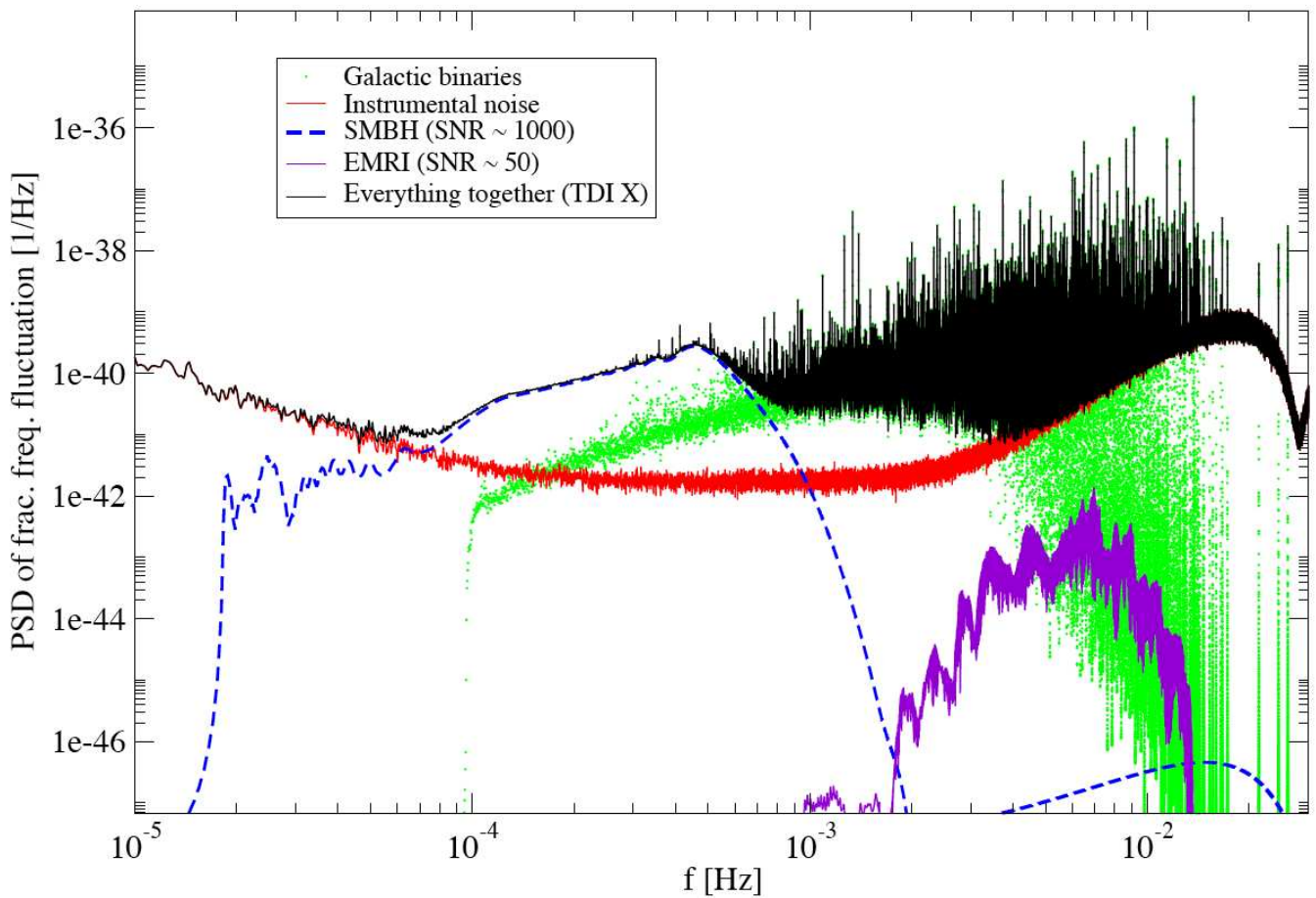}}
\caption{Power spectral density of one of
the unequal arm Michelson TDI channel. It contains 1 MBH inspirals at
luminosity distances of 3.3 Gpc and 1 EMRI at luminosity distances of
2.3 Gpc. The duration of the EMRI was taken to be one and a half years. 
The galactic binary realisation used here was drawn from the
distribution described in \cite{nelemans01}
\label{X-stream}}  
\end{figure}

LISA data analysis is further complicated by the richness of the LISA data
stream.  The motion of LISA in its orbit creates amplitude and phase modulation
of the signals and we have to employ time-delay interferometry
techniques~\cite{aet99,tintolr} in order to remove the laser frequency noise.
Time-delay algorithms applied to the Doppler readouts lead to a rather
complicated response function which depends on the frequency and sky position
of the source under investigation.  It is expected that the detection rates for
EMRIs will fall somewhere between a few tens and a few
thousands~\cite{emriconf,emri04,HA06b}. Additionally, the LISA data will be
very strongly coloured by gravitational wave signals from the foreground of
white dwarf binaries in our galaxy ($\sim 10^7$ sources which create confusion
noise at frequencies below a few
millihertz~\cite{hils90,bender98,nelemans01,farmer}) and signals from a handful
of merging supermassive black hole binaries which might have SNR as high as a
few thousand~\cite{wyithe03,enoki04,sesana05}. All these signals will overlap
in time and frequency. To illustrate this complexity we have simulated a LISA
frequency Doppler shift measurement with $\sim 27$ million Galactic binaries, 1
EMRI and 1 inspiralling MBH binaries. The power spectral density of each
separate source and the total envelope is presented in Figure~\ref{X-stream}, produced
using the methods outlined in \cite{cornish07,ArnaudEtAl07} and software written by MLDC task force
\footnote{\tt{http://sourceforge.net/projects/lisatools}}.
This Figure is primarily illustrative, but it appears that the EMRI signals
will be overwhelmed by the signals from the MBH mergers. However, it should be
possible to identify and remove the high signal-to-noise ratio MBH merger
signals from the data stream before searching for the EMRI signals, so the
situation is not as bad as it may at first appear.

\subsubsection{Current status}
To date, three algorithms for detection of EMRIs in LISA data have been considered. 

The first is a semi-coherent algorithm, which uses a first coherent matched
filtering stage to search for $\sim 3$-week sections of EMRIs, followed by a
second stage where the power is summed incoherently along trajectories through
these sections that correspond to inspirals. This algorithm could detect EMRIs
at redshift $z\sim 1$, which translates to tens to hundreds of LISA events,
depending on the intrinsic astrophysical rate \cite{emri04}. The preliminary
analysis of this algorithm made only limited efforts to optimise its
performance. It is likely that optimisation, such as the addition of extra
stages in the hierarchy, will be able to further improve the reach of search,
but this has not yet been explored.

\begin{figure}
\resizebox{\hsize}{!}{\includegraphics[width=2in,keepaspectratio=true,angle=-90]
{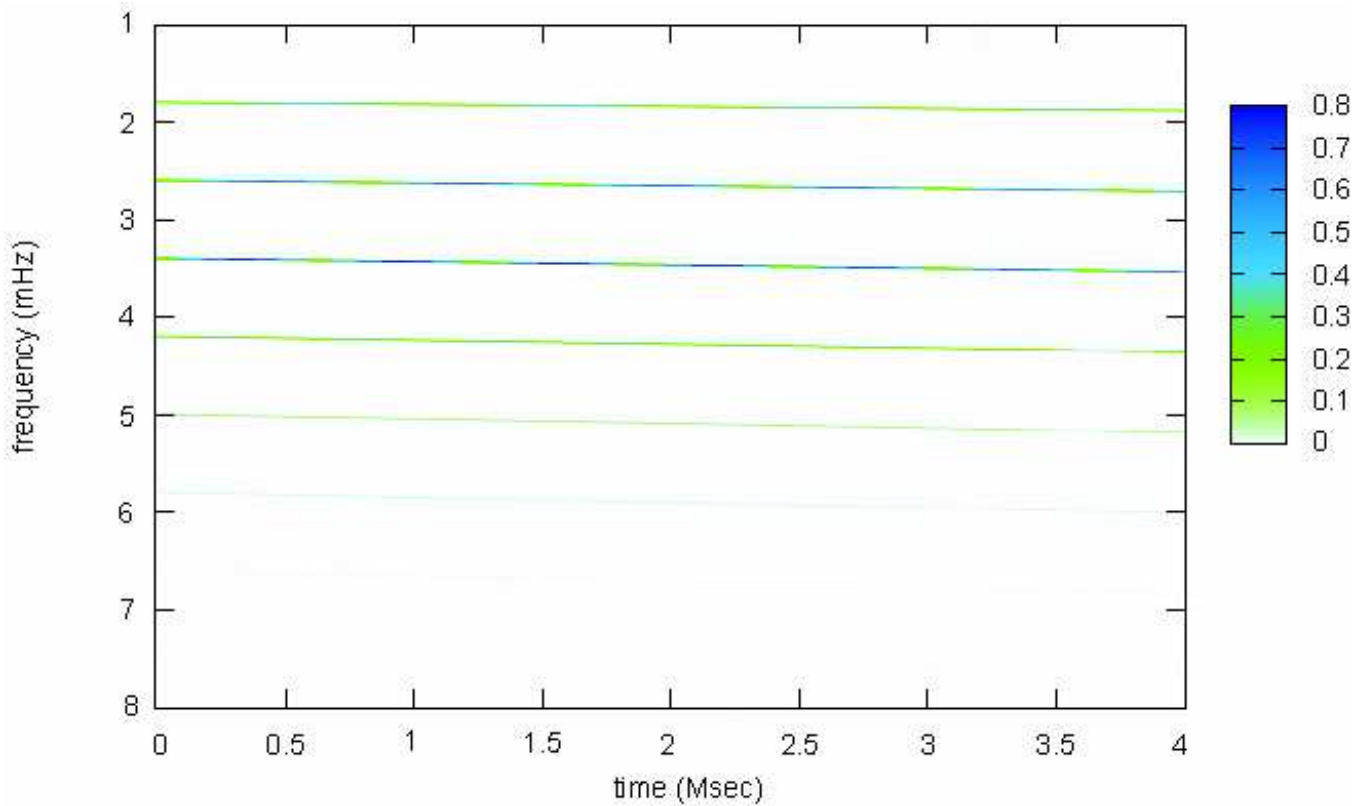}}
\caption{Spectrogram of the signal from an EMRI on an inclined and eccentric orbit. 
One can see several harmonics modulated by orbital precession and 
LISA's orbital motion. 
\label{EMRIspec}}
\end{figure}

A second approach is to use time-frequency methods, i.e., divide the data
stream into segments of a few weeks in length, perform a Fourier transform on
each and then analyse the resulting spectrogram. A simple method that looks for
unusually bright pixels in binned versions of this spectrogram could detect
typical EMRI signals at about half the distance of the semi-coherent search,
but at a tiny fraction of the computational cost \cite{wengair05,gairwen05}. An
improved method that considers clustering of bright pixels in the binned
spectrograms (the Hierarchical Algorithm for Clusters and Ridges), has slightly
further reach, and also more potential for parameter extraction
\cite{gairjones06}. While more work needs to be done, template-free techniques
could detect as many as one tenth of the EMRI sources in the data stream. A
typical spectrogram for an EMRI signal is presented in Fig.~\ref{EMRIspec} for
which the amplitude of the signal at plunge was normalised to one.

The third approach that has been explored is to use Markov Chain Monte Carlo
(MCMC) techniques. The MCMC approach essentially carries out fully coherent matched
filtering, but does so in an intelligent way, reducing the number of waveform
templates that have to be considered. MCMC methods are being explored
extensively for application to all aspects of LISA data analysis
\cite{cornish05,umstatter05,wickham06}. In the context of the EMRI search, the
MCMC approach has been found to work well when searching for a simple model
EMRI signal in a short stretch of LISA data \cite{stroeer06emri}. The exact
reach of the MCMC search has not yet been properly assessed. Given infinite
computing resources, the MCMC would eventually return the posterior probability
function for the source parameters. The ability to correctly identify sources
then depends on the shape of this posterior, which depends on both the
signal-to-noise ratios of the sources and the structure of the waveform template
space. These are the same properties that determine the detection limit of a
fully-coherent matched filtering search, so it is plausible that an MCMC search
with infinite computing power could achieve the same range as the fully
coherent matched filtering approach. However, the MCMC suffers from the same
computational constraints as the fully coherent search, and these will limit
the ability of an MCMC search to sample the posterior. Peaks in the posterior
will have to be larger to be detected, and sources will have to be
correspondingly closer. For these reasons, it is not possible to say how the
performance of the MCMC and the semi-coherent method will compare in practice,
although this should become clear over the next several years.

\subsubsection{Outstanding challenges} The results quoted above for the various search algorithms were obtained using a vastly simplified model of LISA
--- searching for a single EMRI event in coloured Gaussian noise.  However, as
already mentioned, the LISA data stream will actually be source-dominated and in
order to detect EMRIs we might first need to clear the data of signals from
MBH binaries and from resolvable Galactic binaries.  As many as $10^4$
Galactic binaries with frequencies above a few mHz will be individually
resolvable~\cite{farmer} and hopefully removed.  Estimates suggest there could
be tens of coalescing MBH binaries observed each year during the mission
lifetime~\cite{sesana05}. LISA could also see as many as several hundred
individually resolvable EMRI events~\cite{emri04} plus a confusion background
generated by distant EMRI signals~\cite{emriconf}.  In the work on EMRI
searches, the confusion background from unresolvable compact binaries has been
included in an approximate way, but no account has been made of the effect of
interference from the thousands of resolvable sources that will be present in
the data. Research on the problem of resolving individual sources in the rather
complex LISA data is under way within the Mock LISA Data Challenge
effort~\cite{mldc1,mldc2}.

Some of the questions that need to be addressed are as follows:
\begin{itemize}

\item How do the three existing algorithms perform when applied to data streams
containing two, ten or a hundred EMRI events? How does the performance degrade
when there are other sources, e.g., a MBH merger signal, in the data stream?
The semi-coherent and MCMC algorithms are likely to be better at handling
confusion than the time-frequency approach, since the former methods use
matched filtering. Time-frequency analyses will not be readily able to cope
with many sources of comparable brightness that intersect in the
time-frequency plane. Work needs to be done to quantify these statements.

\item If existing algorithms for the extraction of compact binaries and
supermassive black hole mergers from the data stream are used on a data stream
including one or more EMRI signals, how is the EMRI signal affected? Can the
process of ``cleaning'' other sources be modelled merely as an alteration in
the noise properties of the data stream? This will answer the important
question of whether it is necessary to fit simultaneously for all the sources
in the data, or whether the parameters of the different source types can be estimated
sequentially, before following up with a global fit and refinement of the
parameters.  Another concern here is how well we can model MBH merger signals --- the
mismatch between the true signal and the theoretical model could result in rather
high residuals.

\item Markov Chain Monte Carlo (MCMC) techniques in principle can search for
multiple types of source in the data stream, but how well do they perform when
searching for compact binaries, supermassive black hole mergers and EMRIs in
the same data set? How quickly does the chain converge? How complex is the
likelihood function with many different source types in the data stream? It has
recently been shown that MCMC based methods can detect MBH inspirals in the
presence of a Galactic WD binary foreground to a very high accuracy
\cite{cornish06}, which gives reason to hope that this problem will be
surmountable.

\item What are the computational costs of the various approaches? Which of them are
computationally feasible? As mentioned earlier, fully coherent matched
filtering is impossible due to the large number of templates required to tile
the whole parameter space. The semi-coherent algorithm was designed to make
maximum use of expected computational resources, and $3$ weeks was estimated to
be the longest possible ``snippet'' length under that assumption. Markov Chain
Monte Carlo techniques provide a more computationally efficient way to search
high-dimensional parameter spaces. The MCMC search for a simplified EMRI in 1
month of data described in~\cite{stroeer06emri} required the evaluation of
$\sim10^7$ chain states in order to determine the parameters of a source which
had SNR~$\sim10$. This is fairly typical of the number of states required to
accurately recover the posterior in searches for small numbers of
sources~\cite{cornish06}. However, the exact number of states required will
depend on the number of sources in the data (typically a linear scaling), the
complexity of the waveform space and the SNRs of the various sources. This will
likely increase the requirement by several orders of magnitude, although  the
MCMC should be able to obtain reasonable estimates of the parameters of the
loudest sources using many fewer than the $\sim 10^{35}$ templates required to
cover the whole parameter space in a fully coherent matched filtering
search~\cite{emri04}. However, the template at each one of these Markov chain
states still needs to be evaluated as the chain runs. The templates thus either
need to be generated ``on-the-fly'' or a bank of templates needs to be
generated in advance. For the former, we would need quick methods to generate
EMRI waveforms, such as the kludge models. For the latter, we are back to the
necessity that the entire parameter space be covered with EMRI templates. This
is unlikely to be computationally practical, and even if it were, the overhead
associated with accessing such a huge database would be prohibitively high.
Thus, computational costs will have to be considered very carefully when
devising the final EMRI search.

\end{itemize}

The final EMRI search will most likely include a combination of the three
approaches described above, and perhaps some new techniques. The search is
likely to be hierarchical, using inaccurate but quick techniques to get
estimates of the source parameters before refining with more computationally
intensive methods. It is likely that the searches for different types of LISA
source will be somewhat integrated with one another, although one possible
approach might be to estimate the parameters for each source type separately,
before unifying everything in a final global fit.

As described in Section~\ref{sec.nonstd}, there are various ``non-standard''
channels for EMRI formation that may produce comparable numbers of inspirals as
the standard picture. A binary tidal separation or the capture of the core of a
giant star would tend to lead to an EMRI on a circular orbit. The formation of
stars in a disc would tend to lead to an EMRI on a circular and equatorial
orbit. These special types of orbit have less free parameters than a generic
EMRI. It might therefore be worthwhile having three EMRI data analysis
pipelines, focused on circular-equatorial, circular-inclined and
eccentric-inclined orbits respectively. The reduction in the parameter space in
the restricted cases will probably not be sufficient to allow fully coherent
matched filtering to be carried out, but it will allow considerably longer
segments to be used in the first stage of a semi-coherent search. Moreover, the
threshold SNR required for detection of a source will be reduced (since there
are far fewer templates in which a false alarm could be found), increasing the
range of the search. It will be important to find out to what distance a
well-tuned search algorithm targeting circular or circular-equatorial EMRIs can
make a detection, but this has not yet been explored thoroughly.

No algorithms have been examined explicitly in the context of IMRI detection,
although the above algorithms for EMRI detection can all be applied. For the
matched filtering algorithms, the complication is the need for waveform models,
which will be discussed in more detail below. The time-frequency algorithms can
see IMRI events further away, since they are intrinsically brighter, so this
might be a good method to use, but it will depend on the distance to the
nearest likely event. As search techniques for EMRIs and MBH mergers are
further developed in the future, they can be expanded to encompass a search for
IMRIs.

\subsection{Source modelling}
\label{EMRIModel}

Most of the data analysis algorithms outlined above require models of the
source waveforms. Waveform templates will also be essential for parameter
estimation once sources have been detected. Templates for EMRIs could be
constructed in several ways which we discuss here.

\subsubsection{Current status}

\paragraph{Post-Newtonian expansion} The post-Newtonian expansion in powers of
velocity $v/c$ converges poorly when $v/c \gtrsim 0.3$~\cite{brady98}.  Unlike
comparable-mass inspirals, which only spend a few cycles in the regime where
the post-Newtonian approximation breaks down, IMRIs and EMRIs may spend
thousands to millions of cycles in this regime.  Therefore, this expansion is
not useful for EMRI or IMRI waveform modelling.

\paragraph{Numerical relativity} Solving Einstein's equations numerically on a
computer has proven to be a very difficult task, but significant progress has
been made in the past year. Several groups have now successfully modelled the
last orbit, merger and ring-down of a comparable mass binary system
\cite{pretorius05,baker06,camp06, bruegmann}. Numerical techniques are
essential for modelling the highly non-linear dynamics during the last few
orbits and merger of a comparable mass system. However, numerical techniques
are not fast enough to evolve the large number of cycles necessary for EMRI
waveforms. In addition there are technical problems which make numerical
methods unreliable as one goes to higher mass ratios. Fortunately, in EMRI
systems the extreme mass ratio makes it possible to produce templates accurate
over many cycles by perturbative methods expanding in the mass ratio, so
numerical relativity is not needed in this context.

\paragraph{Self-force waveforms} The extreme mass ratio in an EMRI system
allows the waveform to be obtained by perturbation theory. The inspiralling
object can be regarded as a small perturbation on the background space-time of
the central black hole, except very close to the small object. In the vicinity
of the small object, the space-time can be regarded as a point mass moving under the
influence of an external tidal field due to the central body.  Matching these
two regimes allows one to obtain an expression for the self-force acting on the
small body as a result of its motion in the space-time. This self-force can be
thought of as arising from gravitational radiation being generated by the small
object, reflecting off the curvature of the background space-time and then
subsequently acting on the small body. The mathematical theory of this
self-force interaction has been developed over the past ten years (see the
review \cite{poisson} and references therein). In principle, the self-force
formalism will provide accurate EMRI waveforms that can be used for source
characterisation. However, evaluation of the self-force is computationally
difficult. Recently, a new scheme was proposed \cite{barack05} which has
produced results for the self-force acting on circular orbits in the
Schwarzschild space-time~\cite{barack07}. However, it is computationally
expensive to generate the self-force acting even at a single point in an orbit.
A generic {\it inspiral} trajectory and waveform from a particle evolving as a
result of the self-force is still some way in the future.

\paragraph{Adiabatic inspiral waveforms} Evaluation of the self-force acting at
every point on the orbit is necessary to include the `conservative' piece of
the self-force, i.e., the piece that modifies the orbit, but does not dissipate
energy, $E$, angular momentum, $L_z$, or Carter constant, $Q$, (a generalised angular momentum
squared which is the third integral of the motion for orbits in the Kerr
space-time). However, the radiative piece of the self-force can be determined
more easily, by solving the perturbation equations for the background
space-time, with a source that represents a particle moving on a geodesic of
this background. This reduces to integrating the Teukolsky equation
\cite{teuk73,teuk74}. Solutions of the Teukolsky equation encapsulate radiation at
infinity in a single equation for the Weyl scalar $\psi_4$:

\begin{equation}
\psi_4 = \frac1{2}(\ddot{h}_+ - i\ddot{h}_{\times}).
\label{psi4eq}
\end{equation} 

The Teukolsky solution also determines the orbital averaged rates of change of
the orbital constants --- $\left<{\rm d} E/{\rm d} t\right>, \left<{\rm d} L_z/{\rm d} t\right>,
\left<{\rm d} Q/{\rm d} t\right>$ --- from which the value of these constants a short
period of time later can be determined. This allows the construction of `adiabatic waveforms'~\cite{hughes05} --- a sequence of geodesics can be found that represent an inspiral, by solving the Teukolsky equation for a given geodesic, then computing the energy, angular momentum and Carter constant loss rate for that geodesic, neglecting oscillatory terms which average to zero over the orbital period. These are then used to determine the next geodesic
in the sequence.  The corresponding gravitational waveforms generated on each geodesic orbit can then be stitched together to give an adiabatic inspiral waveform. This procedure works provided the timescale over which the orbit is changing is long compared to the orbital period, i.e., the evolution is adiabatic. For special classes of orbits --- eccentric equatorial and circular inclined --- the rate of change of the orbital constants is determined by energy and angular momentum balance. It is possible to extract from $\psi_4$ \erf{psi4eq} the gravitational waveform near the horizon and near infinity, and hence the amount of energy and angular momentum being carried by the waves down the horizon and out to infinity. Equating the loss of energy and angular momentum of the orbit to the sum of the energy and angular momentum fluxes near the horizon and near infinity determines the orbital evolution. This simplification meant that adiabatic inspirals were determined several years ago for both of these special classes of orbit~\cite{kgdk02,hughes00}, although the stitching together of waveforms has only been done for circular inclined orbits \cite{hughes01}.  Recently, `snapshot' waveforms and energy/angular momentum fluxes have also been generated for generic geodesics in the Kerr space-time \cite{drasco06}. Generic geodesics in Kerr have the third integral of the motion, the Carter constant, in
addition to the energy and the polar component of the orbital angular
momentum. It was originally thought that evaluation of the self-force
would be required to correctly evolve the Carter constant. However, it was
recently shown that the evolution of the Carter constant can also be determined
from the same Teukolsky coefficients that are needed for computing the energy
and angular momentum fluxes \cite{mino03,sago05,sago06}.

This should allow the construction of generic adiabatic waveform templates in
the near future. Adiabatic waveforms are accurate except for the omission
of the conservative piece of the self-force. There is some debate in the
literature about how important this omission will be \cite{drasco05,pound05}.
However, the adiabatic waveforms may be accurate enough over a timescale of a
few weeks that they can be used for source detection via the semi-coherent
algorithm. They may also perhaps find a role in source characterisations and are
somewhat less computationally expensive to generate than full self-force
waveforms.

\paragraph{``Kludge'' waveforms}

The adiabatic waveforms, although accurate, are still computationally intensive
to compute. For the purposes of scoping out data analysis algorithms for the
detection of EMRIs with LISA, it is necessary to generate waveforms in large
numbers, e.g., to count the number of templates needed to cover the whole
parameter space with sufficiently high overlap. Perturbative waveforms did not
fit this requirement, which led to the development of two families of
approximate, ``kludge'' waveforms, that capture the main features of true EMRI
waveforms but are much quicker and easier to generate. 

The first family of kludge waveforms use an ``analytic kludge'' (AK).
They are based on waveforms representing emission from a particle on a
Keplerian orbit, as given by Peters and Mathews \cite{PM63,peters64}. The
waveform is augmented by imposing relativistic precession of the orbital
periapsis and orbital plane, plus inspiral (in an analogous way to how the
adiabatic waveforms include inspiral). The precession and inspiral rates are
taken from post-Newtonian results. These waveforms are described in
\cite{BC04}. The AK approach is ``analytic'' since analytic expressions are
known for the Peters and Mathews waveforms. This makes the AK waveforms very
quick to evaluate. However, they are not particularly accurate in the latter
stages of inspiral, since a Keplerian orbit with precessions is not a good
approximation to a true Kerr geodesic close to the central black hole.

The second family of kludge waveforms attempts to address this failing by using
a true geodesic orbit for the inspiralling particle. The geodesic equations
have to be integrated numerically, so the second family is labelled the
``numerical kludge'' (NK). The procedure to compute a numerical kludge waveform
has two stages. Firstly, a phase-space inspiral trajectory is constructed,
i.e., the sequence of geodesics that an inspiral passes through, by integrating
prescriptions for the evolution of the three constants of the motion ---
energy, angular momentum and Carter constant. An initial prescription for this
evolution based on post-Newtonian expansions of the Teukolsky function
\cite{GHK02} was found to exhibit pathologies in certain regimes. By imposing
consistency corrections and augmenting the evolution with higher order
post-Newtonian terms and fits to solutions of the Teukolsky equation, a
considerably improved prescription for the inspiral has now been obtained
\cite{GG06}. This current inspiral prescription is accurate until very close to
the end of the inspiral for circular orbits. It is less accurate for eccentric
orbits, but it should be possible to improve this in the future now that
Teukolsky data for generic orbits is available. Once a phase space trajectory
has been obtained, the Kerr geodesic equations can be numerically integrated,
with the time-dependent constants of the motion inserted. The trajectory of the
inspiralling particle through the Kerr background is then obtained. The second
stage of the NK construction is to construct a waveform based on this
trajectory. This is done by identifying the Boyer-Lindquist coordinates of the
particle trajectory with spherical polar coordinates in flat space, and
applying a weak field gravitational wave emission formula to the pseudo-flat space
trajectory. NK waveforms have been constructed using the standard flat space
quadrupole radiation formula, the quadrupole-octupole formula
\cite{bekenstein73} and the Press formula \cite{press77}, which is derived for
fast-motion but weak field sources. All three prescriptions perform well when
compared to more accurate, adiabatic waveforms, but there is little gain from
using the Press formula in preference to the quadrupole-octupole expression
\cite{babak06}.

\begin{figure}
\resizebox{\hsize}{!}{\includegraphics[scale=1,clip]
{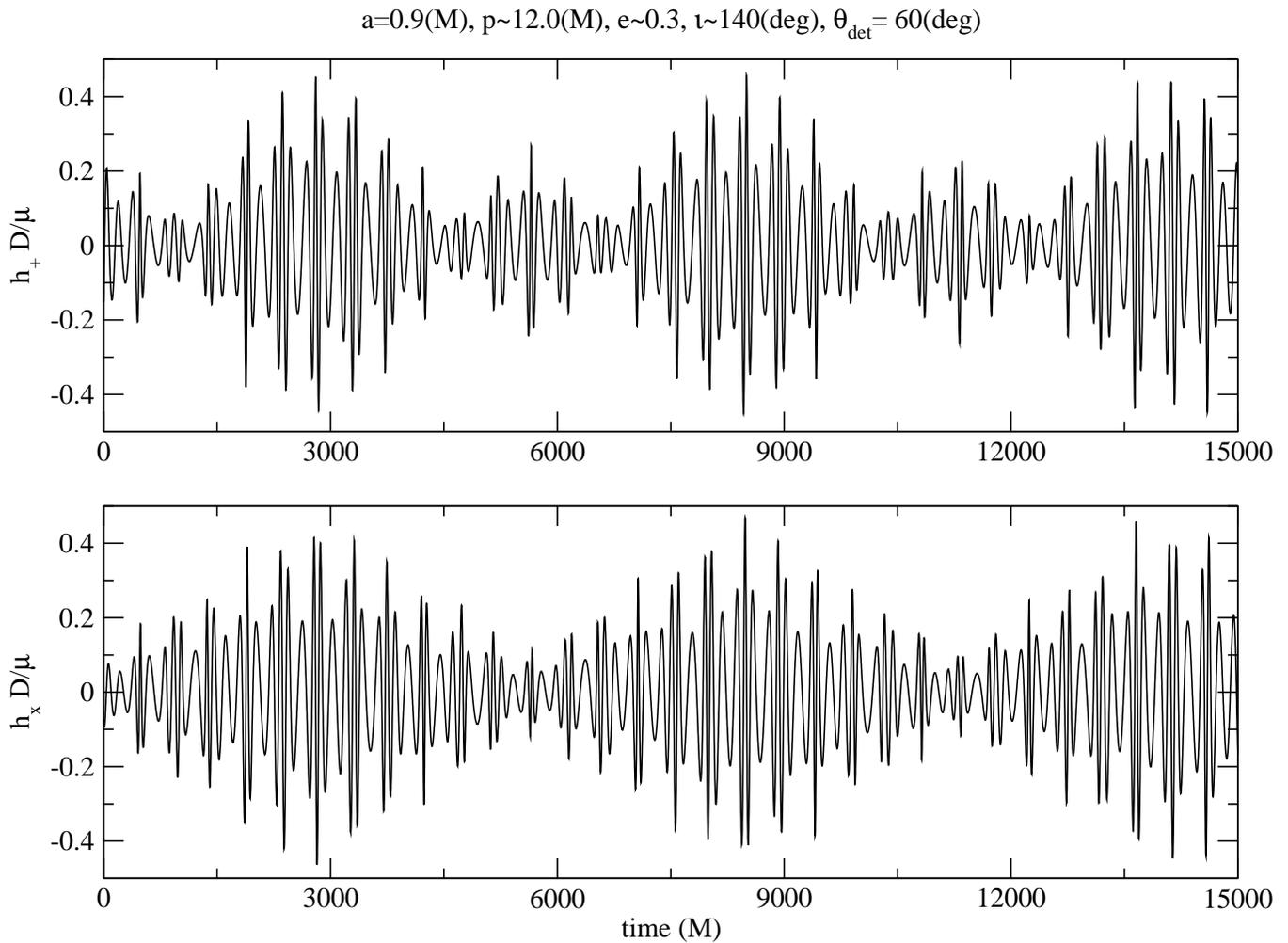}}
\caption{The two polarisations of an NK EMRI waveform with a mass ratio of $10^{-7}$.
The GW amplitude is measured in units of the mass of the compact object over distance ($\mu/D$) and
time is measured in units of MBH mass $M$. The eccentricity is $\sim 0.3$, the semi-latus rectum $p \sim 12M$, the inclination of the plane to the MBH spin axis is 140 degrees
and the detector (observer) is 30 degrees above the azimuthal plane.   
\label{EMRIastrogr}}
\end{figure}

AK waveforms are not particularly ``faithful'' as EMRI templates, i.e., an AK
waveform with a given set of parameters does not have high overlap (noise
weighted inner product) with a more accurately computed waveform with the same
set of parameters. However, they do capture the main features of EMRI
waveforms, which has made them a useful tool for scoping out the semi-coherent
algorithm \cite{emri04} and other studies. They may also be quite ``effectual''
templates, i.e., for any real EMRI waveform, there may be an AK waveform {\it
with different parameters} that has a high overlap with that waveform. The AK
family of waveforms may thus play some role in the final analysis of LISA data.
The NK waveforms are not only effectual but very faithful, because they are
built around true Kerr geodesics. For orbits with periapsis greater than $\sim
5M$ (in geometrical units, where $M$ is the mass of the central black hole),
the NK waveforms have overlaps in excess of $95\%$ with waveforms computed via
solution of the Teukolsky equation. With further improvements (outlined below),
the NK waveforms are likely to be very useful tools in LISA data analysis, not
only for source detection but also for approximate source identification before
subsequent follow-up with more accurate templates.  In Figure~\ref{EMRIastrogr} we show a snapshot of NK EMRI waveform. This
figure serves as an illustration of the structure of the signal.

\subsubsection{Outstanding challenges}

None of these approaches to source modelling is as yet fully developed.
In the case of the self-force formalism, recent progress has been significant,
and the self-force acting on particles in circular orbits in the Schwarzschild
space-time has been computed \cite{barack07}. However, this is only for a small selection of circular orbits. The work must then be
extended to eccentric orbits in the Schwarzschild space-time, then to circular
equatorial orbits in the Kerr space-time before finally moving onto
eccentric-equatorial and ultimately eccentric-inclined orbits in Kerr. Although
this is a non-trivial progression, it should be achieved within the next five
to ten years. By the time LISA flies, it is likely that codes will exist to
compute self-force based waveform templates for arbitrary orbits. However,
these are likely to be computationally expensive, which is why it is necessary
to pursue the alternative models.

Adiabatic waveforms are at a more advanced stage of completion. All that
remains is to compute the evolution of the Carter constant for generic orbits,
using the results of \cite{mino03}, and to ``stitch together'' waveforms for
generic inspiral orbits. There are no technical challenges remaining, although
computational cost is an issue. Generic adiabatic inspiral waveforms should be
available within one to two years. Understanding their range of validity may
take longer, without self-force templates to compare them against. However, the
consideration of conservative corrections outlined below will be important for
developing this understanding.

The NK waveforms can also be improved. One of the reasons that the performance
degrades for small periapsis is that the kludge waveforms do not include tail
contributions, i.e., back-scattering of the radiation from the background
geometry. It should be possible to include this in an approximate way, which is
likely to improve the NK performance for orbits close to the central black
hole. Additionally, the inspiral prescription can be improved by fitting
functions to Teukolsky data for generic orbits. The current inspiral
prescription \cite{GG06} includes fits to Teukolsky data for circular inclined
orbits, and the resulting inspiral trajectories agree very well for that class
of orbit. It is likely that similar accuracy can be obtained for generic orbits
in a similar way. Finally, the NK waveforms can be augmented by inclusion of
conservative corrections. The conservative correction to the phase evolution of
an EMRI is already known in post-Newtonian theory to 3.5 $\PN$ order. By
considering asymptotic observables, namely the rate of change of the orbital,
periapsis precession and orbital plane precession frequencies as functions of
those three frequencies, it is in principle possible to compute the necessary
conservative corrections for inclusion in the NK model. This was demonstrated
for the simple case of circular equatorial orbits in Schwarzschild in
\cite{babak06}. Although the conservative effects could be included this way
only up to a certain $\PN$ order, it is quite plausible that this will suffice,
since conservative effects contribute most significantly to the phase evolution
in the weak field \cite{pound05}, where the $\PN$ expansion is valid. 

This last improvement of the NK model is potentially important, since an issue
that needs to be understood before LISA flies is how much conservative effects
can influence the emitted waveform. If conservative effects are not
significant, then a combination of kludge and adiabatic waveforms will be able
to identify the parameters of EMRI sources in the LISA data stream with quite
high accuracy. If conservative effects are important, the degree to which these
waveform families can constrain the parameter space will be significantly
reduced. Although self-force calculations are now at the point of computing
conservative corrections, this has only been done to date for circular orbits
in Schwarzschild. It is unlikely that generic conservative corrections from
self-force calculations will be available in the near future. The $\PN$ fitting
procedure outlined above will provide approximate results on a much shorter
timescale. This should allow the contribution to the phasing of each $\PN$ order
in the conservative correction to be assessed, which will give some insight
into the importance of including conservative effects. Moreover, the procedure
for including the $\PN$ conservative corrections in the NK model can also be used to
include these effects in the adiabatic waveforms.

As these developments and improvements to each family of waveforms proceed, we
will be able to address three primary groups of questions of great importance:

\begin{itemize}

\item What is the computational cost of evaluating waveforms of each type
(self-force, adiabatic, kludge)? How many self-force templates could be
generated in a reasonable time for a follow-up analysis? How accurately would
kludge/adiabatic templates therefore have to constrain the source parameters
prior to the self-force analysis?

\item What is the overlap of kludge, adiabatic and self-force templates with
one another? How accurately, therefore, can kludge and adiabatic templates
determine the source parameters, assuming we consider ``faithful'' searches
only? Can we compute sufficiently many self-force waveforms to determine
parameter mappings between the families, i.e., can we allow the search to be only
effectual?

\item What is the most computationally efficient way to detect and identify a
source, assuming we are using a multi-stage search employing kludge, adiabatic
and self-force templates at different stages of this search? What limit on
parameter extraction accuracy is set by computational constraints on our
search? Is this limit much worse than the theoretically achievable parameter
measurement accuracy? 

\end{itemize} 

The answers to the first and second itemised points may be incompatible, i.e.,
we may not be able to determine the parameters of the source sufficiently
accurately using kludge or adiabatic templates to allow a self-force follow-up
with reasonable computational cost. In this case, we would have to live with a
potentially larger error in our source parameters, as mentioned in the third
item.

In the above, we have focused on the modelling of waveforms from EMRIs.
However, another outstanding challenge is to develop models for IMRI waveforms.
To date, no models have been developed explicitly for IMRIs, although existing
models for other systems can be easily applied. An IMRI (mass ratio
$\sim1:1000$) lies somewhere between the inspiral of two comparable mass black
holes (a CMRI, mass ratio $\sim1:1$) and an EMRI system (mass ratio $\sim
1:10^6$). For a comparable mass system, the masses and spins of both components
are important, but the system spends very few cycles in the regime where the
velocity of the components is close to the speed of light. The waveforms can
thus be accurately computed from post-Newtonian expansions. In an EMRI system,
the number of cycles spent in the high-velocity regime (scaling as one over the
mass ratio) is $10^6$ times higher, so the post-Newtonian expansion is not
reliable, but the extreme mass ratio allows construction of accurate waveforms
from perturbation theory. An IMRI is somewhere in between, spending $1000$
times longer than a CMRI in the high-velocity regime, but $1000$ times less
than an EMRI. Post-Newtonian results will not be fully reliable for an IMRI,
since they spend so many cycles with $v\sim c$. However, perturbative results
are not fully reliable either, since they are linearised in the mass ratio and
thus omit terms at higher order in $m/M$ (where $m$ is the mass of the
inspiraling object and $M$ is the mass of the central body). In addition, at present we do not have $\PN$ waveforms that include both spins of the two bodies and orbital eccentricity, both of which could be non-negligible for IMRIs.

How quickly do higher-order mass-ratio corrections become important? For
simplicity, we consider circular equatorial orbits, but the following arguments
apply generally. Higher order mass ratio corrections fall into two categories:
(i) corrections to the frequency for an orbit at fixed radius (these
corrections arise from the conservative piece of the self-force and from
spin-orbit interactions due to the spin of the smaller body) and (ii)
corrections to the inspiral rate.  Corrections to the frequency enter at
$O(m/M)$, e.g., the spin of the small object is of order $m^2$ and leads to a
spin-orbit coupling force at this order, with resulting acceleration at order
$m$ \cite{pap51}. However, observationally, we cannot measure the radius of the
orbit, only the orbital frequency, so to lay out accurate templates we in fact
need the rate of change of orbital frequency as a function of the observable
orbital frequency (half the frequency of the fundamental gravitational wave
harmonic). Higher-order mass-ratio corrections just modify the rate of change
of orbital frequency as a function of orbital frequency (as discussed for
conservative self-force corrections in \cite{babak06}). These corrections occur
at order $m/M$ above leading order (which itself is $O(m/M)$ since inspiral
arises from radiation reaction). A change $\Delta \dot{f}$ in the inspiral rate
$\rm d (Mf)/\rm d(t/M)$ leads to a $\Delta N \sim \Delta \dot{f} (T/M)^2$
change in the number of cycles over a dimensionless observation time $T/M$. If
the observation time was fixed, this increases like $O((m/M)^2)$, and therefore
is much bigger for IMRIs than EMRIs. However, LISA has a fixed frequency
bandwidth, and so above a certain mass ratio, inspirals will be observed over a
fixed range of frequency. In that limit, the effective observation time is
proportional to the number of cycles in a fixed frequency range, $T/M \sim
M/m$, which suggests higher order corrections to $\dot{f}$ lead to a phase
shift $\Delta N \sim (m/M)^2 (T/M)^2 \sim 1$, independent of mass ratio. This
argument indicates that perturbative waveform templates will become worse as
the mass ratio increases, but at some point the errors will stabilise.  Of
course, the argument above sweeps many things under the carpet. The orbital
frequency is not the only observable, and we have ignored the changes to the GW
energy spectrum in the above argument. We have also not considered the size (or
post-Newtonian order) of the various corrections to the rate of change of
frequency. This argument can be quantified somewhat by using post-Newtonian
models. It is possible to compute the overlap of a $\PN$ waveform, linearised in
mass ratio, with the full $\PN$ waveform. This calculation will indicate the
relative importance of including higher-order mass ratio terms in perturbative
models, even though the $\PN$ models are not reliable as EMRI templates. Using a
leading order $\PN$ model, describing the last three years of inspiral of two
non-spinning bodies, with central MBH mass of $10^6M_{\odot}$, the mismatch
between the linearised and full waveforms increases from $0.001\%$ (when
$m=0.5M_{\odot}$) to $0.01\%$ ($m=1.4M_{\odot}$) to $1\%$ ($m=10M_{\odot}$) to
$15\%$ ($m=100M_{\odot}$)  to $18\%$ ($m=1000M_{\odot}$). This is consistent
with the above argument. For lower mass central black holes, the mismatches are
likely to be higher since the plunge frequency is correspondingly higher. There
is also likely to be a significant increase in the mismatch when spin is
included.

In summary, perturbative templates without higher order mass ratio corrections
or spin-orbit coupling corrections will probably not be good enough as IMRI
templates. However, it might be possible to construct ``Kludge'' IMRI templates
by including post-Newtonian spin-orbit and other corrections in the current
kludge EMRI models (in the same way that conservative corrections are currently
being included). This needs further investigation.

\section{Testing relativity theory}
\label{testGR}

One of the potentially exciting payoffs from EMRI and IMRI observations made by
LISA is the ability to test aspects of relativity theory. The high mass
ratio ensures the small object acts like a test particle moving in the
background space-time of the central black hole. The emitted gravitational waves
trace out the orbit of the particle, which in turn encodes a map of the
space-time \cite{ryan95}. EMRI events are comparatively ``clean'' systems, and
we know what this map should look like if the inspiral is an inspiral into a
Kerr black hole. Decoding the map then allows us to measure the parameters of
the system to high precision. However, if the inspiral deviates from what we
expect --- an inspiral, described by General Relativity, of a compact object
falling in vacuum into a Kerr black hole --- we should be able to see this
deviation in the emitted gravitational waves.

\subsection{Current status} 

Conceptually, tests of a theory fall into two general categories: comparisons of rival theories (which theory is best supported by the data?) and tests of
consistency (are the data consistent with a given theory?). 

\subsubsection{Comparisons of rival theories}
Currently there are no really plausible rivals to general relativity; rival theories have either been ruled out, or, like Brans-Dicke, can be dialled arbitrarily close to general relativity by adjustment of extra parameters (and so could never be ruled out even if GR were completely correct). Moreover, to test an alternative theory, we need to be able to compute gravitational waveforms for EMRIs in that alternative theory to compare against EMRI gravitational waves from Relativity. This is a very challenging problem both theoretically and computationally. 

Constraining the parameter space that rival theories can occupy can be a useful
exercise, since it is a measure of how close to General Relativity the true
theory must lie. It is possible to compute the leading-order correction to the
gravitational wave phasing for inspiralling objects in Brans-Dicke theory. Work
on this has suggested that LISA observations of neutron stars inspiralling into
$\sim 10^2$--$10^4\Msun$ black holes could put meaningful constraints on the
Brans-Dicke parameter and on the mass of the graviton~\cite{will04,berti05}.
The theoretical waveforms used for this work were rather simple and ignored
important effects such as orbital plane precession due to spin-orbit coupling.
It is known that parameter estimation accuracies for MBH binaries improve
significantly when spin-orbit coupling is included~\cite{vecchio04,lang06}, so
the current results are probably conservative. However, it is unclear whether
the astrophysical rate is sufficiently high that LISA will be likely to see any
of the ``optimal'' events for such an analysis ($1.4\Msun$ + $10^3\Msun$).
Moreover, developing highly accurate waveforms under alternative theories is
difficult, and the constraints that I/EMRI observations will be able to impose
on the parameter space will probably not be significantly tighter than existing
results. For these reasons, most research to date has focussed on the second
type of test --- tests of consistency.

\subsubsection{Tests of consistency within General Relativity} A simple example
of a consistency test would be to divide an observed EMRI signal into several
consecutive pieces, and show that the best-fit parameters from each piece were
consistent with each other, within the error bars.  This would already be a
very strong test of the theory. The complication in regarding EMRI observations
as consistency tests is that taking GR to be the correct theory of gravity is
not the only assumption that goes into generating the waveform. We assume also
that the system is vacuum and that the central body is described by the Kerr
metric. Without requiring an alternative theory, we can regard EMRI
observations as testing the premise that massive compact objects in our
universe are Kerr black holes, rather than some other exotic object (e.g., a
boson star or naked singularity) that is still consistent with Relativity. Any
axisymmetric, vacuum space-time in Relativity can be decomposed into mass
($M_l$) and current ($S_l$) multipole moments \cite{geroch70,hansen74} and it
was demonstrated by Ryan \cite{ryan95} that these multipole moments are
redundantly encoded in gravitational wave observables, namely the periapsis
precession frequency, the orbital plane precession frequency and the gravitational wave energy spectrum for nearly circular, nearly equatorial orbits.
If an object is enclosed by a horizon and there are no time-like curves exterior to the horizon, then the object must be a Kerr black hole (this is the ``no-hair'' theorem), and all of its multipole moments are determined by the mass and spin of the black hole~\cite{hansen74} 
\begin{equation} M_l + {\rm
i} S_l = M\,({\rm i}a)^l \label{nohair}, 
\end{equation} 
where $M$ is a black hole's mass and $a$ is the reduced spin of the black hole $a = S_1/M$. If three moments of the space-time are extracted from the gravitational wave emission, these can be checked for consistency with \erf{nohair}. A boson star
with large self-interaction (one viable alternative to a MBH) is uniquely
characterised by three multipole moments~\cite{ryan97b}, so if four moments are
extracted from the gravitational wave emission, the boson star model could also
be ruled out. This idea of measuring multipole moments is sometimes referred
to as ``testing the no-hair property'' (one sometimes hears the variation
``testing the no-hair theorem'', but this is obviously sloppy wording, since a
true mathematical theorem cannot be invalidated by any experimental test).
Objects with non-Kerr values of higher multipole moments within general
relativity would have to be exotic stars or naked singularities. The no-hair theorem applies only in
Relativity, thus non-Kerr values of the higher multipole moments could also
arise if the object is a black hole, but Relativity is the wrong theory of
gravity. If, for example, the quadrupole moment of the massive object was found
to differ from that of a Kerr black hole, this could, therefore, have several
explanations. We will discuss this in more detail later.

A significant amount of work has been done on quantifying how well LISA could
measure multipole moments and carry out these sorts of tests. Ryan
\cite{ryan97} considered a general axisymmetric space-time, decomposed into
multipole moments and found that a LISA observation of a nearly circular,
nearly equatorial EMRI could measure the mass quadrupole moment to an accuracy
of $\Delta M_2/M^3 \sim 0.0015$ -- $0.015$, depending on the source parameters.
Ryan's approach is somewhat unwieldy, however, since an infinite number of
multipole moments are present in the Kerr space-time. The multipole expansion is
essentially an expansion in $1/r$, where $r$ is the distance from the black
hole. LISA will mostly observe EMRIs that are deep in the strong field region,
very close to the black hole and in that regime all the multipole moments will
be important. Extracting multipole moments one at a time is therefore a rather
inefficient way to characterise a Kerr black hole. Collins and Hughes
\cite{collins04} were the first to suggest an alternative approach to LISA
observations --- regarding the observation as a null-hypothesis test of the
assumption that the EMRI is a Kerr EMRI. By constructing space-times that are
close to Kerr, which Collins and Hughes called ``bumpy black holes'', it is
possible to quantify how large a deviation from the Kerr space-time could be
present while leaving the signal observationally consistent with a Kerr
inspiral \cite{hughes06}. This approach is preferable, since the Kerr space-time
can then be recovered exactly by dialing a small parameter to zero, so we do
not lose much sensitivity to the events that we expect to see by using
detection templates with an additional bumpy parameter. Collins and Hughes
constructed a static space-time that deviates from the Schwarzschild space-time
by a small amount. They did this by using the Weyl metric and adding a
perturbation that represents a pure mass quadrupole asymptotically. They found
that the azimuthal frequency of equatorial orbits with the ``same parameters''
differed by $0.01\%$ (for an orbital periapsis of $\sim 50M$) -- $10\%$ (for
periapsis of $\sim 6M$) when a quadrupole moment perturbation $Q=0.01 M^3$ was
added. 

Babak and Glampedakis \cite{BabGlam06} carried out a similar calculation for
stationary space-times that deviated from Kerr by a small amount, which they
constructed using the Hartle-Thorne approach. They not only considered orbital
frequencies, but also constructed kludge waveforms and found that in the
presence of a $\sim10\%$ deviation in the quadrupole moment of the space-time,
the overlap could degrade by $25\%$ over the radiation reaction timescale for
typical LISA events. In both these analyses, the results were not directly
relevant to observations, since they took no account of the fact that some of
the differences in the waveforms could be mimicked by changing the orbital
parameters (Babak and Glampedakis did comment on this fact in their paper,
however). Recently, Barack and Cutler \cite{BC07} have done an analysis
accounting for parameter correlations, using a waveform model constructed by
including a term representing a non-Kerr value of the quadrupole moment of the
central black hole into their analytic kludge \cite{BC04}. They find that
LISA could measure the quadrupole moment, $Q = -S^2/M$, of the central black
hole to an accuracy $\Delta Q/M^3 \sim 10^{-3}$, while simultaneously measuring the mass and spin to an accuracy of~$\sim 10^{-4}$. We note that spinning boson stars typically have quadrupole moments ten to a hundred times larger than Kerr black holes of the same mass and spin \cite{ryan97b}. This provides an indication of the accuracies required to do meaningful tests. 

The research described above concerns extracting information about the spacetime multipole structure from the inspiral part of the waveform. If a ringdown was also detected, the ringdown frequencies can also be used to measure the multipole structure of the ringing object~\cite{berti}. EMRI-induced ringdowns are unlikely to be detected with sufficient signal-to-noise ratio, but IMRI-induced ringdowns could plausibly be used for such a test. However, more work needs to be done to quantify this, in particular to compute ringdown frequencies for non-Kerr supermassive objects. Moreover, in general the constraints on the multipole moments obtained from the inspiral will be tighter than those obtained from the ringdown due to the large number of wave cycles that can be observed over the inspiral.

There is an analogous source to EMRIs that might be detected by ground-based
gravitational wave detectors, namely the inspiral of a stellar mass neutron
star or black hole into a $\sim 100 M_{\odot}$ IMBH (a ``LIGO IMRI''). The
event rate for such inspirals is somewhat uncertain (see
\cite{HA05,Pfahl05,LIGOrates07} for discussion and further references).
However, if they are detected, these sources have the potential to probe the
strong-field regime with more modest accuracy than LISA's EMRIs, and would be
observed by Advanced LIGO, i.e., a few years before LISA EMRI events are
observed. A significant amount of work has gone into studying these sources
\cite{brown06}, and this work also applies to LISA EMRI/IMRI events. Some of
the applicable results include the generalisation of Ryan's results
\cite{ryan95} to more generic cases. Ryan considered only nearly circular,
nearly equatorial orbits and ignored the effect of tidal coupling. The
generalisation to eccentric but nearly equatorial orbits is straightforward,
although the generalisation to arbitrary orbits is difficult~\cite{brown06,
li}.  An extension of Ryan's theorem to tidal coupling tells us that this
coupling is also encoded in (and could be extracted from) the gravitational
wave observables \cite{brown06, lilovelace}. The inspiralling object distorts the horizon surrounding the central object and orbital energy is lost to gravitational radaition flowing into the horizon as well as out to infinity. These interactions with the horizon can be modelled as a tidal interaction, and characterised in terms of energy being lost to the central body through tidal dissipation. In broad terms, the multipole moments of the space-time can be extracted from observations of the periapsis and orbital plane precession rates, and from these the rate at which energy is being radiated to infinity can be determined.  The rate at which energy is being lost from the orbit can also be measured, by observing the change of the orbital frequency with time. The difference between these two energy fluxes gives the flux of energy going into the central body, which is a measure of the strength of the tidal interaction, and tells us about the structure of the central object \cite{brown06}. 

In another aspect of this effort to study EMRIs, Gair \etal \cite{dynpaper}
have studied the properties of geodesics in other classes of ``bumpy''
space-times. This work has considered two types of nearly Kerr space-times ---
exact solutions in the family of Manko and Novikov \cite{manko92} that deviate
from Kerr in the quadrupole and higher moments, and perturbative solutions
constructed via solving the Teukolsky equation and then applying the
Chrzanowski-Ori procedure to recover the metric \cite{chrz75}. As mentioned
before, the geodesics in the Kerr space-time possess three integrals of the
motion, because the Kerr space-time is one of a special class of GR solutions
for which the Hamilton-Jacobi equation is separable \cite{carter68} and $Q$
arises as the separation constant. Space-times that deviate from Kerr, even by
a small amount, may not be of separable form, and therefore there is no
guarantee that the geodesics will possess a full set of integrals. However, it
turns out that the majority of geodesics in most of the space-times considered
in \cite{dynpaper} have an approximate third invariant, and the geodesics are
tri-periodic to high accuracy: the waveform phase evolution can be decomposed
in terms of harmonics of three fundamental frequencies to an accuracy of one
part in $10^7$ or better. This means that Ryan's theorem can be applied, and
deviations from Kerr show up only in the differences in precession and inspiral
rates. However, in a small subset of cases, the geodesics show apparently
ergodic motion, with no well defined frequency structure. Although this makes
the sources difficult to detect, observations of ergodic dynamics would be a
clear signature that the space-time was not simply Kerr.  This is unlikely in
practice since the ergodic motion only appears for orbits that are very close
to the central object, and this is a regime that is probably not accessible in
an inspiral that begins with the capture of a star on a highly eccentric orbit
some distance from the central object.

\subsection{Outstanding challenges}

There are several questions that still need to be addressed before LISA
will be able to carry out tests of General Relativity.  These include:

\begin{itemize}

\item What are the imprints of deviations from Kerr on the waveforms generated
by EMRIs on generic orbits? As described above, we have a partial answer to
this question already. However, the only work so far that includes the effect
of radiation reaction on generic orbits is \cite{BC07}, which uses a very
simple model.  With further research, it should be possible to make general
statements about how deviations from Kerr manifest themselves in the EMRI
signal, and how the deviations correlate with other parameters. We want to
understand how adding greater complexity in the family of deviations from Kerr
causes the determination of other parameters to degrade, e.g., whether a Kerr
EMRI could also be well described by a ``bumpy'' Kerr EMRI with different
parameters. In \cite{ryan97}, Ryan found that by adding arbitrary multipole
moments (up to $M_{10}$), the accuracy with which the mass of the large body
could be determined in an observation degraded from one part in $10^6$ to one
part in $3$, although most of this degradation was due to the inclusion of
moments up to $S_5$. This is a known problem in data analysis: introducing more
parameters (especially if they are small) causes effective ``noise'' in the
parameter space. This would also apply if we wanted to use EMRI observations to also test alternative theories of gravity. Testing everything simultaneously would yield very poor constraints, since there would likely be correlations between, for example, a non-zero Brans-Dicke parameter and an anomalous quadrupole moment. Thus, it might be necessary to decide {\em a priori} which deviations from Relativity we most want to constrain in order to produce useful statements.

\item Are there any ``smoking gun'' signatures for deviations from Kerr? For
instance, ergodicity in the orbits --- which ``nearly Kerr'' space-times admit
ergodic orbits, and can a star end up on these orbits in practice? What are the
observational signatures?

\item Can matter external to the massive object (e.g., an accretion disc, or
other stars) perturb the EMRI orbit sufficiently to leave a measurable
imprint on the emitted gravitational waves? How would we detect or
recognise such a system? Some work has already been done to estimate
the effects of an accretion disc~\cite{chak96,narayan00,barausse06}, which suggests 
that these effects are unlikely to be measurable unless the accretion disc is very
massive. Such massive discs might be found around MBHs accreting at or
near their Eddington limit, i.e., in active galactic nuclei. As described in Section~\ref{sec.nonstd}, star formation in the disc of
such systems might lead to EMRI events.
. In a
normal nucleus, accretion could happen if a lot of material was recently
dumped in the vicinity of the MBH by, for example, the disruption of a star or a
gas cloud. However, such events occur very infrequently ~\cite{WM04}. If a very tight binary, consisting of a main sequence star and a compact object, were to inspiral into a $10^6\,\Msun$
MBH, the MS star would be disrupted at a distance of $\sim 2\times 10^{-6}\,$pc from
the MBH. Most of the bound stellar material would be accreted within a
few years \cite{Ulmer99,MQ01} but it would take $100-10^4$ years
for the compact object to complete its inspiral. Therefore it seems
unlikely that the material of the disrupted companion can either
perturb the EMRI or create a clear-cut electromagnetic precursor
to it. Nonetheless, if there is a small possibility that such a system
could in principle be observed, it is important that we know how to
detect it and recognise it, not least because an EMRI interacting with a massive accretion disc might lead to an electromagnetic counterpart. An EMRI with a counterpart is a powerful cosmological probe, so it is valuable to maximise our chances of seeing such events.  

Finally, we note that if two EMRIs were occurring simultaneously, this
would almost certainly leave a measurable imprint on the emitted
gravitational waves. However, this is very unlikely as well. A compact
binary of mass $m_{\rm bin}$ tight
enough to survive tidal separation down to a distance $R$ of the MBH, would
have to have a timescale for (self-)merger by GW emission, $\tau_{\rm
merge}$, smaller than $\tau_{\rm merge} <
\tau_{\rm insp}(\MBH/m_{\rm bin})^{2/3}$, 
where $\tau_{\rm insp}$ is the timescale for inspiral into the
MBH on a circular orbit of radius $R$. Therefore there cannot be any
significant number of binaries tight enough to survive until the last
$\sim 100$ years of inspiral. If the binary disrupts earlier it is
likely that both stars will find themselves on orbits with vastly
different inspiral times and when the faster-inspiralling one becomes
detectable the other will still be too dim. However, the distant EMRI might cause a
detectable orbital perturbation on its ex-companion and this has to be assessed.

\item How do we interpret deviations from Kerr if they are observed? There
could be several explanations --- there could be material external to the black
hole; alternatively, the massive object could be some exotic star with a
non-singular distribution of matter (e.g., a boson star); yet another
possibility is that the massive object is a naked singularity, which would
disprove the cosmic censorship conjecture, but would not contradict the no-hair
theorem. How could we distinguish these possibilities in an observation? Over a
long ($\sim$year) observation, it should be relatively straightforward to
distinguish the effect of material outside the black hole (an ``external''
quadrupole perturbation) from a change in structure of the central object (an
``internal'' quadrupole perturbation), since the effect would accumulate
differently over the course of an inspiral. The existence/location of a horizon
might be determined from gravitational wave observations.  While an inspiral
into a Kerr black hole would undergo a rapid plunge from the innermost
stable circular orbit followed by a ring-down, an inspiral into a boson star may
continue to produce inspiral-like waves after the compact object crosses the
stellar surface. The signal-to-noise ratio generated during the plunge and ringdown for an EMRI will be small, so this will almost certainly not be observed. However, if a signal persisted after the ``plunge'', this might be detected by building up the signal-to-noise over several waveform cycles. Thus, we might be able to say whether a signal is still ``present'' or ``absent'', although the resolution of the time at which the signal became absent would not be very high. A horizon could be inferred by comparing the approximate time of plunge with the plunge time predicted based on parameters measured in the
early part of the inspiral.  If a horizon is found to be absent, the object might be an extended mass distribution (e.g., a boson star) or a naked singularity. If emission after the object's path started to intersect the boson star material was observed with sufficiently high signal-to-noise, the features of this emission might allow us to distinguish between these two possibilities~\cite{kesden05}. Only if the massive object were found to have a non-Kerr quadrupole moment that was not due
to the presence of exterior matter {\it and} a horizon would there be firm evidence that the system did not have the no-hair property; however, proving that any horizon completely surrounds the body and that no closed time-like curves exist in the exterior may be impractical.  While these ideas give us some hope that interpretation will be possible, further research is needed on all of these topics.

\item How do we detect deviations from GR in practice? The need to use matched
filtering for EMRI detection makes it difficult to detect signals that look
very different from our template models. It also makes it hard to detect small
deviations in the model. The simplest thing we can do is to look only for
inspirals into Kerr black holes. If our observations are consistent with this
model, then we have tested the theory to high precision. The existing research
programme then allows us to make statements such as ``this observation is
consistent with a Kerr inspiral, with agreement in the quadrupole moment to
$x$\%''. To do this in practice we would perform Monte Carlo simulations to find the maximum ``$x$'' such that the gravitational waveform emitted during an inspiral into a non-Kerr object with a quadrupole moment that differed by $x$ from the Kerr value still had sufficiently high overlap (for a suitable definition of ``high'') with a waveform emitted during an inspiral into a Kerr black hole with some (not necessarily the same) parameters. A more sophisticated analysis could look at segments of the inspiral
separately, and check for consistency in the parameters estimated for each
segment. We could also look for characteristic signatures of a deviation from
Kerr, for instance a transition from regular to ergodic motion in the orbits or
the existence/location of the horizon inferred by the plunge time. This might
be done by dividing the end of the inspiral into short segments, in which the
signal should have SNR large enough so that we can say with high confidence
whether the signal is present or absent. The resolution of the plunge time is
thus likely to be poor. If the EMRI is close enough to be loud (SNR $\gtrsim 50$),
it might be possible to detect the signal in a time-frequency analysis. This
would not only make it easier to measure things like the plunge point, but
would allow us to detect signals that deviate significantly from Kerr
inspirals. Finally, it might be possible to do a more generic analysis using
templates parametrised by space-time multipole moments, e.g., the family
employed by Ryan \cite{ryan97}. Such a technique would not be particularly
sensitive to Kerr inspirals, but if it was used in conjunction with a matched
filtering search for Kerr EMRIs it might be a useful diagnostic. As EMRI data
analysis techniques are developed, techniques for space-time mapping will need
to be properly explored.
\end{itemize}

\section{EMRI science}
\label{EMRIScience}

It is clear from the discussion of EMRI detection above that, while much is
already known, there is still some work to be done before LISA flies. However,
the scientific payoffs if we detect and characterise a large number of EMRI
events could be very significant. From a single EMRI observation, we can
measure the parameters of the system to very high precision \cite{BC04}. The
mass and spin of the central black hole, the mass of the inspiralling object,
and the orbit's eccentricity (at some fiducial instant) can all be determined
to a part in $10^4$, typically, while the cosine of the orbit's inclination
angle (roughly, the angle between the MBH's spin vector and the orbital angular
momentum of the CO) can typically be determined to $\sim 10^{-3}$--$10^{-2}$.
The luminosity distance to the source can be measured to an accuracy of $\sim
5\%$, and the sky position to a resolution of $10^{-3}$sr (a few square
degrees). The accuracies achievable with LISA IMRI observations should be significantly better. In addition, for each of these observed systems we will be able to test the black hole hypothesis to high accuracy (e.g., constrain the mass quadrupole moment to a fraction of a percent), as described
in Section~\ref{testGR}.

LISA may detect as many as several hundred EMRIs out to a redshift of
$z\sim1-2$ \cite{BarackEtAl03,emri04}. The first estimates of signal-to-noise ratios for EMRIs were done by Finn and Thorne~\cite{finn00}. They considered circular equatorial inspirals only, and found that at a distance of $1$Gpc, one year before plunge, the inspiral of a $10\Msun$ object into a rapidly spinning $10^6\Msun$ black hole would have signal-to-noise ratio of $\sim100$ in a gravitational wave frequency bandwidth equal to the frequency. To obtain EMRI rate estimates, updated signal-to-noise ratios were computed for inclined and eccentric orbits, using kludged waveforms~\cite{GG06,babak06} and including a more accurate model of the LISA response provided by the Synthetic LISA simulator~\cite{vallis05}. Table~\ref{detectvol} shows the results of those calculations --- the signal to noise ratio of a variety of systems at a fiducial distance of 1Gpc~\cite{BarackEtAl03,emri04}. These signal to noise ratios assume that the LISA mission lasts five years and that the satellite is fully functional for the whole time, so that the optimal combination of TDI data streams can be used. These results are consistent with Finn and Thorne~\cite{finn00} when one accounts for the gravitational wave bandwidth remaining one year from plunge,  the fact that Finn and Thorne compute the SNR in one LISA Michelson channel only and the fact that these orbits are eccentric.

\begin{table}
\begin{tabular}{|cc|c|ccc|}
\hline M/$\Msun$ & m/$\Msun$&SNR at 1Gpc&$z_{max}$&$M_i/\Msun(z_{max})$&$m_i/\Msun(z_{max})$ \\ \hline \hline &0.6&18&0.13&$2.7\times10^5$&0.53 \\ $3\times 10^5$&10&73&0.44&$2.1\times10^5$&6.9 \\ &100&620&2.5&$8.5\times10^4$&29 \\ \hline &0.6&30&0.21&$8.3\times10^5$&0.50 \\ $1\times10^6$&10&210&1.0&$4.9\times10^5$&4.9 \\ & 100&920&3.5&$2.2\times10^5$&22 \\ \hline &0.6&25&0.17&$2.6\times10^6$&0.51 \\ $3\times10^6$&10&270&$1.3$&$1.3\times10^6$&4.4 \\ &100&1500&5.2&$4.8\times10^5$&16 \\ \hline
\end{tabular}
\caption{This table shows the signal-to-noise ratio (SNR) at a distance of 1Gpc for systems with a variety of observed masses $M$ and $m$. Also shown is the maximum redshift at which such a source could be detected, $z_{max}$, and the intrinsic masses of the system, $M_i = M/(1+z_{max})$ and $m_i = m/(1+z_{max})$, that a source at redshift $z_{max}$ would need to have in order to have apparent red-shifted masses $M$ and $m$. The SNRs were computed assuming the optimal TDI combination of LISA data streams could be constructed for five years of observation. All sources have MBH spin of $S/M^2=0.8$, inclination of $45^{{\rm o}}$ and eccentricity at plunge of $0.25$. The waveforms were computed using the numerical kludge model~\cite{GG06,babak06} and the LISA response was included using the Synthetic LISA simulator~\cite{vallis05}. These results were used for computing event rate estimates using the semi-coherent search~\cite{emri04}.}
\label{detectvol}
\end{table}

To translate these SNRs into a maximum detectable distance, we first need to specify a detection threshold. The estimated detection threshold for the semi-coherent algorithm described earlier is $\sim 30$, although optimisation of this method may be able to reduce this threshold somewhat. A GW source at a redshift $z$ with masses $M$ and $m$ looks like the same type of gravitational wave source at a Euclidean distance equal to the luminosity distance to redshift $z$, $D_L(z)$, but with red-shifted masses $(1+z)M$ and $(1+z)m$. Table~\ref{detectvol} also shows the redshift at
which the source would have SNR of $30$ (computed by setting SNR(1Gpc)$/30$ =
$D_L(z)/1$Gpc) and the intrinsic masses $M_i$ and $m_i$ that a source at that
redshift would have to have in order to give the appropriate observed masses. This Table serves to illustrate the typical distances to which sources can be
detected.

\begin{figure}
\resizebox{\hsize}{!}{\includegraphics[width=4in,angle=-90]{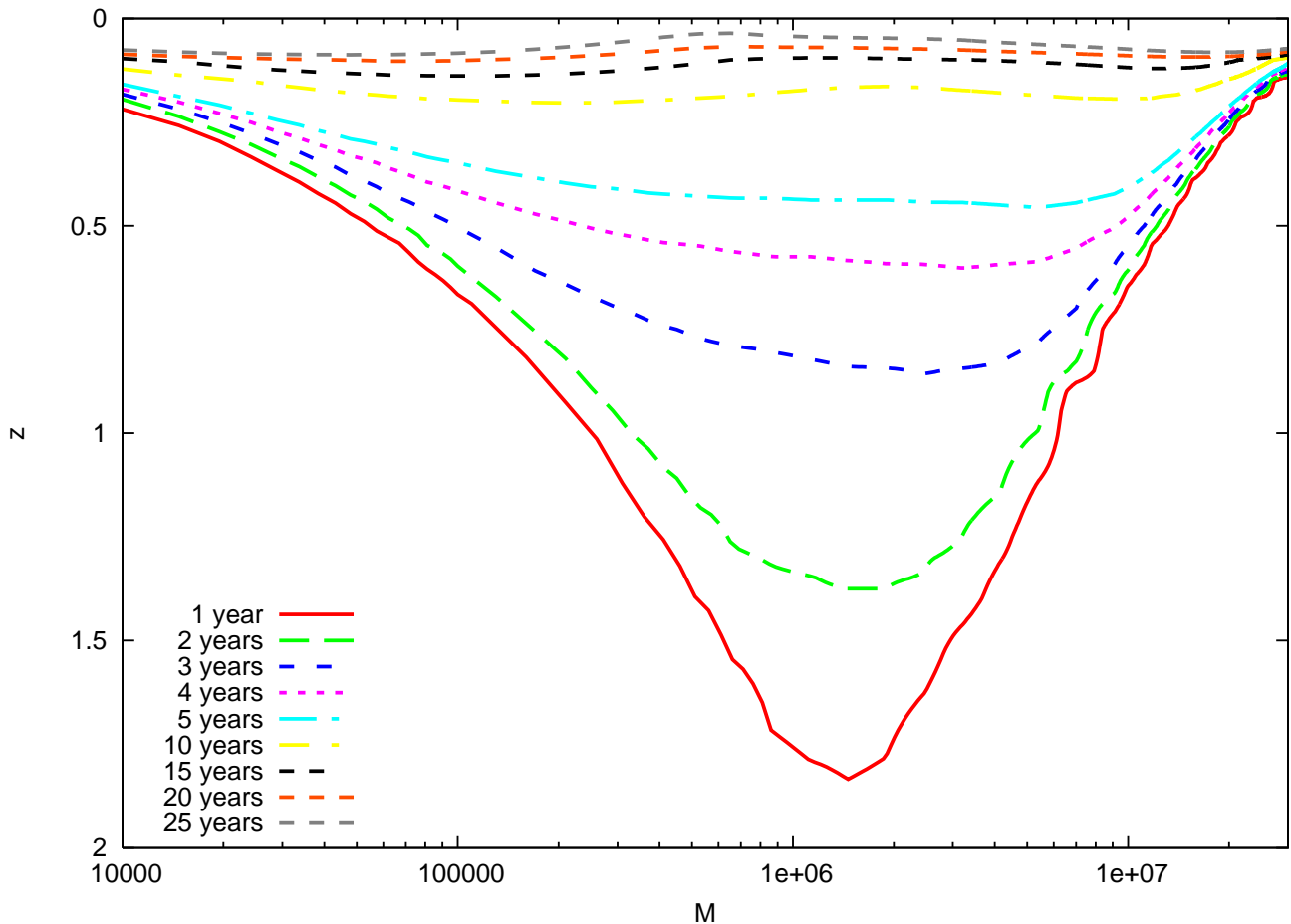}}
\caption{Contours of constant ``detectable lifetime'' (as defined in the text) for the circular-equatorial inspiral of a $10\Msun$ black hole into a MBH with spin $a=0.99$, as a function of MBH mass $M$ and redshift $z$. \label{rangeplota}}
\end{figure}

\begin{figure}
\resizebox{\hsize}{!}{\includegraphics[width=4in,angle=-90]{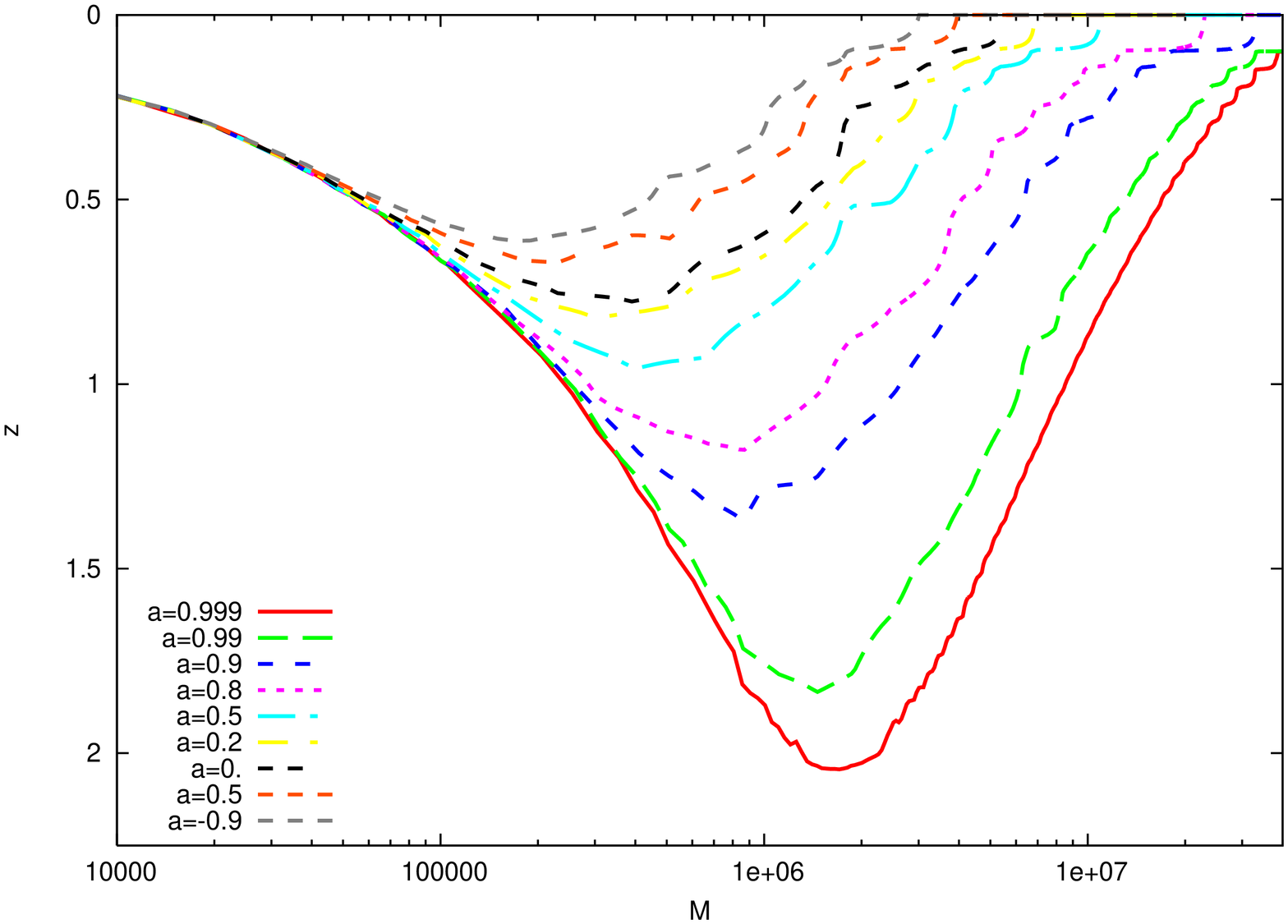}}
\caption{As for Figure~\ref{rangeplota}, this figure shows contours of constant ``detectable lifetime'', $\tau$, for the circular-equatorial inspiral of a $10\Msun$ black hole into a MBH. Here we show $\tau=1$~year contours for several different spins of the central black hole, as labelled in the key. Negative spins indicate retrograde circular-equatorial inspirals into black holes of the same spin magnitude. 
\label{rangeplotb}}
\end{figure}

The mass red-shifting makes the process of computing the range for a given
source a more complicated procedure than a simple linear distance scaling. Figures~\ref{rangeplota} and~\ref{rangeplotb} show preliminary results of a more careful calculation~\cite{gair07}. The plots show contours of constant ``detectable lifetime'' for circular-equatorial EMRI sources as a function of the mass of the central black hole. The figures plot the
intrinsic mass of the MBH on the x-axis, and redshift on the y-axis.  The
lines on the plot are contours of equal observable lifetime, $\tau$. A source
will be detectable only if the signal-to-noise ratio accumulated over the LISA
mission exceeds the necessary threshold (assumed to be $30$ for this plot). This requirement will only be satisfied if the source is in a certain
range of phases of the inspiral at the moment that LISA starts taking data. The
observable lifetime is the length of that acceptable range of phases, as
measured at the source. If EMRIs of this type start once every $T$ years in any
galaxy, the expected observed number of events would be $\tau/T$ per galaxy.
For Figure~\ref{rangeplota}, we consider only prograde inspirals into a central black hole with spin $a=0.99$ and plot contours for various values of $\tau$. In Figure~\ref{rangeplotb} we show $\tau = 1$yr contours for various black hole spins (NB a negative spin indicates a retrograde inspiral into a black hole with spin of the same magnitude). In both plots we have assumed a constant mass for the compact object of
$10M_{\odot}$, and are considering circular-equatorial inspirals only. These results were computed using the flux data tabulated in Finn and Thorne~\cite{finn00}, assuming a five-year LISA observation that uses both Michelson channels, averaging over the sky position and orientation of the source, and taking the LISA noise spectral density as given in~\cite{BC04}, with the assumption that the white-dwarf background has been subtracted using five years of LISA data. It is clear from this figure that we can see EMRI events out to fairly large distances, but that this distance is very spin dependent. There is also a significant spin-dependence of the mass to which LISA has maximal reach. The fraction of the total energy radiated that is radiated in a circular-equatorial inspiral between a Boyer-Lindquist radius $r$ and plunge at the innermost stable circular orbit, $r_{\rm isco}$, effectively depends on spin only through the ratio $r/r_{\rm isco}$. The ISCO radius decreases as the black hole spin increases, and the total energy radiated increases. This means that inspirals into rapidly spinning black holes radiate more energy, having, therefore, higher total signal-to-noise ratios, and this radiation is emitted at higher frequencies for a given central black hole mass. LISA will be most sensitive to systems that radiate most of their energy in the detector's optimal sensitivity range of $\sim 3$--$10$mHz. The frequency of the radiation decreases as the black hole mass increases, but increases as the spin increases. The mass that ensures radiation at $\sim 5$mHz will, therefore, be larger for higher black hole spins, as illustrated in the figure. It is also clear from Figure~\ref{rangeplotb} that the sensitivity to low mass central black hole systems is effectively independent of spin. For these systems, LISA will observe the middle part of the inspiral, while the final plunge occurs at higher frequencies that are out of band. LISA will, therefore, observe a phase of the inspiral where the radius is large, at which distances the orbit does not ``feel'' the effect of the black hole spin.

These results illustrate the LISA range to typical EMRIs, but are only for circular-equatorial orbits, and assume a simplifed model of the LISA response. Eccentric orbits in general will lose more energy in the LISA band, and hence should have larger signal-to-noise ratio. Treating the LISA data stream more carefully, using a TDI analysis (e.g., by using Synthetic LISA to include the response function~\cite{vallis05}), will also tend to increase the signal-to-noise, especially for sources that generate significant radiation at high frequencies. These general expectations are supported by the results in Table~\ref{detectvol} and by further calculations using numerical kludge waveforms~\cite{GG06,babak06} for eccentric-inclined orbits.

For all of the systems that LISA can see, we will obtain accurate parameter
estimates.  LISA will thus provide data on a large sample of black holes in the
relevant mass and redshift range, which can be used for astrophysics. For the
discussion sessions at the meeting that inspired this review, we divided the
scientific questions on the astrophysical benefits and consequences of LISA
EMRI and IMRI observations into five categories. We summarise these in the
following sections, along with an outline of the answers that LISA might give
us.

\subsection{What can we learn from the characterisations of
EMRI/IMRI dynamics, i.e., the observed eccentricities etc. of the orbits?} 

The observed eccentricities will carry information about the capture mechanism:
significant eccentricities are indicative of the direct capture scenario via
two-body relaxation, while negligible eccentricities suggest capture via binary
tidal disruption or, possibly, tidal stripping of giants 
(such captures occur at a higher periapsis and have time to
circularise before entering the LISA band).  The orbital inclination is also
informative: random inclinations are expected in the standard scenario with a
spherical cluster, while formation in the disc would be manifested by  a
significant fraction of EMRIs having near-zero inclinations  (although this could be made more complicated by disc warpage). The event rates
themselves would be interesting, although it may be very difficult to deconvolve the
various uncertain parameters that influence event rates (e.g., the MBH density
and mass function, the density and distribution of compact objects near the
MBH, etc.).

\subsection{What can we learn about the inspiralling compact objects from
EMRIs/IMRIs?} We will learn the CO mass to fractional accuracy $\sim 10^{-4}$.
From the distribution of masses we will obtain 
information about the relative numbers of  the WD, NS, and BH populations close to the MBH
(within $\sim 0.01$ pc). Although it is difficult to deconvolve mass
segregation from the initial mass function or from the mass-dependence in
capture rates, unexpected results such as the under-representation of more
massive COs could be very intriguing. If BHs with masses in the range $20-1000
M_{\odot}$ are found, this is already an important discovery, since such
objects are not yet firmly known to exist (and, if $1000 M_{\odot}$ IMBHs do
exist, it is not clear how well they can sink to the centre). 
Even the precise measurement of the mass of a few stellar BHs 
(in the range $3-20\,\Msun$) would be of great interest to constrain models 
of stellar evolution and collapse.
CO spins will
probably not be measurable for EMRIs, but we guess they might be measurable
to $\lesssim 10\%$ for IMRIs. This would tell us
about the formation mechanism for IMRIs (has mass been accumulated mostly via accretion, or via mergers?), just as for the MBH spin. 

\subsection{What can we learn about the MBHs from EMRIs/IMRIs?} From
EMRIs/IMRIs, we will learn the masses and spins of MBHs to fractional accuracy
$\sim 10^{-4}$. Rapid spins will imply that much of the MBH mass was built up
by gas accretion from a disc (unless the gas arrives in randomly oriented
events \cite{KP06,KP07,RV07}), moderate spins will imply the MBH was built as a result of
a major merger of comparable-mass black holes \cite{hb02,Miller02}, 
while low spins will
imply the MBH was mostly built from a sequence of minor mergers with smaller
objects coming in from random directions (since then the spin angular momentum
increases only in a random-walk fashion). The boundaries that separate these
three spin regimes are somewhat uncertain. Spins $S/M^2 > 0.9$ are undoubtedly
``high'', but lower spins might also arise from accretion. The spins resulting
from major mergers depend on the magnitude and direction of the spins of the
two components prior to merger and could be anywhere in the range $0.4 \lesssim
S/M^2 \lesssim 0.9$. Spins $S/M^2 < 0.2$ are undoubtedly ``low'', but the
boundary between this regime and the major merger regime is unknown.  Once
again, decoding the spin observations must be done carefully.  The prograde
inspiral of a black hole into a rapidly spinning MBH has much greater SNR at a
given distance than a retrograde inspiral or an inspiral into a MBH of lower
spin. Our observations will therefore be weighted towards more rapidly spinning
black holes. Since we can compute the relative SNR of systems with different
spins, it will not be too difficult to account for bias when interpreting the
observations, but this must be done carefully.

\subsection{What can we learn about cosmology and early structure formation from EMRIs/IMRIs?}

An EMRI or IMRI observation will give a very accurate measure of the luminosity
distance to a source, but not an independent value for the redshift at which
the source is located. However, it is possible that an electromagnetic
counterpart to an EMRI event will be observed (the sky position accuracy of a
few square degrees, although poor, is not beyond the reach of survey
telescopes). The EMRI will be observed for sufficiently long before plunge that
the plunge time can be accurately estimated, and advance warning of the moment
of coalescence supplied to other telescopes. However, it is not entirely clear
what mechanisms could give rise to an electromagnetic signature at coalescence
that would be sufficiently bright to be seen at cosmological distances. If we
were lucky, and a single EMRI event was observed with an electromagnetic
counterpart, it would provide an estimate of the Hubble constant that is not
tied to the local distance scale.  We could thus measure the Hubble constant to
$\sim 2\%$, compared to the current systematic error from the Hubble Key
Project of $\sim 10\%$. With $N$ observations, the error decreases like
$1/\sqrt{N}$.

The inspiral of a $10\,\Msun$ object into a $10^5-10^6\,\Msun$ MBH
should be detectable to a redshift of about $2$, corresponding to a
time when the Universe was less than 4\,Gyr old. Although MBH-MBH
mergers are visible to much larger redshift, offering the potential to
probe the whole early history of MBH formation and growth, there
are considerable uncertainties about their predicted detection rate
\cite{Haehnelt94,MHN01,wyithe03,sesana05}. Also the merger rate for
LISA-detectable systems probably peaks at $z>2$.  EMRIs therefore
offer an alternative and complementary way to probe the relatively
late evolution of MBHs with masses below a few $10^6\,\Msun$.

\subsection{How can EMRIs/IMRIs be used to test GR, or (assuming GR is correct)
that the central massive object is a Kerr BH?} As discussed above, what we can
perform are consistency checks (i.e., is the signal consistent with GR
predictions?) or compare the Kerr null hypothesis to straw man alternatives. In
the simplest case, if we find that the observed EMRI waveforms agree with the
models predicted by GR, for reasonable physical parameters, this would be an
impressive verification of the theory. Matching strong-field EMRI waveforms to
one cycle in $10^5$ will be compelling evidence in support of GR. Beyond this,
one can further check consistency in the EMRI by checking, for example, that
different pieces of the waveform yield the same consistent estimates for the
physical parameters of the system.  This amounts to verifying the consistency
of relativity in the strong field region in the immediate vicinity of MBHs, a
regime in which it has not been verified to date (though one hopes this regime
will be probed by LIGO before LISA flies, albeit with less accuracy). These
consistency checks are not testing the theory against an alternative; however,
this is familiar from other areas of physics. For instance, in particle
physics, tests of the standard model are not based on comparison with serious
rivals: one measures that the W mass or top quark mass is roughly where it is
predicted, and this is trumpeted as a substantial  validation of the theory.
Also, the idea of embedding GR/Kerr predictions in a somewhat larger,
phenomenological theory (e.g., with non-zero graviton mass, or non-standard
quadrupole moment for the BH), is similar to the current status of GR tests
with binary pulsars, in which the orbital motion is first fit to a
phenomenological set of Keplerian and post-Keplerian parameters, and radio
astronomers then show that the fitted values are consistent with the
predictions of GR (twenty years ago binary pulsar measurements killed a host of
alternatives to GR, but now that they are dead, radio astronomers are also
stuck with demonstrating consistency with GR). 

One important issue is whether, if systems that differ from
the Kerr hypothesis exist, we will actually be able to detect them, given that our data analysis will rely on matched
filtering. As mentioned above, if the deviations are small, this will not be a
problem, since the source will remain consistent with our templates for long
enough to be detected. The point at which deviations start to appear will then
be a probe of the nature of the deviations present (e.g., the
existence/location of a horizon as determined by the frequency of plunge, if
any). If the deviations are large, then matched filtering might fail entirely,
and we would have to rely on a source being close enough to show up in a
time-frequency analysis or other template-free technique.

If we see a single event that differs from the Kerr model, this will be very
weak evidence against the black hole hypothesis (no doubt many possible
explanations would appear in the literature over time). If every observation
differed, then the evidence would be difficult to refute. Explaining the
observations would be a difficult, but extremely interesting task.

\section{Conclusions} \label{summ} 

Black hole binaries with large mass ratios are uniquely important sources for
planned space-based gravitational wave detectors such as LISA.  These EMRIs or
IMRIs will provide information about the stellar dynamics of galactic nuclei
that will be difficult, if not impossible, to obtain any other way.  Single
events will yield precise measurements of the masses and spins of supermassive
black holes in a mass range extremely difficult to observe electromagnetically.
In addition, the high mass ratios mean that inspiral events will map the
space-time around black holes and test predictions of general relativity in the
strong field.  These enticing prospects have led to a recent surge in interest
in the astrophysical, general relativistic, and data analysis aspects of EMRIs
and IMRIs, and we have given here an overview the state of the art in these
three areas.  Although much analytical and numerical work remains, the level of
progress in the last few years suggests that when LISA flies in roughly a
decade, the community will be ready to maximise the scientific return from
observations of these events.

Currently, the most discussed mechanism for the production of EMRIs involves
the gradual evolution of the orbits of stellar-mass black holes and other
compact objects via two-body relaxation.  The estimated rates for a galaxy such
as the Milky Way are in the range of $\sim 10^{-8}-10^{-6}$~yr$^{-1}$.  The
further technical development of $N-$body codes will be essential to reducing the
uncertainty of these rates, as well as to proper inclusion of effects such as
mass segregation and resonant relaxation.  There are additional qualitatively
different mechanisms that have been proposed recently, including tidal
separation of binary stellar-mass black holes, and formation or capture of
black holes in accretion discs around the MBH, that could lead to an increase
in the estimated EMRI rate.  In addition, whereas standard EMRIs are likely to
have high eccentricity and random inclination in the LISA frequency range,
tidal separations would lead to circularised orbits with random inclination,
and disc processes to circularised orbits in the spin plane of the MBH.  The
distinct waveforms from these different mechanisms suggest that they will be
distinguishable in data, and hence will carry important information about
different properties in galactic nuclei.

Detection of EMRI and IMRI signals in the LISA data stream is a difficult task.
In Section~\ref{EMRIDA} we described three existing algorithms for EMRI
detection and discussed some of the outstanding issues in EMRI data analysis.
The current algorithms might be able to detect as many as several hundred EMRIs
in the LISA data stream --- the reach of the best searches (semi-coherent and
Markov Chain Monte Carlo) is out to $z\sim1-2$. However, the performance of
each of these algorithms has so far been analysed only for the detection of a
single EMRI in noisy data. The LISA data stream will be source-dominated, and
the need to simultaneously identify and extract all these signals puts severe
demands on data analysis algorithms. Understanding how to extract EMRIs in the
presence of source confusion is the key data analysis issue that must be
addressed in the future.

Several of the proposed data analysis algorithms employ some variant of matched
filtering, for which models of the signals present in the data must be known.
The extreme mass ratio means that accurate EMRI waveform templates can be
constructed using black hole perturbation theory. However, this is
computationally intensive. Various approximate EMRI waveform models have been
developed, and these were described in Section~\ref{EMRIModel}. Comparison to
perturbative results suggests that these models might be good enough for LISA
data analysis, but more work needs to be done. A key uncertainty is in the
computational costs. A plausible data analysis strategy would use an
approximate waveform model to get estimates for the source parameters, before
carrying out a follow up search with more accurate waveforms. This hierarchical
approach is subject to constraints on computing power. It may turn out that it
is not possible to constrain the source parameters with sufficient accuracy in
the first stage of the search to perform the follow-up search in a reasonable
time. If that is the case, we will end up with larger errors on observed source
parameters, but this needs to be quantified as waveform models are further
developed in the future. Modelling of IMRI waveforms has not yet been
considered in detail. While EMRI or comparable mass binary models or a
combination of the two might be applied, this needs further investigation.

One of the key science goals of EMRI observations is to test general relativity
theory in the strong field. The extreme mass ratio means that the inspiralling
object effectively acts like a test-body in the space-time of the central
object. The emitted gravitational waves encode a map of the space-time of the
central object. If we can decode that map, we will be able to tell to high
precision if the central body is indeed a Kerr black hole, or some other
object.  Understanding how LISA will be able to test relativity in practice is
a subject of much current research, which we summarised in
Section~\ref{testGR}. The main outstanding issues include producing waveforms
for generic inspirals in bumpy (modified Kerr) space-times to test detection and
data analysis strategies; finding efficient ways to search for deviations from
a Kerr black hole; and interpreting such deviations to determine their origin
(e.g., an accretion disc around a Kerr black hole, a boson star, a naked
singularity with non-Kerr higher-order multipole moments).

If we do manage to detect many EMRI and IMRI events with LISA, we stand to
learn a lot about astrophysics, and we summaries some of this discovery space
in Section~\ref{EMRIScience}. The EMRI eccentricity distribution can tell us
about the capture mechanisms, while the distribution of inclinations can shed
light on whether compact objects are formed in a disc around the central black
hole.  The distribution of compact object masses could provide information on
their populations, and any IMRI detection would be exciting in demonstrating
that intermediate-mass black holes exist.  Spin measurements of massive black
holes will enlighten our understanding of their formation history.  If any
electromagnetic counterparts to EMRIs are observed, it could yield an improved
measurement of the Hubble constant.  EMRIs could also be used to confirm
whether massive objects are indeed Kerr black holes, as generally assumed, and
to test strong-field general relativity.

\begin{acknowledgments}

This work is a result of the meeting {\tt LISA Astro-GR@AEI}$^{\dag}$ organised
by PAS which took place at the Max-Planck Institut f{\"u}r Gravitationsphysik
(Albert Einstein-Institut) from 18th-22nd September 2006.  We thank all
participants for the useful discussions during the meeting which made possible
this article.  The work of PAS has been supported in the framework of the Third
Level Agreement between the DFG and the IAC (Instituto de Astrof\'\i sica de
Canarias). JG's work was supported by St.Catharine's College. The work of MF is
funded through the PPARC rolling grant at the Institute of Astronomy (IoA) in
Cambridge. MCM acknowledges the National Science Foundation grant AST0607428.
IM would like to thank the Brinson Foundation, NASA grant NNG04GK98G and NSF
grant PHY-0601459 for financial support. The work of CC was carried out at the
Jet Propulsion Laboratory, California Institute of Technology and was sponsored
by the National Aeronautics and Space Administration. 

\noindent
PAS is indebted to all the members of the Astrophysical
Relativity group of the AEI and especially with Ute Schlichting for her help in
the organisation, which was crucial for the success of the meeting.
\vspace{0.3cm}
\\ $^{\dag}$For online talks and slides please see\\ {\tt http://www.aei.mpg.de/$\sim$pau/LISA\_Astro-GR@AEI}
\end{acknowledgments}

\end{document}